\documentclass[seceq]{ptptex}
\usepackage{wrapft}
\usepackage{graphicx}

\newcommand{\ket}[1]{| {#1} \rangle}     
\newcommand{\kket}[1]{| {#1} \rangle\!\rangle}     
\newcommand{\rket}[1]{| {#1} )}     
\newcommand{\dket}[1]{|\!| {#1} \rangle}     
\newcommand{\dkket}[1]{|\!| {#1} \rangle\!\rangle}     
\newcommand{\wtilde}[1]{\widetilde{#1}} 

\def\beq{\begin{eqnarray}}
\def\eeq{\end{eqnarray}}
\def\bsub{\begin{subequations}}
\def\esub{\end{subequations}}
\def\b{\begin{equation}}

\title{
Many-Quark Model with $su(4)$ Algebraic Structure
}
\subtitle{
An Example of Analytically Soluble Many-Fermion System
}    

\author{
Yasuhiko {\sc Tsue},$^{1}$ Constan\c{c}a {\sc Provid\^encia},$^{2}$ 
Jo\~ao da {\sc Provid\^encia}$^{2}$  and Masatoshi {\sc Yamamura}$^{3}$  
}

\inst{
$^{1}$Physics Division, Faculty of Science, Kochi University, Kochi 780-8520, 
Japan\\
$^{2}$Departamento de F\'{i}sica, Universidade de Coimbra, 3004-516 Coimbra, 
Portugal\\
$^{3}$Department of Pure and Applied Physics, 
Faculty of Engineering Science,\\
Kansai University, Suita 564-8680, Japan
\\

}

\recdate{
\today
}

\abst{
With the use of a certain type of the Schwinger boson representation of 
the $su(4)$ algebra, the Bonn model for many-quark systems and an 
extension of the model, in which the energies of 
the various exact states are controled, namely, the energies are enhanced or 
de-enhanced, are investigated. 
Especially, the triplet formation and the pairing correlation in 
many-quark states are discussed. 
In the boson space, all the results are analytically expressed in 
the exact forms. 
}

\begin{document}

\maketitle

\section{Introduction}

The description of multi-nucleon systems in terms of their QCD constituents 
is a topic of great current interest. 
Concerning that viewpoint, a model worthy mentioning is the many-quark 
model proposed by Petry et al., also known as the Bonn model, which 
describes the nucleus as a system of interacting quarks.\cite{a} 
An important ingredient in the Bonn model is an attractive pairing force, 
acting between quarks of different colors, that suppresses physically 
undesirable degeneracies of the many-quark system. 
This is very similar to the case of many nucleons. 
This model was originally devised as a model for the formation of color 
neutral triplets. 
This is quite remarkable since the involved interaction is a two-body force, 
which is naturally associated with two-body correlations, but not with 
three-body correlations. 
Under the name of the quark shell model, this model has accounted 
qualitatively for some features of nuclear physics and helped to understand 
certain properties of hadron physics connected with its symmetries. 
However, through the investigation of nuclear magnetic moments, it has been 
pointed out that the Bonn model in its most simple form is 
incompatible with the traditional treatment due to the lack of clustering 
effects and these effects are induced by configuration mixing in the 
$j$-$j$ coupling quark shell model.\cite{2,3} 
But, this statement does not mean that by this investigation, the framework 
of the Bonn model was rejected. 
In fact, Ref.\citen{2} mentions that, in any case, the quark shell model 
may give us a convenient and practical basis for treating the nucleus 
as a relativistic many-quark system.

The above argument makes us recognize that the original framework 
of the Bonn model does not disappear. 
However, we must mention that Petry et al. gave only the wave functions 
of the color neutral triplets and their energies. 
Therefore, two questions should be noted. 
Does the Bonn model generate the colored states or not~? 
If it 
generates them, how can the energies and the wave functions be expressed~? 
This is the first question. 
In QCD, there are no colored states. 
Therefore, this question is quite important. 
The second is related to the pairing correlation. 
The Bonn model was inspired by the famous seniority model of nuclear physics 
and, as was already mentioned, the quarks interact by a certain type 
of the pairing force. 
It indicates that the triplet formation appears as certain complicated 
superposition of the pairing correlation, and then, there may exist 
the pairing correlation which does not belong to the triplet formation. 
Especially, if the treatment for the Bonn model is limited to the 
triplets, we describe only the system in which the quark number is 
multiplet of 3. 
But, there is no necessity for the Bonn model to be restricted to such a 
system. 
By this reason, 
it is natural to investigate what types of the pairing 
correlations are generated by the Bonn model. 
Conventional seniority model is related to the $su(2)$ algebra and, 
as will be discussed in \S 2, the Bonn model is related to the 
$su(4)$ algebra. 
Therefore, it may be indispensable for the Bonn model to clarify 
in which aspects both pairing schemes resemble or are different from 
each other. 
This is the second question. 
In the study of the Bonn model, the interest lies in the triplet 
formation, but the pairing correlation, 
which does not belong to the triplet formation, for example, such as the 
quark-pair, 
is also interesting.

In addition to the Bonn model, we know that there exist at least three kinds 
of $su(4)$ algebraic models for many-fermion system: 
(1) Isospin vector and scalar pairing model for many-nucleons,\cite{b} 
(2) $su(4)$-algebraic model for high temperature superconductivity\cite{5} 
and (3) four-single particle level Lipkin model for many identical 
nucleons.\cite{6} 
In these three, (1) and (2) are the models extended from the 
$so(5)$-models for the isospin vector pairing correlation\cite{7} 
and for the high temperature superconductivity,\cite{8} respectively. 
The models (1) and (2) consist of four kinds of fermion-pair 
operators which are relevant to constructing the orthogonal set. 
For example, in the model (1), we have four kinds of nucleon-pairs 
(isospin $=1$ with $z$-component $-1,\ 0,\ 1$ and isospin $=0$). 
Therefore, mathematically, both are equivalent to each other. 
The model (3) is extended from the Lipkin model consisting of 
two-single particle levels.\cite{9} 
In this model, particle-hole pair-operators play a central role 
for constructing the orthogonal set and the number of the pair-operators 
depends on the choice of free vacuum, i.e., the lowest single-particle 
level is occupied or the lowest and the next one are occupied. 
The numbers of the pair-operators are 3 and 4, respectively.\cite{6} 
On the contrary, the Bonn model treats three kinds of the 
pair-operators composed of the quarks occupying colors 2 and 3, 3 and 1, and 
1 and 2. 
Therefore, the treatment becomes different from the above three.

It is well known that with the aid of boson operators, we can describe 
various phenomena of nuclear and hadron physics successfully and 
in many cases the boson operators are introduced through the idea of 
boson realization of Lie algebra if the original many-fermion system 
obeys a Lie algebra.\cite{10} 
The simplest one may be the Schwinger boson representation for the 
$su(2)$ algebra,\cite{11} which governs the seniority model. 
The $su(2)$ generators can be expressed in the bilinear forms for two 
kinds of bosons. 
The boson Hamiltonian obtained by the boson realization can be easily 
diagonalized and the energy eigenvalues and the eigenstates are obtained 
in analytically exact form. 
But, in the process of the boson realization, an unforgettable 
point is how to express the number of total available single-particle 
states and the fermion number in the boson space. 
We know that the obtained energy eigenvalues with their eigenstates serve us 
for understanding of physics of the seniority model, i.e., 
the $su(2)$ pairing model. 
Of course, they are used widely for testing certain classes of approximation 
technique.

For the boson realization of more complicated algebras, the present 
authors (J. da P. and M. Y.) with Kuriyama proposed the Schwinger boson 
representation for the $su(M+1)$ algebra in the frame of 
$(M+1)(N+1)$ kinds of boson operators.\cite{c} 
With the use of these bosons, we can construct the Schwinger boson 
representation of the $su(N,1)$ algebra and any $su(N,1)$ generator commutes 
with any $su(M+1)$ generator. 
Further, the present authors (M. Y.) with Kuriyama and Kunihiro 
proposed the Schwinger boson representation of the $su(4)$ algebra 
which are suitable for treating the $so(5)$, the $so(4)$ and the 
$su(2)\otimes su(2)$ algebras.\cite{13} 
This representation corresponds to the case $(M=3,\ N=1)$. 
The explicit form can be found in the relation (2$\cdot$17) of 
Ref.\citen{13}. 
With the aid of this representation, we will perform the boson realization 
of the Bonn model. 
Through the discussion in \S 2.3, the reason why we adopt this 
representation may be clear. 
Of course, in order to respond to the two questions already mentioned, 
we must give analytically exact solutions. 
The reason is simple: 
The energy eigenvalues of the color neutral triplets obtained by 
Petry et al. are analytically given and, further, we know the various 
expressions of the seniority model in analytical forms. 
We must compare our results with the above two cases. 
Through the process for obtaining analytical expressions, 
the Schwinger boson representation 
of the $su(1,1)$ algebra 
will play also a central role. 
The $su(1,1)$ algebraic model has been investigated as a typical 
example of quantum dissipative systems.\cite{14,e} 
This model has been also extensively investigated by the present 
authors (Y. T., J. da P. and M. Y.) with Kuriyama.\cite{16}

As was already mentioned, the Bonn model obeys the $su(4)$ algebra, 
which is composed of fifteen generators. 
They are three kinds of quark-pair creation operators, their annihilation 
operators and nine bilinear forms of single quark creation and 
annihilation. 
This algebra contains a sub-algebra: 
the $su(3)$ algebra. 
The Hamiltonian of the Bonn model is expressed in terms of simple 
sum of the products of the quark-pair creation and annihilation operators. 
It is characterized by the following points: 
The Hamiltonian consists of a kind of two-body force 
and any generator of the $su(3)$ (sub)algebra commutes with the Hamiltonian. 
In this paper, we treat not only the Bonn model but also a modified one. 
The modified Hamiltonian is expressed in the form of the Bonn model 
Hamiltonian plus the Casimir operator of the $su(3)$ algebra with arbitrary 
force strength. 
Even if this modification is done, the above-mentioned two charcteristic 
point are not altered. 
Through a boson realization of the $su(4)$ algebra shown in Eq.(2$\cdot$17)  
of Ref.\citen{13}, we obtain a boson Hamiltonian. 
By this realization, we can describe the case of full and partial symmetric 
representation for the $su(4)$ algebra. 
Including asymmetric representation, 
these two are defined in \S 2.1. 
Of course, this realization is of the Schwinger type. 
In addition to this realization, we also give the expressions 
for the number of total available single-particle states 
and the fermion number in the boson space. 
Furthermore, we make two re-formulations for the present formalism. 
One is a re-formulation from the side that all single-particle 
states are occupied by the quarks and the other is a rewriting 
of the present $su(4)$ algebraic system in terms of the 
Schwinger boson representation of the $su(1,1)$ algebras. 
With the aid of the form, we analyze the present system for 
the two cases: 
the triplet formation and the pairing correlation 
which does not belong to the triplet formation. 
Hereafter, we will call the second case simply as the pairing 
correlation. 
All the results are obtained in the analytically exact form.

Concerning the triplet formation, our main results are as follows: 
In our treatment, not only color neutral but also colored states appear, 
even if the new force is switched off and in certain conditions, energetically 
the colored states are lower than the color neutral ones. 
However, if the new force is switched on, the situations change. 
Energetically, the positions of the color neutral states do not change, but 
the colored states are influenced by the new force. 
Therefore, the position of the colored states are controlled by the 
force strength. 
This is interesting in relation to QCD. 
For the case of pairing correlation, also, 
the colored eigenstates often have 
the lower energies than those of the color neutral ones in the original Bonn 
model. 
However, the situation is similar to the case of the triplet formation 
as is mentioned above. 
Under a certain condition, the energy of the state, which is 
identical to the energy of the color neutral triplet one,  
is unchanged even if the new force is switched on. 
Then, the energy of colored states raises together with the new positive 
force strength. 
Thus, the color neutrality is retained energetically in 
our modified Bonn model. 
Furthermore, it is shown that 
the structure of the ground state in the case of pairing 
correlation is changed with respect to the particle number $N$. 
It seems that these phenomena indicate the phase change or phase transition.

This paper is organized as follows: 
In \S 2, a general framework of many-quark model with the $su(4)$ algebraic 
structure and its boson realization are given with some 
supplementary arguments. 
In \S 3, the triplet formation is treated with the help of the $su(1,1)$ 
algebraic framework. 
In \S 4, the case of the pairing correlation is discussed. 
In \S 5, various features obtained in \S\S 3 and 4 are presented 
with some numerical results. 
As a final remark, in \S 6, it is discussed that the asymmetric 
representation of the $su(4)$ algebra is meaningless 
in the case of the Bonn model and its modification. 
Finally, \S 7, future problem is mentioned.

\section{The $su(4)$ algebra for many-quark system}
\subsection{Basic framework}

As was mentioned in \S 1, there exist at least four forms of the 
$su(4)$ algebraic model for many-fermion systems. 
In this section, we formulate the $su(4)$ algebra in a form suitable 
for the description of a many-quark system with $su(3)$ color symmetry. 
More precisely, we will investigate the 
Bonn model by Petry et al. that describes the nucleus 
as a MIT bag\cite{a} and its possible modification.

This model is formulated in terms of the generators of 
the $su(4)$ algebra and is essentially equivalent to a three-level 
shell-model under a certain interaction among the constituents. 
The levels are specified as $i=1,2,3$, which denote colors 1,2,3.  
Each level has the degeneracy 
$2\Omega$ (here, $2\Omega=2j_s+1$ and $j_s$ is a half integer). 
An arbitrary single-particle state is specified as $(i,m)$, with 
$i=1,2,3$ and $m=-j_s,-j_s+1,\cdots,j_s-1,j_s$, and is created and 
annihilated by the fermion operators $c^*_{im},c_{im}.$ 
For simplicity, we neglect the degrees of freedom related to the 
isospin. 
We define 
the following bilinear forms:
\begin{eqnarray}\label{2-1}
&&\tilde S^1=\sum_m c^*_{2m}c^*_{3\tilde m},\quad \tilde
S^2=\sum_mc^*_{3 m}c^*_{1\tilde m},\quad \tilde
S^3=\sum_mc~^*_{1m}c^*_{2\tilde m},\nonumber\\
&&\nonumber\tilde S_1^2=\sum_mc^*_{2m}c_{1m},\quad \tilde
S_2^3=\sum_mc^*_{3m}c_{2m},\quad \tilde
S_3^1=\sum_mc^*_{1m}c_{3m}, \nonumber\\
&&\nonumber\tilde
S^1_1=\sum_m(c^*_{2m}c_{2m}+c^*_{3m}c_{3m})-2\Omega,\quad
\tilde S^2_2=\sum_m(c^*_{3m}c_{3m}+c^*_{1m}c_{1m})-2\Omega,\\
&&\nonumber\tilde
S^3_3=\sum_m(c^*_{1m}c_{1m}+c^*_{2m}c_{2m})-2\Omega,\quad\tilde
S_1=(\tilde S^{1})^*,\quad \tilde S_2=(\tilde S^{2})^*,\quad \tilde
S_3=(\tilde S^{3})^*,\\&& \tilde S^1_2=(\tilde S^2_1)^*,\quad \tilde
S^2_3=(\tilde S^3_2)^*,\quad\tilde S^3_1=(\tilde S^1_3)^*\label{1} \ .
\end{eqnarray}
Here, $c^*_{i\tilde m}=(-1)^{j_s-m}c^*_{i,~-m}.$ 
The operators in 
the definition (\ref{2-1}) are generators of the $su(4)$ algebra: 
\begin{eqnarray}\label{2-2}
&&
\tilde S^*_i=\tilde S^i,\quad (\tilde S^i_j)^*=\tilde S^j_i,\quad
\nonumber [\tilde S^i,\tilde S^j]=0,\quad  [\tilde S^i,\tilde
S_j]=\tilde S^j_i\\&&  [\tilde S^j_i,\tilde S^k]=\delta_{ij}\tilde
S^k+\delta_{jk}\tilde S^i, \quad  [\tilde S^j_i,\tilde
S^k_l]=\delta_{jl}\tilde S^k_i-\delta_{ik}\tilde S^j_l.
\end{eqnarray}
The Casimir operator ${\wtilde {\mib P}}^2$ for the $su(4)$ algebra reads 
\begin{eqnarray}\label{2-3}
{\wtilde {\mib P}}^2
&=&\sum_{i=1}^3\left({\wtilde S}^i{\wtilde S}_i+{\wtilde S}_i{\wtilde S}^i
\right)
+\sum_{i,j=1}^3{\wtilde S}_j^i{\wtilde S}_i^j
-\frac{1}{4}\left(\sum_{i=1}^3 {\wtilde S}_i^i\right)^2 \nonumber\\
&=&
2(\wtilde S^1\wtilde S_1+\wtilde S^2\wtilde S_2+\wtilde S^3\wtilde
S_3)+2(\wtilde S^2_1\wtilde S_2^1+\wtilde S^3_1\wtilde S_3^1+\wtilde
S^3_2\wtilde S_3^2)\nonumber\\
& &+(\wtilde S^1_1)^2+(\wtilde
S_2^2)^2+(\wtilde S^3_3)^2-\frac{1}{4}\left(\wtilde S^1_1+\wtilde S_2^2+\wtilde
S^3_3\right)^2+\left(-3\wtilde S^1_1-\wtilde S_2^2+\wtilde S^3_3\right)\ .\ \ 
\end{eqnarray}
The fermion number operators with $i=1,2,3$ read, respectively, 
\begin{subequations}\label{2-4}
\begin{eqnarray}\label{2-4a}
& &\wtilde N_1=\Omega-\frac{1}{2}
\left(\wtilde S^1_1-\wtilde S^2_2-\wtilde
S^3_3\right),\quad 
\wtilde N_2=\Omega-\frac{1}{2}\left(\wtilde S^2_2-\wtilde
S^3_3-\wtilde S^1_1\right),\nonumber\\ 
& &\wtilde N_3=\Omega-\frac{1}{2}\left(\wtilde
S^3_3-\wtilde S^1_1-\wtilde S^2_2\right)
\end{eqnarray}
and the total quark number is
\begin{equation}\label{2-4b}
\wtilde N= \wtilde N_1+ \wtilde N_2+ \wtilde
N_3=3\Omega+\frac{1}{2}\left(\wtilde S^1_1+\wtilde S^2_2+\wtilde
S^3_3\right).
\end{equation}
\end{subequations}
Conversely, the generators ${\wtilde S}_i^i$ are expressed as
\begin{equation}\label{2-5}
\wtilde S_1^1=\wtilde N_2+\wtilde N_3-2\Omega,\quad 
\wtilde S_2^2=\wtilde N_3+\wtilde
N_1-2\Omega,\quad 
\wtilde S_3^3=\wtilde N_1+\wtilde
N_2-2\Omega.
\end{equation}
As a sub-algebra, the $su(4)$ algebra contains the $su(3)$ algebra 
which is 
generated by
\begin{equation}\label{2-6}
\wtilde S_1^2,\;\wtilde S_2^1,\;\wtilde S_2^3,\;\wtilde S_3^2,\;\wtilde
S_3^1,\; \wtilde S_1^3,\;\frac{1}{2}(\wtilde S_2^2-\wtilde
S_3^3),\;\wtilde S_1^1-\frac{1}{2} (\wtilde S_2^2+\wtilde S_3^3). 
\end{equation}
The Casimir operator ${\wtilde {\mib Q}}^2$ for the $su(3)$ algebra reads 
\begin{eqnarray}\label{2-7}
{\wtilde {\mib Q}}^2&=&
\sum_{i\neq j}{\wtilde S}_j^i{\wtilde S}_i^j
+2\left(\frac{1}{2}\left({\wtilde S}_2^2-{\wtilde S}_3^3\right)\right)^2
+\frac{2}{3}\left({\wtilde S}_1^1-\frac{1}{2}\left({\wtilde S}_2^2
+{\wtilde S}_3^3\right)\right)^2 
\nonumber\\
&=&
2(\wtilde S^2_1\wtilde S^1_2+\wtilde S^3_1\wtilde S_3^1+\wtilde S^3_2\wtilde
S_3^2)+2\left(\left(\frac{1}{2}(\wtilde S^2_2-\wtilde S_3^3)\right)^2
-\frac{1}{2}(\wtilde
S^2_2-\wtilde S_3^3)\right)
\nonumber\\
& &+\frac{2}{3}\left(\wtilde
S_1^1-\frac{1}{2}(\wtilde S^2_2+\tilde S_3^3)\right)^2-2 \left(\wtilde
S_1^1-\frac{1}{2}(\wtilde S^2_2+\wtilde S_3^3)\right)\ .
\end{eqnarray}

The Bonn model is defined by the Hamiltonian 
\begin{equation}\label{2-8}
{\wtilde H}=-\left({\wtilde S}^1{\wtilde S}_1+{\wtilde S}^2{\wtilde S}_2
+{\wtilde S}^3{\wtilde S}_3\right) \ , 
\end{equation}
where the coupling constant has been omitted. 
Usually one consider $G{\wtilde H}$, ($G<0$). 
We observe that ${\wtilde H}$ is color neutral: 
\begin{equation}\label{2-9}
[\ {\wtilde H}\ , \ {\wtilde S}_j^i\ ]=0 \ , \qquad
i,j=1,2,3. 
\end{equation}
In this paper, by modifying ${\wtilde H}$, we also treat the following form: 
\begin{equation}\label{2-10}
{\wtilde H}_m={\wtilde H}+\chi{\wtilde {\mib Q}}^2\ . \qquad
(\chi\ : \ \hbox{\rm a\ real\ parameter})
\end{equation}
This modification conserves the characteristics of the Bonn model: 
(1) It obeys the $su(4)$ algebra. (2) The interaction is of the two body 
force with the pairing plus the particle-hole type. 
(3) It satisfies the same condition as that shown in the relation (\ref{2-9}): 
\begin{equation}\label{2-11}
[\ {\wtilde H}_m\ , \ {\wtilde S}_j^i\ ]=0 \ , \qquad
i,j=1,2,3. 
\end{equation}

The investigation of the present model requires the construction of 
an orthogonal set of states. 
For this aim, let us assume that there 
exists, in the fermion space, a unique state $|m)$ such that 
\begin{equation}\label{2-12} 
\wtilde S_i|m)=0,\quad i=1,2,3,\qquad \wtilde
S^1_2|m)= \tilde S^1_3|m)= \tilde S^2_3|m)=0.
\end{equation}
The state $|m)$ is called a minimum weight state. 
By acting with 
$\wtilde S^i_i$ on both sides of Eq.(\ref{2-12}) and keeping in mind 
the relation (\ref{2-2}) we find 
\begin{equation}\label{2-13}
\wtilde S_k\cdot\wtilde S^i_i|m)=0\ , \quad 
\wtilde S_k^l\cdot\wtilde
S^i_i|m)=0, \quad \hbox{\rm for}\ \ l<k\ .
\end{equation}
From the relation (\ref{2-13}) and the assumption that $|m)$ is unique, 
we find 
\begin{equation}\label{2-14}
\wtilde S^i_i|m)=-2\sigma_i|m),\quad i=1,2,3,
\end{equation}
where $\sigma_i$ denotes a $c$-number, so that $|m)$ is an 
eigenstate of ${\wtilde S}^i_i$. 
Clearly, $|m)$ satisfies Eq.(\ref{2-12}).

We observe that
\begin{eqnarray}
& &(m|\wtilde S_i\cdot \wtilde S^i|m)=(m|[\wtilde S_i,\wtilde
S^i]|m)=-(m|\wtilde S_i^i|m)=2\sigma_i\geq0\ , 
\label{2-15}\\
& &(m|\wtilde S_i^j\cdot \wtilde S^i_j|m)=(m|[\wtilde S_i^j,\wtilde
S^i_j]|m)=(m|\wtilde S_i^i- \wtilde
S_j^j|m)=2(\sigma_i-\sigma_j)\geq 0 ,
\nonumber\\
& &\quad\quad\quad\quad\quad\quad\quad{\rm~~for~~}
i>j. 
\label{2-16}
\end{eqnarray}
It follows that 
\begin{equation}\label{2-17}
\sigma_1\geq\sigma_2\geq\sigma_3\geq 0\ .
\end{equation}
Obviously, $\sigma_i=0$ if $\wtilde S_i^i|m)=0$ and 
$\sigma_1>\sigma_2=\sigma_3$ if $\wtilde S^3_2|m)=0$. 
Moreover, 
$\sigma_1=\sigma_2=\sigma_3$ if $\wtilde S^2_1|m)=\wtilde
S^3_1|m)=\wtilde S^3_2|m)=0.$ 
From the relation (\ref{2-4a}), it follows that the 
number $n_i$ of quarks with color $i$ in $|m)$ reads 
\begin{equation}\label{2-18}
n_1=\Omega+\sigma_1-\sigma_2-\sigma_3,\quad
n_2=\Omega+\sigma_2-\sigma_3-\sigma_1,\quad
n_3=\Omega+\sigma_3-\sigma_1-\sigma_2.
\end{equation}
Conversely, it follows that 
\begin{equation}\label{2-19}
\sigma_1=\Omega-{1\over2}(n_2+n_3),\quad
\sigma_2=\Omega-{1\over2}(n_3+n_1),\quad
\sigma_3=\Omega-{1\over2}(n_1+n_2).\quad
\end{equation}
From the relations (\ref{2-17}) and (\ref{2-18}) we obtain 
\begin{equation}\label{2-20}
n_1\geq n_2\geq n_3,\qquad
{1\over2}(n_1+n_2)\leq\Omega .
\end{equation}
It is well known that the case 
$\sigma_1=\sigma_2=\sigma_3\;(n_1=n_2=n_3)$ is associated with the full 
symmetric representation of the $su(4)$ algebra. 
In this paper, besides the case $\sigma_1=\sigma_2=\sigma_3$, we will 
discuss the case $\sigma_1 > \sigma_2=\sigma_3$ ($n_1 > n_2=n_3$), 
which we call partial symmetric representation. 
At several occasions, we will call these two the symmetric representation 
collectively. 
The case $\sigma_1 > \sigma_2 > \sigma_3$ will be called the 
asymmetric representation. 
The reason why we restrict ourselves to the symmetric representation 
will be mentioned in \S 2.3. 
The following notation is introduced:
\begin{eqnarray}
& &|m)=|m_1),~~{\rm
if}~~\sigma_2=\sigma_3=\sigma_0,\quad{\rm i.e.},\quad
n_2=n_3=n_0, 
\label{2-21}\\
& & |m)=|m_0),~~{\rm
if}~~\sigma_1=\sigma_2=\sigma_3=\sigma_0,\quad{\rm i.e.} ,\quad
n_1=n_2=n_3=n_0. 
\label{2-22}
\end{eqnarray}
The states $|m_0)$ and $|m_1)$ satisfy
\begin{eqnarray}
& &\wtilde S_1|m_1)=\wtilde S_2|m_1)=\wtilde S_3|m_1)=0,\quad
\wtilde S_2^1|m_1)=\wtilde S_3^1|m_1)=\wtilde S_3^2|m_1)=\wtilde S_2^3|m_1)
=0,\nonumber\\
& &\wtilde S_1^1|m_1)=-2\sigma_1|m_1) \ , 
\qquad
\wtilde S_2^2|m_1)=\wtilde S_3^3|m_1)=-2\sigma_0|m_1),
\label{2-23}\\
& &\wtilde S_1|m_0)=\wtilde S_2|m_0)=\wtilde S_3|m_0)=0,\quad
\wtilde
S_2^1|m_0)=\wtilde S_3^1|m_0)=\wtilde S_3^2|m_0)=0,\nonumber\\
& &
\wtilde S_1^2|m_0)=\wtilde S_1^3|m_0)=\wtilde S_2^3|m_0)=0,\quad
\wtilde
S_1^1|m_0)=\wtilde S_2^2|m_0)=\wtilde S_3^3|m_0)=-2\sigma_0|m_0).\qquad
\label{2-24}
\end{eqnarray}
Therefore, we have 
\begin{eqnarray}
& &{\wtilde {\mib P}}^2\rket{m_1}=
(3\sigma_1^2-4\sigma_1\sigma_0+4\sigma_0^2+6\sigma_1)|m_1)\ , 
\label{2-25}\\
& &\wtilde{\mib Q}^2|m_1)=
\frac{4}{3}(\sigma_1-\sigma_0)(2(\sigma_1-\sigma_0)+3)|m_1). 
\label{2-26}
\end{eqnarray}

In the next section, following Ref.\citen{b}, we will introduce a 
boson realization of the $su(4)$ algebra, of the Schwinger type. 
Correspondingly, operators and state vectors will be denoted as 
$\hat O$ and $|\phi\rangle$, instead of the notation previously 
adopted, $\wtilde O$ and $|\phi)$.

\subsection{A Schwinger boson realization}

We can prove that the following set of operators obeys the $su(4)$ algebra: 
\begin{equation}\label{2-27}
\hat S^i=\hat a^*_i\hat b-\hat a^*\hat b_i,\quad
\hat S_i=\hat b^*\hat a_i-\hat b^*_i\hat a,\quad \hat S^j_i=(\hat
a^*_i\hat a_j-\hat b^*_j\hat b_i)+\delta_{ij}(\hat a^*\hat a-\hat
b^*\hat b),
\end{equation}
where $\hat a_i, \hat a_i^*, \hat b_i, \hat b_i^*,  \hat a,\hat
a^*,\hat b,\hat b^*\ (i=1,2,3)$ denote boson operators. 
The form (\ref{2-27}) is obtained by replacing $({\hat a},{\hat a}^*)$ 
in the relation (2$\cdot$17) of Ref.\citen{13} with 
($-{\hat a}, -{\hat a}^*$). 
In \S 2.3, the reason why we adopt this form will be clear. 
As will be shown in the relations (\ref{2-31}) and (\ref{2-32}), 
the form (\ref{2-27}) enables us to describe the case of the 
symmetric representation. 
Associated with the form (\ref{2-27}), we define the 
$su(1,1)$ algebra: 
\begin{eqnarray}\label{2-28}
& &\hat T_+=\hat b^*\hat a^*+\sum_{i=1}^3\hat b_i^*\hat
a_i^*,\quad \hat T_-=\hat a\hat b+\sum_{i=1}^3\hat a_i\hat b_i,\nonumber\\
& &\hat T_0={1\over2}(\hat b^*\hat b+\hat a^*\hat
a)+{1\over2}\sum_{i=1}^3(\hat b_i^*\hat b_i+\hat a_i^*\hat a_i)+2,
\end{eqnarray}
which satisfy 
\begin{equation}\label{2-29}
[\hat T_+,\hat T_-]=-2\hat T_0, \quad [\hat T_0,\hat T_\pm]=\hat
T_\pm.
\end{equation}
We have 
\begin{equation}\label{2-30}
[\hat T_\mu,\hat S_i]=[\hat T_\mu,\hat S^i]=[\hat T_\mu,\hat
S_i^j]=0,\quad\mu=\pm,\ 0 \ .
\end{equation}
This representation belongs to a special class of the Schwinger 
representation for the $su(M+1)$ and the $su(N,1)$ algebra 
($M=3, N=1$).\cite{c}
In the Schwinger representation, the counterpart of
the state $|m_1)$ shown in the relation (\ref{2-21}) is the state
\begin{equation}\label{2-31}
|m_1\rangle=(\hat b^*_1)^{2(\sigma_1-\sigma_0)}(\hat
b^*)^{2\sigma_0}|0\rangle
\end{equation}
and the counterpart of the state $|m_0)$ in the relation 
(\ref{2-22}) is the state
\begin{equation}\label{2-32}
|m_0\rangle=(\hat
b^*)^{2\sigma_0}|0\rangle.
\end{equation}
These states are minimum weight states of the $su(4)$ and of the
$su(1,1)$ algebras:
\begin{eqnarray}
& &\hat T_-|m_1\rangle=0,\qquad
\hat
T_0|m_1\rangle=(\sigma_1+2)|m_1\rangle,
\label{2-33}\\
& &\hat T_-|m_0\rangle=0,\qquad\hat
T_0|m_0\rangle=(\sigma_0+2)|m_0\rangle.
\label{2-34}
\end{eqnarray}
In this paper, we will omit any numerical factor such as normalization 
constant for any state.

In connection with the application of the Schwinger representation to 
the description of a many-fermion system such as the Bonn model, the 
problem arises of the identification of the level degeneracy 
$\Omega$ and the fermion number $N$. 
Here, $\Omega$ is related to the number of the total available 
single-particle states. 
This point has been stressed in \S 1. 
The case of the $su(2)$ algebra 
is instructive. 
The Schwinger representation of the $su(2)$ algebra reads 
\begin{equation}\label{2-35}
\hat S_+=\hat a^*\hat b,\qquad \hat S_-=\hat b^*\hat
a,\qquad \hat S_0={1\over2}(\hat a^*\hat a-\hat b^*\hat b).
\end{equation}
The minimum weight state may be expressed as 
\begin{equation}\label{2-36}
\hat S_-|m\rangle=0,\qquad \hat S_0|m\rangle=-S|m\rangle,
\qquad |m\rangle=(\hat
b^*)^{2S}|0\rangle.
\end{equation}
The fermion pairing model is expressed in terms of the $su(2)$
generators 
\begin{equation}\label{2-37}
\wtilde S_+=\frac{1}{2}\sum_mc_m^*c_{\tilde m}^*,
\qquad 
\wtilde S_-=\frac{1}{2}\sum_mc_{\tilde
m}c_m,\qquad 
\wtilde
S_0=\frac{1}{2}\sum_m(c_m^*c_{m}-\Omega),  
\end{equation}
and its minimum weight state satisfies 
\begin{equation}\label{2-38}
\wtilde S_-|m)=0, \qquad \wtilde
S_0|m)=-S|m).
\end{equation}
From the form (\ref{2-37}) it follows that the fermion number operator reads 
\begin{equation}\label{2-39}
\wtilde N=\Omega+2\wtilde S_0,
\end{equation}
so that 
\begin{equation}\label{2-40}
\wtilde N|m)=n_0|m)=(\Omega-2S)|m).
\end{equation}
From Eq.(\ref{2-40}) we have 
\begin{equation}\label{2-41}
n_0=\Omega-2S,\qquad S={1\over2}(\Omega-n_0),\qquad (n_0\leq
\Omega), 
\end{equation}
where $n_0$ is the seniority number of the pairing model.

In the fermion space, the level degeneracy $\Omega$ is a parameter 
of the model. 
However, in the boson space it is associated with an 
operator denoted $\hat\Omega$ which should commute with 
$\hat S_+,\hat S_-,\hat S_0.$ 
It is natural to postulate 
\begin{equation}\label{2-42}
\hat\Omega=x+y(\hat a^*\hat a+\hat b^*\hat b),
\end{equation}
where $x,y$ are $c$-numbers. 
From the expression (\ref{2-39}) 
it is natural to write the fermion number operator as 
\begin{equation}\label{2-43}
\hat N=\hat \Omega+2\hat S_0=x+(y+1)\hat a^*\hat
a+(y-1)\hat b^*\hat b.
\end{equation}
Acting with $\hat N,\hat \Omega$ on $|m\rangle$ we find 
\begin{equation}\label{2-44}
\hat\Omega|m\rangle=\Omega|m\rangle=(x+2yS)|m\rangle,\qquad
\hat N|m\rangle=n_0|m\rangle=(x+2(y-1)S)|m\rangle,
\end{equation}
which implies 
\begin{equation}\label{2-45}
S={1\over2}(\Omega-n_0),\qquad x=(1-y)\Omega+yn_0,
\end{equation}
which recovers the relation (\ref{2-41}). 
Thus, the form (\ref{2-42}) reduces 
\begin{equation}\label{2-46}
\hat \Omega=(1-y)\Omega+y(n_0+\hat a^*\hat a+\hat b^*\hat
b). 
\end{equation}
It is not natural that the operator 
$\hat\Omega$ should depend on its eigenvalue. 
Therefore, we set 
$y=1,\;x=n_0,$ which leads to 
\begin{equation}\label{2-47}
\hat \Omega=n_0+\hat a^*\hat a+\hat b^*\hat b,
\qquad \hat N=n_0+2\hat a^*\hat a.
\end{equation}

We return now to the Bonn model, searching for the operators 
associated with the level degeneracy and the fermion number. 
Operating with $\wtilde N_i$ in the expression 
(\ref{2-4a}) on $|m_1)$, we find 
$n_1=\Omega+(\sigma_1-2\sigma_0),\;n_0=\Omega-\sigma_1,$ so that 
\begin{equation}\label{2-48}
\sigma_1=\Omega-n_0,\qquad \sigma_0=\Omega-{1\over2}(n_0+n_1).
\end{equation}
In the boson space, it may be natural to postulate 
\begin{equation}\label{2-49}
\hat \Omega=x+{y\over2}\left(\sum_i(\hat a^*_i
\hat a_i+\hat b^*_i\hat b_i)+\hat a^*\hat a+\hat b^*\hat
b\right).
\end{equation}
This operator commutes with all generators 
in the form (\ref{2-27}). 
By analogy with the form (\ref{2-4a}), we write 
\begin{equation}\label{2-50}
\hat N_i=\hat \Omega-\hat S_i^i+{1\over2}\sum_{j=1}^3\hat
S_j^j,\qquad i=1,2,3.
\end{equation}
Thus, we have 
\begin{eqnarray}\label{2-51}
\hat N_i&=&x+\frac{y}{2}\left(\sum_{j=1}^3(\hat a^*_j
\hat a_j+\hat b^*_i\hat b_i)+\hat a^*\hat a+\hat b^*\hat
b\right)
\nonumber\\
& &-(\hat a^*_i\hat a_i-\hat b^*_i\hat
b_i)+\frac{1}{2}(\hat a^*\hat a-\hat b^*\hat
b)+\frac{1}{2}\sum_{j=1}^3(\hat a^*_j\hat
a_j-\hat b^*_j\hat b_j
),\quad i=1,2,3 .\ \ 
\end{eqnarray}
Acting with $\hat \Omega,\;\hat N_i$ on the state 
$|m_1\rangle$ we obtain $\Omega=x+\sigma_1,\;
n_1=x+(y+1)\sigma_1-2\sigma_0,\;n_0=x+(y-1)\sigma_1$, so that 
\begin{equation}\label{2-52}
\sigma_1=\Omega-n_0,\qquad\sigma_0=\Omega-{1\over2}(n_0+n_1),\qquad
x=(1-y)\Omega+y n_0.
\end{equation}
In order that $\hat\Omega$ does not depend on its eigenvalue 
$\Omega$, we set $y=1,$ thus obtaining, 
\begin{equation}\label{2-53}
\hat \Omega=n_0+\frac{1}{2}\left(\sum_{i=1}^3(\hat a^*_i
\hat a_i+\hat b^*_i\hat b_i)+\hat a^*\hat a+\hat b^*\hat
b\right).
\end{equation}
The number of color $i$ quarks and the total 
quark number read 
\begin{eqnarray}
& &\hat N_i=n_0+\hat a^*\hat a+(\hat b^*_i\hat b_i-\hat
a^*_i\hat a _i)+\sum_{j=1}^3\hat a^*_j\hat a_j,
\label{2-54}\\
& &\hat N=3n_0+3\hat a^*\hat a+2\sum_{j=1}^3\hat a^*_j\hat
a_j+\sum_{j=1}^3\hat b^*_j\hat b_j.
\label{2-55}
\end{eqnarray}
We observe that the bosons $\hat a^*,\;\hat a^*_i$ and $\hat b^*_i$ 
carry the fermion numbers 3, 2 and 1, respectively, while the boson 
$\hat b^*$ carries fermion number 0. 
Then, we may classify the 
present system into two cases. 
In case (i), $\sum_i\hat b^*_i\hat a^*_i$ and $\hat b^*\hat a^*$ carry the 
fermion number 3. 
In case (ii), $\hat S^i=\hat a_i^*\hat b-\hat a^*\hat b_i$ 
carries the fermion number 2. 
Therefore, the frameworks given in the cases (i) and (ii) may be 
convenient for the descriptions of the triplet formation and 
the pairing correlation, respectively.

\subsection{Reformulation of the $su(4)$ algebra: hole picture}

Let us consider the operators
\begin{equation}\label{2-56}
\breve{S}^i=-\tilde S_i,\quad \breve{S}^j_i=-\tilde
S^i_j,\qquad i,j=1,2,3.
\end{equation}
It may be easily seen that the set 
$\breve{S}_i,\breve{S}^i,\breve{S}_i^j,$ also generates a $su(4)$ 
algebra. 
We also introduce the operators 
\begin{equation}\label{2-57}
\breve{N}_i=\Omega-\breve{S}_i^i+{1\over2}\sum_{j=1}^3\breve{S}_j^j,\qquad
i=1,2,3.
\end{equation}
From the relation (\ref{2-56}), it follows that 
\begin{equation}\label{2-58}
\breve{N}_i=2\Omega-\wtilde{N}_i,\qquad
i=1,2,3.
\end{equation}
%
%
Here, $\breve{N}_i$ is identical to 
the hole number operator with color index $i$.
Clearly, the total hole number 
operator reads 
\begin{equation}\label{2-59}
\breve{N}=\sum_{i=1}^3\breve{N}_i=3\Omega+{1\over2}\sum_{i=1}^3\breve{S}_i^i.
\end{equation}
The Hamiltonian ${\wtilde H}_m$ satisfies 
\begin{equation}\label{2-60}
\wtilde{H}_m=\breve{H}_m+\breve{K},\qquad
\breve{H}_m=-\sum_{i=1}^3\breve{S}^i\breve{S}_i+\chi\breve{\mib Q}^2, \qquad
\breve{K}=\sum_{i=1}^3\breve{S}_i^i=2(\breve{N}-3\Omega).
\end{equation}
The reformulated $su(4)$ generators allow the description of the 
present model in terms of the hole picture.

It is instructive to discuss the $su(2)$ pairing model defined by 
the relation (\ref{2-37}), in the hole picture. 
In this case, we have 
\begin{equation}\label{2-61}
\breve{S}_+=-\wtilde{S}_-,\qquad
\breve{S}_-=-\wtilde{S}_+,\qquad
\breve{S}_0=-\wtilde{S}_0={1\over2}(\breve{N}-\Omega)
\end{equation}
and
\begin{equation}\label{2-62}
\breve{N}=2\Omega-\wtilde{N}.
\end{equation}
The Hamiltonian $\wtilde H$ reads 
\begin{equation}\label{2-63}
\wtilde{H}=\breve{H}+\breve{K}=-\wtilde{S}_+\wtilde{S}_-,\qquad
\breve{H}=-\breve{S}_+\breve{S}_-,\qquad
\breve{K}=2\breve{S}_0=\breve{N}_0-\Omega.
\end{equation}
The eigenvalues and eigenstates of $\wtilde{H}$ read 
\begin{eqnarray}\label{2-64}
& &E_{N,n_0}=-{1\over2}(N-n_0)\left(
\Omega-{1\over2}n_0+1-{1\over2}N\right) \ , 
\nonumber\\
& &|N,n_0)=(\wtilde{S}_+)^{(N-n_0)/2}|n_0) \ , 
\end{eqnarray}
where $|n_0)=|m)$, for  $|m)$ the minimum weight state defined in
the form (\ref{2-38}). 
Here, $N$ and $n_0$ denote, respectively, the particle 
and seniority numbers. 
In the hole picture, the analogous result reads 
\begin{eqnarray}\label{2-65}
& &|N',n_0))=(\breve{S}_+)^{(N'-n_0)/2}|n_0)),\quad
\breve{S}_-|n_0))=0,\quad
\breve{S}_0|n_0))=\frac{1}{2}(\Omega-n_0)|n_0)),\nonumber\\
& &{\cal E}_{N',n_0}=-\frac{1}{2}(N'-n_0)
\left(\Omega-{1\over2}n_0+1-{1\over2}N'\right)+(N'-\Omega).
\end{eqnarray}
Here, $N'$ denotes the hole number, $N'=2\Omega-N$. 
It follows that 
\begin{equation}\label{2-66}
{\cal E}_{N',n_0}=E_{N,n_0}.
\end{equation}
The above result is not surprising because $|n_0))$ plays the role 
of the maximum weight state of the $su(2)$ algebra, since 
$\tilde{S}_+|n_0))=0,\;\tilde{S}_0|n_0))=\frac{1}{2}(\Omega-n_0)|n_0))=0
$. 
One might conjecture that an analogous result holds for the present 
$su(4)$ model. 
Starting from the $N=0$ side or from the $N=6\Omega$ 
side one would arrive at equivalent results. 
In \S\S 3 and 4 we will show that this is not true.

The present reformulation has been performed under the replacement
\begin{equation}\label{2-67}
c^*_{im}=d_{i\tilde m}=(-1)^{j_s-m}d_{i ~-m},
\end{equation}
where $d_{i m},d^*_{i\tilde m}$ denote hole operators. 
The Schwinger boson representation reads 
\begin{eqnarray}\label{2-68}
& &\check{S}^i=-\hat{S}_i=\hat{b}^*_i\hat{a}-\hat{b}^*\hat{a}_i , \nonumber\\
& &\check{S}^j_i=-\hat{S}_j^i=\hat{b}^*_i\hat{b}_j-\hat{a}^*_j\hat{a}_i
+\delta_{ij}(\hat{b}^*\hat{b}-\hat{a}^*\hat{a}).
\end{eqnarray}
It is found that in the Schwinger boson representation the role of 
the $\hat{a}$-type and $\hat{b}$-type bosons is reversed. 
The reason why we adopt the form (\ref{2-27}) comes from the property 
(\ref{2-68}). 
Of course, the form (\ref{2-27}) corresponds to the case 
($M=3, N=1$) in Ref.\citen{c}. 
The hole number operators read 
\begin{equation}\label{2-69}
\check{N}_i=\hat{\Omega}-\check{S}_i^i+\frac{1}{2}\sum_{j=1}^3\check{S}_j^j,
\qquad
i=1,2,3 
\end{equation}
or, using the relation (\ref{2-53}), 
\begin{equation}\label{2-70}
\check{N}_i=n_0+\hat{b}^*\hat{b}+(\hat{a}^*_i\hat{a}_i-\hat{b}^*_i\hat{b}_i)
+\sum_{j=1}^3
\hat{b}^*_j\hat{b}_j,\qquad i=1,2,3.
\end{equation}
The total hole number operator reads 
\begin{equation}\label{2-71}
\check{N}=3n_0+3\hat{b}^*\hat{b}+\sum_{j=1}^3\hat{a}^*_j\hat{a}_j
+2\sum_{j=1}^3
\hat{b}^*_j\hat{b}_j,\qquad i=1,2,3.
\end{equation}
Again, the reversion of the role of the 
$\hat{a}$-type and $\hat{b}$-type  bosons is found.

The reformulated version of 
\begin{equation}
\hat{H}_m=
-\sum_{j=1}^3\hat{S}^j\hat{S}_j+\chi{\hat {\mib Q}}^2
\nonumber
\end{equation}
reads 
\begin{equation}\label{2-72}
\hat{H}_m=\check{H}_m+\check{K},
\qquad
\check{H}_m=
-\sum_{j=1}^3\check{S}^j\check{S}_j+\chi\check{\mib Q}^2,\qquad
\check{K}=\sum_{j=1}^3\check{S}^j_j=2(\check{N}-3\hat{\Omega}).
\end{equation}
We observe that in the relations (\ref{2-71}) and (\ref{2-55}), 
the roles of 
the boson operators ${\hat a}$, ${\hat b}$ are reversed. 
It is easily verified that $\check{\mib Q}^2$ is equal to 
${\hat {\mib Q}}^2$.

\subsection{Two $su(1,1)$ algebras}

We pay attention to the relation (\ref{2-11})((\ref{2-9})). 
This relation suggests that ${\hat H}_m\ ({\hat H})$ may be a function 
of certain sets of operators which commute with ${\hat S}_j^i$. 
As for these operators, we introduce 
\begin{eqnarray}
& &{\hat t}_+={\hat b}^*{\hat a}^* \ , \quad
{\hat t}_-={\hat a}{\hat b} \ , \quad 
{\hat t}_0=\frac{1}{2}({\hat b}^*{\hat b}+{\hat a}^*{\hat a})+\frac{1}{2} \ ,
\label{2-73}\\
& &{\hat \tau}_+=\sum_{i=1}^3{\hat b}_i^*{\hat a}_i^* \ , \quad
{\hat \tau}_-=\sum_{i=1}^3{\hat a}_i{\hat b}_i \ , \quad 
{\hat \tau}_0=\frac{1}{2}\sum_{i=1}^3
({\hat b}_i^*{\hat b}_i+{\hat a}_i^*{\hat a}_i)+\frac{3}{2} \ .
\label{2-74}
\end{eqnarray}
In association with the above, further, we introduce 
\begin{eqnarray}
& &{\hat t}^0=\frac{1}{2}({\hat b}^*{\hat b}-{\hat a}^*{\hat a})+\frac{1}{2}
\ , \label{2-75}\\
& &{\hat \tau}^0=\frac{1}{2}\sum_{i=1}^3
({\hat b}_i^*{\hat b}_i-{\hat a}_i^*{\hat a}_i)+\frac{3}{2}
\ . \label{2-76}
\end{eqnarray}
It is easily verified that the above operators commute with 
${\hat S}_j^i$ defined in the relation (\ref{2-27}), and further, 
it may be interesting to see that ${\hat t}^0$ and ${\hat \tau}^0$ 
commute with ${\hat t}_\mu$ and ${\hat \tau}_\mu$, $\mu=\pm,0$. 
Since $\{{\hat t}_\mu\}$ and $\{{\hat \tau}_\mu\}$ satisfy the 
commutation relations as those shown in the relation (\ref{2-29}), 
they form $su(1,1)$ algebras and 
$\{{\hat T}_\mu\}$ defined in the relation (\ref{2-28}) is expressed as 
\begin{equation}\label{2-77}
{\hat T}_\mu ={\hat t}_\mu +{\hat \tau}_\mu \ , \qquad
\mu=\pm, 0 \ . 
\end{equation}
The Casimir operators for these $su(1,1)$ generators are given as 
\begin{eqnarray}
& &{\hat {\mib t}}^2={\hat t}_0^2-{\hat t}_0-{\hat t}_+{\hat t}_- \ , 
\label{2-78}\\
& &{\hat {\mib \tau}}^2={\hat \tau}_0^2-{\hat \tau}_0
-{\hat \tau}_+{\hat \tau}_- \ , 
\label{2-79}\\
& &{\hat {\mib T}}^2={\hat T}_0^2-{\hat T}_0-{\hat T}_+{\hat T}_- \ . 
\label{2-80}
\end{eqnarray}

With the use of the above $su(1,1)$-generators, ${\hat H}$ and 
${\hat {\mib Q}}^2$ can be expressed in the form 
\begin{eqnarray}
& &{\hat H}=-({\hat {\mib T}}^2-{\hat {\mib t}}^2-{\hat {\mib \tau}}^2)
+2{\hat t}^0{\hat \tau}^0 \ , 
\label{2-81}\\
& &{\hat {\mib Q}}^2=2{\hat {\mib \tau}}^2+\frac{2}{3}({{\hat \tau}^0})^2
-2{\hat \tau}^0 \ . 
\label{2-82}
\end{eqnarray}
Certainly, ${\hat H}_m\ ({\hat H})$ are expressed in terms of the 
operators which commute with ${\hat S}_j^i$. 
From the above argument, our problem reduces to the eigenvalue problem 
of the $su(1,1)$ algebras. 
Of course, the $su(3)$ algebra plays also a central role in our problem.

Now, let us determine the minimum weight state $\ket{m_2}$. 
The state $\ket{m_2}$, at least, should satisfy the conditions 
\begin{eqnarray}
& &{\hat S}_2^1\ket{m_2}={\hat S}_3^1\ket{m_2}={\hat S}_3^2\ket{m_2}=0 \ , 
\qquad {\hat \tau}_-\ket{m_2}=0 \ , 
\label{2-83}\\
& &{\hat t}_-\ket{m_2}=0 \ . 
\label{2-84}
\end{eqnarray}
As a possible choice, we adopt the following form :
\begin{equation}\label{2-85}
\ket{m_2}=({\hat a}_3^*)^{2\lambda}({\hat b}_1^*)^{2\tau-3-2\lambda}
({\hat b}^*)^{2t-1}\ket{0} \ .
\end{equation}
Then, we have 
\begin{eqnarray}
& &\frac{1}{2}({\hat S}_2^2-{\hat S}_3^3)\ket{m_2}=-\lambda\ket{m_2} \ , 
\nonumber\\
& &\left({\hat S}_1^1-\frac{1}{2}({\hat S}_2^2+{\hat S}_3^3)\right)\ket{m_2}
=-((2\tau-3)-\lambda)\ket{m_2} \ , \nonumber\\
& &{\hat \tau}_0\ket{m_2}=\tau\ket{m_2} \ , 
\label{2-86}\\
& &{\hat t}_0\ket{m_2}=t\ket{m_2} \ . 
\label{2-87}
\end{eqnarray}
Certainly, $\ket{m_2}$ is an eigenstate of 
$({\hat S}_2^2-{\hat S}_3^3)/2$, ${\hat S}_1^1-({\hat S}_2^2+{\hat S}_3^3)/2$, 
${\hat \tau}_0$ and ${\hat t}_0$. 
Here, $t,\ \tau$ and $\lambda$ take the values 
\begin{eqnarray}\label{2-88}
& &t=\frac{1}{2},\ 1,\ \frac{3}{2},\ \cdots \ , \nonumber\\
& &\tau=\frac{3}{2},\ 2,\ \frac{5}{2},\ \cdots \ , \nonumber\\
& &\lambda=0,\ \frac{1}{2},\ 1,\ \cdots , \tau-\frac{3}{2} \ .
\end{eqnarray}
Further, we have 
\begin{eqnarray}
& &{\hat \tau}^0\ket{m_2}=(\tau-2\lambda)\ket{m_2} \ , 
\label{2-89}\\
& &{\hat t}^0\ket{m_2}=t\ket{m_2} \ . 
\label{2-90}
\end{eqnarray}
However, we should note that there exists another possibility for the choice 
of $\ket{m_2}$:
\begin{equation}\label{2-91}
\ket{m_2}=({\hat a}_3^*)^{2\lambda}({\hat b}_1^*)^{2\tau-3-2\lambda}
({\hat a}^*)^{2t-1}\ket{0} \ .
\end{equation}
This form also satisfies the condition (\ref{2-83}) and (\ref{2-84}). 
The condition discriminating from the form (\ref{2-85}) is as follows: 
\begin{equation}\label{2-92}
{\hat t}^0\ket{m_2}=(1-t)\ket{m_2} \ . 
\end{equation}
In the case $t=1/2$, both coincide with each other. 
Then, hereafter, we use the notations $\ket{m_2;b}$ and $\ket{m_2;a}$ 
for the forms (\ref{2-85}) and (\ref{2-91}), respectively, 
if necessary. 
As is mentioned above, there are two minimum weight states, 
$\ket{m_2;b}$ and $\ket{m_2;a}$. 
This situation is originated from that of the $su(1,1)$ algebraic 
structure in the Schwinger representation, 
which will be investigated in the next section (see, \S\S3.1 and 
3.3). 
It was pointed out, at the first time, that the two minimum weight states 
exist and both are necessary to describe the system under consideration 
completely in our previous paper.\cite{e}

For $\ket{m_2}$, the quark numbers in the color $i=1, 2, 3$ are given as 
\begin{equation}\label{2-93}
n_1=n_0+(2\tau-3)\ , \qquad n_2=n_0+2\lambda\ , \qquad
n_3=n_0 \ . 
\end{equation}
The relation (\ref{2-88}) gives us 
$n_1 \geq n_2 \geq n_3$.

\section{Triplet formation}

\subsection{Reformulation in terms of the $su(1,1)$ algebra}

In this section, we will investigate the case in which the fermion numbers 
change by units of 3 (triplet type). 
To reach this goal, the Schwinger boson representation is essential, 
and its reformulation developed in \S 2.4 plays a relevant role. 
The reason is as follows: 
As was already mentioned in \S 2.2, the expression of ${\hat N}$ 
(the relation (\ref{2-55})) 
suggests us that ${\hat t}_+={\hat b}^*{\hat a}^*$ and 
${\hat \tau}_+=\sum_i {\hat b}_i^*{\hat a}_i^*$ carry the fermion number 3. 
In \S 2.4, we presented a possible form of the minimum weight state 
$\ket{m_2}$ shown in the expression (\ref{2-85}). 
This state can be decomposed to 
\begin{eqnarray}
& &\ket{m_2;b}=\dket{\lambda,\tau}\otimes \kket{t;b} \ , 
\label{3-1}\\
& &\dket{\lambda,\tau}=({\hat a}_3^*)^{2\lambda}
({\hat b}_1^*)^{2\tau-3-2\lambda}\ket{0}=({\hat a}_3^*)^{2\lambda}
\dket{\tau-\lambda} \ , 
\label{3-2}\\
& &\kket{t;b}=\kket{t}=({\hat b}^*)^{2t-1}\ket{0} \ .
\label{3-3}
\end{eqnarray}
We can treat the whole space as the tensor product of two separate spaces, 
the $t$-space, constructed from ${\hat b}^*$, ${\hat a}^*$, and 
the $\tau$-space, constructed from ${\hat b}_1^*$, ${\hat a}_1^*$, 
${\hat b}_2^*$, ${\hat a}_2^*$, ${\hat b}_3^*$, ${\hat a}_3^*$. 
We call the whole space T-space.

The orthogonal set in $t$-space is easily obtained as the eigenstates of 
${\hat {\mib t}}^2$ and ${\hat t}_0$ constructed on $\kket{t}$, 
which reads 
\begin{eqnarray}
& &\kket{tt_0}=({\hat t}_+)^{t_0-t}\kket{t} \ , \qquad
{\hat t}_-\kket{t}=0 \ , \qquad
{\hat t}_0\kket{t}=t\kket{t} \ , 
\label{3-4}\\
& &{\hat {\mib t}}^2\kket{tt_0}=t(t-1)\kket{tt_0} \ , \qquad
{\hat t}_0\kket{tt_0}=t_0\kket{tt_0} \ , 
\label{3-5}\\
& &{\hat t}^0\kket{tt_0}=t\kket{tt_0} \ .
\label{3-6}
\end{eqnarray}
Here, 
\begin{equation}\label{3-7}
t=\frac{1}{2},\ 1,\ \frac{3}{2},\ \cdots , \qquad 
t_0=t,\ t+1,\ t+2,\cdots .
\end{equation}
The above treats the eigenvalue problem in $t$-space.

The eigenvalue problem in $\tau$-space is much more complicated 
because it involves six kinds of bosons. 
Then, for its description, six mutually commuting hermitian operators are 
needed. 
The operators ${\hat {\mib \tau}}^2$, ${\hat \tau}_0$ and ${\hat \tau}^0$ 
are natural candidates. 
Further, we may include ${\hat {\mib R}}^2$, ${\hat R}_0$ and 
${\hat Q}_0$ defined in Appendix A. 
The Casimir operator ${\hat {\mib Q}}^2$ is not needed because of relation 
(\ref{2-82}). 
Noting the state (\ref{a7}), we introduce the state 
\begin{eqnarray}
& &\dket{\lambda\mu\nu\nu_0,\tau\tau_0}={\hat Q}_+(\lambda\mu\nu\nu_0)
\dket{\lambda,\tau\tau_0} \ , 
\label{3-8}\\
& &\dket{\lambda,\tau\tau_0}=({\hat \tau}_+)^{\tau_0-\tau}
\dket{\lambda,\tau} \ . \quad (\tau_0=\tau,\ \tau+1,\cdots )
\end{eqnarray}
The state $\dket{\lambda\mu\nu\nu_0,\tau\tau_0}$ ($=\dket{\tilde \tau}$) 
satisfies 
\begin{eqnarray}
& &{\hat {\mib \tau}}^2\dket{\tilde \tau}=\tau(\tau-1)\dket{\tilde \tau} \ , 
\qquad
{\hat \tau}_0\dket{\tilde \tau}=\tau_0\dket{\tilde \tau} \ , 
\label{3-10}\\
& &
{\hat \tau}^0\dket{\tilde \tau}=(\tau-2\lambda)\dket{\tilde \tau} \ , 
\label{3-11}\\
& &{\hat {\mib R}}^2\dket{\tilde \tau}=\nu(\nu+1)\dket{\tilde \tau} \ , 
\qquad
{\hat R}_0\dket{\tilde \tau}=\nu_0\dket{\tilde \tau} \ , 
\label{3-12}\\
& &
{\hat Q}^0\dket{\tilde \tau}=[3\mu-((2\tau-3)-\lambda)]\dket{\tilde \tau} \ , 
\label{3-13}
\end{eqnarray}
Comparison with the relation (\ref{a12}) gives us 
\begin{equation}\label{3-14}
\kappa=(2\tau-3)-\lambda \ .
\end{equation}
The extra quantum number $\alpha$ is in the present case $\alpha=\tau_0$. 
Operation of ${\hat {\mib Q}}^2$ on $\dket{\tilde \tau}$ is as follows: 
\begin{eqnarray}\label{3-15}
{\hat {\mib Q}}^2\dket{\tilde \tau}&=&
\left[2\tau(\tau-1)+\frac{2}{3}(\tau-2\lambda)
(\tau-2\lambda -3)\right]\dket{\tilde \tau} \nonumber\\
&=&\left[2\lambda(\lambda+1)+\frac{2}{3}\kappa(\kappa+3)\right]
\dket{\tilde \tau} \ .
\end{eqnarray}
We enumerate conditions which the quantum numbers specifying 
$\dket{\lambda\mu\nu\nu_0,\tau\tau_0}$ obey. 
The Clebsch-Gordan coefficient $\langle \lambda\lambda_0\mu\mu_0
\ket{\nu\nu_0}$ gives 
\begin{equation}\label{3-16}
|\lambda-\mu| \leq \nu \leq \lambda+\mu \ , \qquad
-\nu \leq \nu_0 \leq \nu \ .
\end{equation}
Further, we have 
\begin{equation}\label{3-17}
2(\lambda+\mu)\leq 2\tau-3 
\ . 
\end{equation}
In addition to the former shown already in the relation (\ref{2-88}), 
the later is also necessary. 
The proof is given in Appendix B. 
Concerning $\tau$ and $\tau_0$, the conditions are shown in the 
relations (\ref{2-88}) and (\ref{3-7}).

Finally, we consider the orthogonal set in T-space. 
Following Ref.\citen{d}, we investigate the eigenvalue problem for 
${\hat T}_+$, ${\hat T}_-$, ${\hat T}_0$, which reduces essentially to 
the addition of two $su(1,1)$ spins. 
Let us assume that we know the minimum weight state 
$\dkket{\lambda\mu\nu\nu_0;t\tau T}$. 
Then, we have 
\begin{eqnarray}
& &\dkket{\lambda\mu\nu\nu_0;t\tau TT_0}=({\hat T}_+)^{T_0-T}
\dkket{\lambda\mu\nu\nu_0;t\tau T} \ , 
\label{3-18}\\
& &{\hat {\mib T}}^2
\dkket{\lambda\mu\nu\nu_0;t\tau TT_0}
=T(T-1)\dkket{\lambda\mu\nu\nu_0;t\tau TT_0} \ , 
\label{3-19}\\
& &{\hat {T}}_0
\dkket{\lambda\mu\nu\nu_0;t\tau TT_0}
=T_0\dkket{\lambda\mu\nu\nu_0;t\tau TT_0} \ , 
\label{3-20}\\
& &\qquad 
T=t+\tau,\ t+\tau+1,\ t+\tau+2, \cdots \ , 
\quad
T_0=T,\ T+1,\ T+2,\cdots \ .
\label{3-21}
\end{eqnarray}
The minimum weight state for ${\hat T}_+$, ${\hat T}_-$, ${\hat T}_0$ 
may be expressed as 
\begin{eqnarray}\label{3-22}
\dkket{\lambda\mu\nu\nu_0;t\tau T}&=&
({\hat O}_+(t\tau))^{T-(t+\tau)}{\hat Q}_+(\lambda\mu\nu\nu_0)
\dket{\lambda,\tau}\otimes \kket{t} \ , 
\nonumber\\
{\hat O}_+(t\tau)&=&
{\hat t}_+({\hat t}_0+t+\epsilon)^{-1}-{\hat \tau}_+
({\hat \tau}_0+\tau+\epsilon)^{-1} \ , 
\end{eqnarray}
where $\epsilon$ is an infinitesimal parameter such that 
$({\hat t}_0+t+\epsilon)^{-1}$ and $({\hat \tau}_0+\tau+\epsilon)^{-1}$ 
exist. 
By expanding $({\hat O}_+(t\tau))^{T-(t+\tau)}$, we obtain 
\begin{eqnarray}\label{3-23}
& &\dkket{\lambda\mu\nu\nu_0;t\tau T}\nonumber\\
&=&\sum_{t_0\tau_0}{}'\frac{(-1)^{\tau_0-\tau}\Gamma(T-(t+\tau)+1)\Gamma(2t)
\Gamma(2\tau)}{\Gamma(t_0-t+1)\Gamma(\tau_0-\tau+1)
\Gamma(t_0+t)\Gamma(\tau_0+\tau)}
\dket{\lambda\mu\nu\nu_0,\tau\tau_0}\otimes \kket{tt_0} \ , \quad
\end{eqnarray}
where $\sum'_{t_0\tau_0}$ means that the sum is restricted to 
$t_0=t, \ t+1,\cdots$ and $\tau_0=\tau,\ \tau+1,\cdots$ under the 
condition $t_0+\tau_0=T$ for a given $T$.

The state (\ref{3-8}) can be rewritten down as 
\begin{eqnarray}
& &\dkket{\lambda\mu\nu\nu_0;t\tau TT_0}=
({\hat T}_+)^{T_0-T}{\hat Q}_+(\lambda\mu\nu\nu_0)\dkket{\lambda\tau tT} \ , 
\label{3-24}\\
& &\dkket{\lambda\tau tT}=({\hat O}_+(t\tau))^{T-(t+\tau)}
\dket{\lambda,\tau}\otimes \kket{t}\ . 
\label{3-25}
\end{eqnarray}

As is clear from the forms (\ref{2-81}) and (\ref{2-82}), 
${\hat H}_m$ can be expressed in terms of ${\hat {\mib T}}^2$, 
${\hat {\mib t}}^2$, ${\hat {\mib \tau}}^2$, ${\hat t}^0$ and 
${\hat \tau}^0$, which commute with ${\hat T}_+$ and 
${\hat Q}_+(\lambda\mu\nu\nu_0)$.
Therefore, in order to get the eigenvalues of ${\hat H}_m$, it may be 
enough to take into account only $\dkket{\lambda\tau tT}$. 
However, the following point should be pointed out: 
We are interested in the operators which have their counterparts 
in the original fermion space. 
Such operators commute with 
$\{{\hat T}_\mu\}$. 
Then, the orthogonal set, which treats such operators, is closed 
in the sub-space, for example, with the condition $T_0=T$. 
This means that it may be enough to investigate the present system 
in the frame of 
$\{ \dkket{\lambda\mu\nu\nu_0;t\tau T}={\hat Q}_+(\lambda\mu\nu\nu_0)
\dkket{\lambda\tau tT}\}$. 
Since ${\hat H}_m$ commutes with ${\hat Q}_+(\lambda\mu\nu\nu_0)$, 
the eigenvalue of ${\hat H}_m$ is treated in the frame of 
$\{ \dkket{\lambda\tau tT}\}$.

\subsection{Energy eigenvalue}

Following the procedure discussed in \S 3.1, we investigate the energy 
eigenvalue obtained by ${\hat H}_m$. 
First, we note that the eigenvalues of ${\hat {\mib T}}^2$, 
${\hat {\mib t}}^2$, ${\hat t}^0$, ${\hat {\mib \tau}}^2$ and ${\hat \tau}^0$ 
for $\dkket{\lambda\tau tT}$ are given as 
\begin{equation}\label{3-26}
{\hat {\mib T}}^2 : \ T(T-1) \ , \quad 
{\hat {\mib t}}^2 : \ t(t-1) \ , \quad 
{\hat t}^0 : \ t \ , \quad 
{\hat {\mib \tau}}^2 : \ \tau(\tau-1) \ , \quad 
{\hat \tau}^0 : \ \tau-2\lambda \ .
\end{equation}
Therefore, the eigenvalue of ${\hat H}_m$, $E_{Tt\tau\lambda}^{(m)}$, is 
given in the form 
\begin{equation}\label{3-27}
E_{Tt\tau\lambda}^{(m)}=E_{Tt\tau\lambda}+\chi F_{\tau\lambda}^{(t)} \ , 
\qquad\qquad\qquad\qquad\qquad\qquad\qquad\quad
\end{equation}
\vspace{-0.8cm}
\begin{subequations}\label{3-28}
\begin{eqnarray}
& &{\hat H}\dkket{Tt\tau\lambda}=E_{Tt\tau\lambda}\dkket{Tt\tau\lambda} \ ,
\label{3-28a}\\
& &{\hat {\mib Q}}^2\dkket{Tt\tau\lambda}=F_{\tau\lambda}^{(t)}
\dkket{Tt\tau\lambda} \ , 
\qquad\qquad\qquad\qquad\qquad\qquad\qquad
\label{3-28b}
\end{eqnarray}
\end{subequations}
\vspace{-0.8cm}
\begin{subequations}\label{3-29}
\begin{eqnarray}
E_{Tt\tau\lambda}&=&-(T(T-1)-t(t-1)-\tau(\tau-1))+2t(\tau-2\lambda) 
\nonumber\\
&=&-(T-t-\tau)(T+t+\tau-1)-4t\lambda\ , 
\label{3-29a}\\
F_{\tau\lambda}^{(t)}&=&
2\tau(\tau-1)+\frac{2}{3}(\tau-2\lambda)(\tau-2\lambda-3) \nonumber\\
&=&2\lambda(\lambda+1)+\frac{2}{3}((2\tau-3)-\lambda)((2\tau-3)
-\lambda+3) \ . 
\label{3-29b}
\end{eqnarray}
\end{subequations}
Here, we used the relations (\ref{2-81}), (\ref{2-82}), (\ref{3-25}) 
and (\ref{3-26}).

Next, we rewrite $E_{Tt\tau\lambda}$ in terms of $\Omega$ and $N$. 
From the relations (\ref{2-53}) and (\ref{2-55}) with (\ref{2-28}), 
(\ref{2-75}) and (\ref{2-76}), we find the following relations: 
\begin{eqnarray}
& &{\hat \Omega}=n_0-2+{\hat T}_0 \ , 
\label{3-30}\\
& &{\hat N}=3n_0-3+3{\hat T}_0-3{\hat t}^0-{\hat \tau}^0 \ , 
\label{3-31}
\end{eqnarray}
so that 
\begin{eqnarray}
& &{\hat T}_0={\hat \Omega}+2-n_0 \ , 
\label{3-32}\\
& &{\hat t}^0={\hat \Omega}+1-\frac{1}{3}{\hat \tau}^0-\frac{1}{3}{\hat N}
\ . 
\label{3-33}
\end{eqnarray}
Operation of $\dkket{\lambda\tau tT}$ on ${\hat \Omega}$ and ${\hat N}$ 
leads us to 
\begin{eqnarray}
& &{T}={\Omega}+2-n_0 \ , 
\label{3-34}\\
& &{t}={\Omega}+1-\frac{1}{3}({\tau}-2\lambda)-\frac{1}{3}{N}
\ . 
\label{3-35}
\end{eqnarray}
By substituting the relations (\ref{3-34}) and (\ref{3-35}) into 
the expression (\ref{3-29a}), $E_{Tt\tau\lambda}$ can be rewritten as 
\begin{eqnarray}\label{3-36}
E_{Tt\tau\lambda}&=&
-\left(\frac{1}{3}N-n_0-\frac{1}{3}(2\tau-3)-\frac{2}{3}\lambda\right) 
\nonumber\\
& &\times \left(2\Omega+3-\frac{1}{3}N-n_0+\frac{1}{3}(2\tau-3)+
\frac{2}{3}\lambda\right)
\nonumber\\
& &-4\lambda\left(\Omega+\frac{1}{2}-\frac{1}{3}N-\frac{1}{6}
(2\tau-3)+\frac{2}{3}\lambda\right) \ .
\end{eqnarray}
With the use of the relations (\ref{3-27}), (\ref{3-29b}) and (\ref{3-36}), 
the energy eigenvalue $E_{Tt\tau\lambda}^{(m)}$ is obtained.

In the quantum numbers specifying $E_{Nn_0\tau\lambda}^{(m)}$, 
we know that $\lambda$ can take the values shown in the relation (\ref{2-88}), 
but, we do not know which values are permitted in the cases $N$, $n_0$ 
and $\tau$. 
We will discuss this problem. 
The relation (\ref{3-34}) shows that $T$ is an integer, because 
$\Omega$ and $n_0$ are integers, and, further the relation (\ref{3-21}) 
teaches the relation $T-(t+\tau)=$integer. 
Therefore, for $t$ and $\tau$, the following two cases are permitted: 
(1) both are half-integers and (2) both are integers. 
Since $T \geq t+\tau$, the relations (\ref{3-34}) and (\ref{3-35}) give us 
\begin{equation}\label{3-37}
N \geq N_{\rm min} \ , \qquad N_{\rm min}=3n_0+(2\tau-3)+2\lambda \ . 
\end{equation}
The term $(2\tau-3)$ is integer, and then, $N_{\rm min}$ is an integer. 
From the relations (\ref{3-34}) and (\ref{3-35}), we can derive 
\begin{equation}\label{3-38}
N=N_{\rm min}+3k \ , \qquad
k=\Omega-n_0-(t+\tau)+2 \ . 
\end{equation}
Since $t+\tau=$integer, certainly, $k$ is integer. 
The relation (\ref{3-38}) indicates that $N$ increases from $N_{\rm min}$ 
in unit 3. 
This suggests us that the Hamiltonian ${\hat H}_m$ generates the 
triplet formation. 
There exists the upper limit of $N$, which we denote 
$N_{\rm max'}$ ($k_{\rm max'}$). 
The reason why we use $N_{\rm max'}$ $(k_{\rm max'}$) instead of 
$N_{\rm max}$ $(k_{\rm max}$) will be later clarified. 
In the case (1), the condition $t\geq 1/2$ and the relation (\ref{3-35}) 
give us 
\begin{subequations}\label{3-39}
\begin{eqnarray}
& &N\leq N_{\rm max'} \ , \qquad N_{\rm max'}=3\Omega-\left(\tau-\frac{3}{2}
\right)+2\lambda\ , 
\label{3-39a}\\
& &k\leq k_{\rm max'} \ , \qquad k_{\rm max'}=\Omega-n_0-
\left(\tau-\frac{3}{2}\right)\ . 
\label{3-39b}
\end{eqnarray}
\end{subequations}
In the case (2), the condition $t\geq 1$ and the relation (\ref{3-35}) lead 
us to the relations 
\begin{subequations}\label{3-40}
\begin{eqnarray}
& &N\leq N_{\rm max'} \ , \qquad N_{\rm max'}=3\Omega-\tau+2\lambda\ , 
\label{3-40a}\\
& &k\leq k_{\rm max'} \ , \qquad k_{\rm max'}=\Omega-n_0-
(\tau-1)\ . 
\label{3-40b}
\end{eqnarray}
\end{subequations}
Since $k_{\rm max'}\geq 0$, $\tau$ is restricted to 
\begin{subequations}\label{3-41}
\begin{eqnarray}
& &(1)\ \ \frac{3}{2} \leq \tau \leq \tau_{\rm max'} \ , 
\qquad \tau_{\rm max'}=\Omega-n_0+\frac{3}{2}\ , 
\label{3-41a}\\
& &(2)\ \ 2 \leq \tau \leq \tau_{\rm max'} \ , 
\qquad \tau_{\rm max'}=\Omega-n_0+1\ . 
\label{3-41b}
\end{eqnarray}
\end{subequations}
Of course, (1) $3/2 \leq\tau_{\rm max'}$ and (2) $2\leq\tau_{\rm max'}$ 
and we have 
\begin{subequations}\label{3-42}
\begin{eqnarray}
& &(1)\ \ n_0 \leq\Omega \ , 
\label{3-42a}\\
& &(2)\ \ n_0 \leq\Omega -1\ . 
\label{3-42b}
\end{eqnarray}
\end{subequations}

\subsection{The energy eigenvalue: supplementary development}

We observe that the state (\ref{3-23}) depends on the bosons 
$({\hat a}, {\hat a}^*)$ and $({\hat b}, {\hat b}^*)$ only 
through $\kket{tt_0}$, which has been introduced through 
the relations (\ref{3-4})$\sim$(\ref{3-7}). 
By interchanging $({\hat a}, {\hat a}^*)$ with 
$({\hat b}, {\hat b}^*)$, we are led to introduce the state 
\begin{equation}\label{3-43}
\kket{t;a}=\kket{1-t}=({\hat a}^*)^{2t-1}\kket{0} \ . 
\end{equation}
The reason for the notation $\kket{1-t}$ will become clear later. 
Let us consider the state 
\begin{equation}\label{3-44}
\kket{1-t\ t_0}
=({\hat t}_+)^{t_0-t}\kket{1-t} \ , \qquad
{\hat t}_-\kket{1-t}=0 \ , \qquad
{\hat t}_0\kket{1-t}=t\kket{1-t} \ , 
\end{equation}
which satisfies 
\begin{eqnarray}
& &{\hat {\mib t}}^2\kket{1-t\ t_0}=t(t-1)\kket{1-t\ t_0}\ , \qquad
{\hat {t}}_0\kket{1-t\ t_0}=t_0\kket{1-t\ t_0}\ , 
\label{3-45}\\
& &{\hat {t}}^0\kket{1-t\ t_0}=(1-t)\kket{1-t\ t_0}\ . 
\label{3-46}
\end{eqnarray}
We observe that 
\begin{equation}\label{3-47}
t(t-1)=(1-t)((1-t)-1) \ , \qquad
\kket{1-t}|_{t=\frac{1}{2}}=\kket{0} \ .
\end{equation}
Since ${\hat t}^0\kket{tt_0}=t\kket{tt_0}$, we have 
$\langle\!\langle tt_0\kket{1-t\ t_0}=0$ for $t> 1/2$. 
We present now a reformulation of the $su(4)$ many fermion algebra based on 
$\kket{1-t\ t_0}$. 
The new development is obtained from the one of the previous section 
through the replacement of $t$ by $(1-t)$. 
We observe, moreover, that $({\hat a}, {\hat a}^*)$ participates in the 
formalism on an equal footing with $({\hat b}, {\hat b}^*)$, 
so that, any new result obtained from a previous one by interchanging 
$({\hat a}, {\hat a}^*)$ with 
$({\hat b}, {\hat b}^*)$, is also valid.

Replacing $t$ by $(1-t)$ in $E_{Tt\tau\lambda}$ (Eq.(\ref{3-29a})), we find 
\begin{equation}\label{3-48}
E_{T1-t\tau\lambda}=-(T+t-\tau-1)(T-t+\tau)-4(1-t)\lambda \ .
\end{equation}
Clearly, Eq.(\ref{3-34}) remains unchanged, while, if in Eq.(\ref{3-35}) we 
replace $t$ by $(1-t)$, we find 
\begin{equation}\label{3-49}
t=-\Omega+\frac{1}{3}(\tau-2\lambda)+\frac{1}{3}N \ .
\end{equation}
Using the relations (\ref{3-34}) and (\ref{3-44}) to replace $T$ and $t$ 
in the result (\ref{3-48}), 
an expression is obtained for $E_{T1-t\tau\lambda}$, in terms of 
$N$, $n_0$, $\lambda$ and $\tau$. 
The result is the same as $E_{Tt\tau\lambda}$ shown in the 
relation (\ref{3-36}). 
Therefore, the energy eigenvalue $E_{T1-t\tau\lambda}^{(m)}$ is 
the same as $E_{Tt\tau\lambda}^{(m)}$.

Our next task is to rewrite $E_{T1-t\tau\lambda}$ in terms of 
$\Omega$ and $N$. 
The idea is the same as the previous one. 
The cases (1) and (2) are unchanged for $t$ and $\tau$: 
(1) both are half-integers and (2) both are integers. 
From the condition $T \geq t+\tau$, we have 
\begin{equation}\label{3-50}
N \leq N_{\rm max} \ , \qquad 
N_{\rm max}=6\Omega-3n_0-2(2\tau-3)+2\lambda \ . 
\end{equation}
With the use of the integer $k$ given in the relation (\ref{3-38}), 
$N$ can be expressed as 
\begin{equation}\label{3-51}
N=N_{\rm max}-3k \ .
\end{equation}
The quantity $N_{\rm max}$ is an integer and $N$ decreases from $N_{\rm max}$ 
in unit 3, which suggests the triplet formation. 
In this case, there exists the lower limit of $N$, which we denote 
$N_{\rm min'}$. 
In the case (1), $N_{\rm min'}$ is determined by the condition $t\geq 1/2$:
\begin{equation}\label{3-52}
N\geq N_{\rm min'} \ , \qquad 
N_{\rm min'}=3\Omega-\left(\tau-\frac{3}{2}\right)+2\lambda\ . 
\end{equation}
It should be noted that we have $N_{\rm max'}=N_{\rm min'}$ and $k$, 
which gives $N_{\rm min'}$, is equal to $k_{\rm max'}$ shown in the relation 
(\ref{3-39b}). 
The relation $N_{\rm min'}=N_{\rm max'}$ gives us the reason why we used 
the notations $N_{\rm max'}$ and $N_{\rm min'}$. 
At the point giving $N_{\rm min'}=N_{\rm max'}$, 
$\kket{1/2;b}=\kket{1/2;a}=\ket{0}$ and we 
can observe that at this point $\ket{m_2}$ defined in the form 
(\ref{2-85}) changes to $\ket{m_2}$ defined in the form (\ref{2-91}). 
In the case (2), the condition $t\geq 1$ gives us 
\begin{equation}\label{3-53}
N\geq N_{\rm min'} \ , \qquad
N_{\rm min'}=3\Omega-\tau+2\lambda+3\ . 
\end{equation}
In this case, also, it should be noted that $N_{\rm max'}=N_{\rm min'}-3$ 
and $k$, which gives $N_{\rm min'}$, is equal to $k_{\rm max'}$ shown 
in the relation (\ref{3-40b}). 
At the points $N=N_{\rm max'}$ and $N=N_{\rm min'}$, we have 
$\kket{1;b}={\hat b}^*\ket{0}$ and $\kket{1;a}={\hat a}^*\ket{0}$, 
respectively. 
In this case, also, we can see the change of $\ket{m_2}$.

Since both ranges intersect, it follows that the energy eigenvalue is valid 
in the whole range 
\begin{equation}\label{3-54}
3n_0+(2\tau-3)+2\lambda \leq N \leq 6\Omega-3n_0-2(2\tau-3)+2\lambda \ .
\end{equation}
The energy eigenvalue $E_{Nn_0\tau\lambda}^{(m)}=E_{Nn_0\tau\lambda}
+\chi F_{\tau\lambda}^{(t)}$ is given as 
\begin{subequations}\label{3-55}
\begin{eqnarray}
E_{Nn_0\tau\lambda}&=&
-\left(\frac{1}{3}N-n_0-\frac{1}{3}(2\tau-3)-\frac{2}{3}\lambda\right) 
\nonumber\\
& &\times \left(2\Omega+3-\frac{1}{3}N-n_0+\frac{1}{3}(2\tau-3)
+\frac{2}{3}\lambda\right) \nonumber\\
& &-4\lambda\left(\Omega+\frac{1}{2}-\frac{1}{3}N-\frac{1}{6}(2\tau-3)
+\frac{2}{3}\lambda\right) \ , 
\label{3-55a}\\
F_{\tau\lambda}^{(t)}&=&
2\lambda(\lambda+1)+\frac{2}{3}\left((2\tau-3)-\lambda\right)\left(
(2\tau-3)-\lambda+3\right) \ . 
\label{3-55b}
\end{eqnarray}
\end{subequations}

However, the whole story is not yet finished, 
since the range (\ref{3-54}) does 
not cover all possible $N$ values for given $n_0$, $\tau$ and $\lambda$. 
Let us study our system starting from the side 
$N=6\Omega$, as is presented in \S 2.3. 
The reformulation is straightforward. 
The energy eigenvalue reads 
\begin{equation}\label{3-56}
{\cal E}_{N'n_0\tau\lambda}^{(m)}=E_{N'n_0\tau\lambda}^{(m)}
+2(N'-3\Omega) \ , 
\qquad
N'=6\Omega-N \ . 
\end{equation}
Therefore, ${\cal E}_{N'n_0\tau\lambda}^{(m)}
=E_{6\Omega-Nn_0\tau\lambda}^{(m)}+2(3\Omega-N)$ is also an eigenvalue 
of the $N$ quark system. 
Replacing in the range (\ref{3-54}) $N$ by $N'$, we find 
\begin{equation}\label{3-57}
3n_0+(2\tau-3)+2\lambda \leq N'= 6\Omega-N \leq 
6\Omega-3n_0-2(2\tau-3)+2\lambda \ .
\end{equation}
Thus, the eigenvalue of the type ${\cal E}_{6\Omega-Nn_0\tau\lambda}^{(m)}$ 
occurs for 
\begin{equation}\label{3-57-2}
3n_0+2(2\tau-3)-2\lambda \leq N \leq 6\Omega-3n_0-(2\tau-3)-2\lambda \ .
\end{equation}

The quantities appearing in the relations (\ref{3-54}) and (\ref{3-57-2}) 
obey the same conditions as those shown in the relations 
(\ref{3-41}) and (\ref{3-42}): 
\begin{eqnarray}\label{3-58}
& &0\leq \lambda \leq \tau-\frac{3}{2} \ , \nonumber\\
& &\frac{3}{2} \leq \tau \leq \Omega-n_0+\frac{3}{2} \ , \qquad
n_0 \leq\Omega \ , \qquad (\tau\ :\ \hbox{\rm half-integer})
\nonumber\\
& &2\leq \tau \leq \Omega-n_0+1 \ , \qquad
n_0\leq \Omega-1 \ . \qquad (\tau\ : \ \hbox{\rm integer})
\end{eqnarray}
Both types of the eigenvalues occur for 
\begin{eqnarray}\label{3-59}
& &3n_0+2(2\tau-3)-2\lambda \leq N \leq 6\Omega-3n_0-2(2\tau-3)+2\lambda \ , 
\quad \left(0\leq 2\lambda < \tau-\frac{3}{2} \right) \nonumber\\
& &3n_0+\frac{3}{2}(2\tau-3) \leq N \leq 6\Omega-3n_0
-\frac{3}{2}(2\tau-3) \ , 
\quad \left(2\lambda = \tau-\frac{3}{2} \right) \nonumber\\
& &3n_0+(2\tau-3)+2\lambda \leq N \leq 6\Omega-3n_0-(2\tau-3)-2\lambda \ . 
\nonumber\\
& &\qquad\qquad\qquad\qquad\qquad\qquad\qquad\qquad\qquad\qquad
\left(\tau-\frac{3}{2} < 2\lambda \leq 2\left(\tau-\frac{3}{2}\right) 
\right)
\end{eqnarray}

Finally, we will give a short comment. 
The state $\dkket{\lambda\tau tT}$ and $\dkket{\lambda\tau 1-t T}$ 
are obtained by operating ${\hat O}_+(t\tau)$ defined 
in the relation (\ref{3-22}) on the states 
$\dket{\lambda,\tau}\otimes \kket{t;b}$ and 
$\dket{\lambda,\tau}\otimes \kket{t;a}$ for $(T-(t+\tau))$ times, 
respectively. 
Therefore, we can call ${\hat O}_+(t\tau)$ as the triplet generating 
operator. 
Here, $T-(t+\tau)=(N-N_{\rm min})/3$ and 
$(N_{\rm max}-N)/3$ for $\kket{t;b}$ and $\kket{t;a}$, respectively. 
On the other hand, $\dket{\lambda,\tau}\otimes \kket{t;b}$ can be 
rewritten as 
\begin{equation}\label{3-61-add}
\dket{\lambda,\tau}\otimes \kket{t;b}
=({\hat S}^3)^{2\lambda}\cdot ({\hat b}_1^*)^{2\tau-3-2\lambda}
({\hat b}^*)^{2(t+\tau)-4}\ket{0} \ . 
\end{equation}
The form (\ref{3-61-add}) indicates that ${\hat S}^3$ plays a role 
similar to ${\hat S}_+$ in the $su(2)$-pairing model. 
By exchange ${\hat a}$ and ${\hat b}$, we obtain the same form 
as the above in the case of 
$\dket{\lambda,\tau}\otimes \kket{t;a}$. 
The details of the results obtained in this section will be discussed 
in \S 5, especially, the comparison with the results by Petry et. al\cite{a} 
will be performed.

\section{Pairing correlation}

\subsection{Orthogonal set for diagonalizing the Hamiltonian}

In \S 3, we presented a form of the orthogonal set which enables us to 
describe the triplet formation. 
For making preparation for describing the case of pairing correlation, first, 
we summarize the orthogonal set given in \S 3 in a slightly 
general form. 
Let the orthogonal set under investigation specified by the quantum numbers 
$(\lambda, \mu, \nu, \nu_0, x, y, T, T_0)$ : 
$\ket{\lambda\mu\nu\nu_0;xyTT_0}$. 
Here, $(\lambda, \mu, \nu, \nu_0, T, T_0)$ are identical to those used in 
\S 3. 
As for $x$ and $y$, in \S 3, $x=\tau-2\lambda$ and $y=t$ or $(1-t)$ are used. 
In this section, we will search another set of $(x,\ y)$. 
With the use of the notations defined in Appendix A, 
the state $\ket{\lambda\mu\nu\nu_0;xyTT_0}$ is expressed as 
\begin{equation}\label{4-1}
\ket{\lambda\mu\nu\nu_0;xyTT_0}=({\hat T}_+)^{T_0-T}
{\hat Q}_+(\lambda\mu\nu\nu_0)\ket{\lambda xyT} \ .
\end{equation}
Here, $\ket{\lambda xyT}$ is the minimum weight state of the 
$su(3)\otimes su(1,1)$ algebra: 
\begin{eqnarray}
& &{\hat Q}_{-\frac{1}{2}}\ket{\lambda xyT}={\hat Q}_{\frac{1}{2}}\ket{
\lambda xyT}=0 \ , \qquad
{\hat R}_{-}\ket{\lambda xyT}={\hat T}_{-}\ket{\lambda xyT}=0 \ , 
\label{4-2}\\
& &{\hat R}_0\ket{\lambda xyT}=-\lambda\ket{\lambda xyT} \ , \qquad
{\hat T}_0\ket{\lambda xyT}=T\ket{\lambda xyT}\ , 
\label{4-3}
\end{eqnarray}
In the case of the triplet formation, ${\hat \tau}^0$ and ${\hat t}^0$ 
were used for specifying $\tau$ and $t$: 
\begin{equation}\label{4-4}
{\hat \tau}^0\ket{\lambda xyT}=(\tau-2\lambda)\ket{\lambda xyT}\ , 
\qquad
{\hat t}^0\ket{\lambda xyT}=(t\ {\rm or}\ 1-t)\ket{\lambda xyT}\ .
\end{equation}

As was  already mentioned, our aim is to find a possible orthogonal set 
suitable for describing the pairing correlation, which makes the quark 
number change in unit 2. 
According to the discussion performed immediately after the relation 
(\ref{2-55}), in the present case, the use of the operator ${\hat S}^i$ 
is necessary for constructing the orthogonal set. 
Further, in \S 3, the minimum weight state $\ket{m_2}$ plays a central 
role. 
In the present case, $\ket{m_1}$ given in the form (\ref{2-31}) may be 
useful.

We start from the state $\ket{m_1}$. 
It can be written as 
\begin{equation}\label{4-5}
\ket{m_1}=({\hat b}_1^*)^{2(\sigma_1-\sigma_0)}
({\hat b}^*)^{2\sigma_0}\ket{0}=\ket{\sigma_0,\sigma_1}=\ket{\sigma_0,T} \ .
\end{equation}
In the case where we make comparison with the state discussed in \S 3, 
we use the quantum number $T$ instead of $\sigma_1$. 
The quantum number $T$ is the eigenvalue of ${\hat T}_0$: 
\begin{equation}\label{4-6}
{\hat T}_0\ket{\sigma_0,T}=T\ket{\sigma_0,T}\ , \qquad
T=\sigma_1+2 \ .
\end{equation}
The state $\ket{\sigma_0,T}$ satisfies 
\begin{equation}\label{4-7}
{\hat R}_{\pm}\ket{\sigma_0,T}=0 \ , \qquad 
{\hat R}_0\ket{\sigma_0,T}=0 \ ,
\end{equation}
which means that $\ket{\sigma_0,T}$ is a state with $R$-spin$=0$. 
Further, we have the relation 
\begin{equation}\label{4-8}
[\ {\hat R}_- \ , \ {\hat S}^3\ ]=0 \ , \qquad
[\ {\hat R}_0 \ , \ {\hat S}^3\ ]=-\frac{1}{2}{\hat S}^3 \ ,
\end{equation}
which leads us to 
\begin{equation}\label{4-9}
[\ {\hat R}_- \ , \ ({\hat S}^3)^{2\lambda}\ ]=0 \ , \qquad
[\ {\hat R}_0 \ , \ ({\hat S}^3)^{2\lambda}\ ]=-\lambda({\hat S}^3)^{2\lambda} 
\ .
\end{equation}
Of course, ${\hat S}^3$ commutes with $\{{\hat T}_\mu\}$. 
Next, we note that ${\hat S}^1$, ${\hat S}_1^1$, 
$({\hat S}_2^2+{\hat S}_3^3)/2$ and 
$({\hat S}^2{\hat S}_1^2+{\hat S}^3{\hat S}_1^3)$ are scalar for 
$\{{\hat R}_\mu\}$, that is, they commute with $\{{\hat R}_\mu\}$. 
With the use of these scalar operators and real parameters $\xi$ and 
$\eta$, 
we define the operator ${\hat S}^4$ in the following form: 
\begin{equation}\label{4-10}
{\hat S}^4={\hat S}^1\left({\hat S}_1^1-\frac{\xi}{2}
({\hat S}_2^2+{\hat S}_3^3)\right)+
\eta({\hat S}^2{\hat S}_1^2+{\hat S}^3{\hat S}_1^3) \ . 
\end{equation}
Of course, ${\hat S}^4$ commutes with $\{{\hat R}_\mu\}$ and 
$\{{\hat T}_\mu\}$. 
With the use of ${\hat S}^3$ and ${\hat S}^4$, we set up the state 
$\ket{\lambda\rho\sigma_0 T}$ in the form 
\begin{equation}\label{4-11}
\ket{\lambda\rho\sigma_0 T}=({\hat S}^3)^{2\lambda}
({\hat S}^4)^{2\rho}\ket{\sigma_0,T} \ .
\end{equation}
Clearly, $\ket{\lambda\rho\sigma_0 T}$ satisfies 
\begin{eqnarray}
& &{\hat R}_-\ket{\lambda\rho\sigma_0 T}
={\hat T}_-\ket{\lambda\rho\sigma_0 T}=0 
\ , 
\label{4-12}\\
& &{\hat R}_0\ket{\lambda\rho\sigma_0 T}
=-\lambda\ket{\lambda\rho\sigma_0 T} \ , \qquad
{\hat T}_0\ket{\lambda\rho\sigma_0 T}
=T\ket{\lambda\rho\sigma_0 T} \ . 
\label{4-13}
\end{eqnarray}
The quantum numbers $\rho$ and $\sigma_0$ are related to the operators 
${\hat S}_1^1$ and $({\hat S}_2^2+{\hat S}_3^3)/2$ through the relation 
\begin{eqnarray}\label{4-14}
& &{\hat S}_1^1\ket{\lambda\rho\sigma_0 T}=
-2(\sigma_1-2\rho-\lambda)\ket{\lambda\rho\sigma_0 T} \ , 
\nonumber\\
& &\frac{1}{2}({\hat S}_2^2+{\hat S}_3^3)\ket{\lambda\rho\sigma_0 T}
=-2\left(\sigma_0-\rho-\frac{3}{2}\lambda\right)\ket{\lambda\rho\sigma_0 T} \ .
\end{eqnarray}
For deriving the relation (\ref{4-14}), the following formula is useful: 
\begin{equation}\label{4-15}
[\ {\hat S}_1^1\ , \ {\hat S}^4\ ]=2{\hat S}^4 \ , \qquad
\left[\ \frac{1}{2}({\hat S}_2^2+{\hat S}_3^3)\ , \ {\hat S}^4\ \right]
={\hat S}^4 \ . 
\end{equation}
The relations (\ref{4-12})$\sim$(\ref{4-14}) correspond to the latter of the 
relation (\ref{4-2}) and the relations (\ref{4-3}) and (\ref{4-4}). 
Of course, its correspondence is independent of the values of 
$\xi$ and $\eta$ 
contained in ${\hat S}^4$. 
In order to obtain complete correspondence, further, we must require the 
former of the relation: 
\begin{equation}\label{4-16}
{\hat Q}_{-\frac{1}{2}}\ket{\lambda\rho\sigma_0 T}=
{\hat Q}_{\frac{1}{2}}\ket{\lambda\rho\sigma_0 T}=0 \ .
\end{equation}
We determine the values of $\xi$ and $\eta$ so as to fulfill this 
requirement. 
Since $[{\hat Q}_m , {\hat S}^3]=0$ ($m=\pm 1/2$), 
the requirement (\ref{4-16}) is equivalent to 
\begin{equation}\label{4-17}
{\hat Q}_{-\frac{1}{2}}({\hat S}^4)^{2\rho}\ket{\sigma_0, T}=
{\hat Q}_{\frac{1}{2}}({\hat S}^4)^{2\rho}\ket{\sigma_0, T}=0 \ .
\end{equation}
For the understanding of the requirement (\ref{4-17}), the following 
relation is useful: 
\begin{eqnarray}
& &[\ {\hat Q}_{-\frac{1}{2}}\ , \ {\hat S}^4 \ ]
=-{\hat S}^2{\hat D}+\left(1+\frac{\xi}{2}\right)
{\hat S}^1{\hat Q}_{-\frac{1}{2}}-\eta({\hat S}^2{\hat R}_0
+{\hat S}^3{\hat R}_+)\ , \nonumber\\
& &[\ {\hat Q}_{\frac{1}{2}}\ , \ {\hat S}^4 \ ]
={\hat S}^3{\hat D}+\left(1+\frac{\xi}{2}\right)
{\hat S}^1{\hat Q}_{\frac{1}{2}}-\eta({\hat S}^3{\hat R}_0
-{\hat S}^2{\hat R}_-)\ , 
\label{4-18}\\
& &{\hat D}=(1-\eta){\hat S}_1^1-\frac{\xi-\eta}{2}
({\hat S}_2^2+{\hat S}_3^3) \ . 
\label{4-19}
\end{eqnarray}
Let ${\hat D}$ be the null operator. 
Then, if noting $[{\hat R}_\mu , {\hat S}^4]=0$ and 
${\hat Q}_m\ket{\sigma_0,T}$\break
$={\hat R}_\mu\ket{\sigma_0,T}=0$, the relation 
(\ref{4-18}) leads us to 
\begin{equation}\label{4-20}
{\hat Q}_m({\hat S}^4)^{2\rho}\ket{\sigma_0,T}=
\left(1+\frac{\xi}{2}\right)\sum_{r=0}^{2\rho-1}
({\hat S}^4)^{2\rho-r}{\hat S}^1\cdot {\hat Q}_m({\hat S}^4)^{r}
\ket{\sigma_0,T} \ . 
\end{equation}
By mathematical induction, the relation (\ref{4-20}) leads us to the 
relation (\ref{4-17}), i.e., (\ref{4-16}). 
The condition that ${\hat D}$ is null operator gives us 
\begin{equation}\label{4-21}
\xi=\eta=1\ .
\end{equation}
Then, the form(\ref{4-10}) is reduced to 
\begin{equation}\label{4-22}
{\hat S}^4={\hat S}^1\left(
{\hat S}_1^1-\frac{1}{2}({\hat S}_2^2+{\hat S}_3^3)\right)
+({\hat S}^2{\hat S}_1^2+{\hat S}^3{\hat S}_1^3) \ . 
\end{equation}
Thus, we obtain the orthogonal set for the present aim: 
\begin{eqnarray}
& &\ket{\lambda\mu\nu\nu_0;\rho\sigma_0 TT_0}
=({\hat T}_+)^{T_0-T}{\hat Q}_+(\lambda\mu\nu\nu_0)
\ket{\lambda\rho\sigma_0 T}\ , 
\label{4-23}\\
& &\ket{\lambda\rho\sigma_0 T}=({\hat S}^3)^{2\lambda}({\hat S}^4)^{2\rho}
\ket{\sigma_0,T} \ . 
\label{4-24}
\end{eqnarray}
In order to obtain the energy eigenvalues of the Hamiltonian 
${\hat H}_m={\hat H}+\chi{\hat {\mib Q}}^2$, only 
$\ket{\lambda\rho\sigma_0 T}$ is necessary.

\subsection{Properties of the state $\ket{\lambda\rho\sigma_0\sigma_1}$}

In this subsection, we will discuss two points: 
(1) to find the conditions governing the quantum numbers $\lambda$ and 
$\rho$ appearing in $\ket{\lambda\rho\sigma_0\sigma_1}$ ($T=\sigma_1+2$) 
and (2) to investigate the relevance to the symmetric representation $(
\sigma_1=\sigma_0)$. 
Let us start with (1). 
For completing the point (1), we will provide a representation of 
$\ket{\lambda\rho\sigma_0\sigma_1}$ in the framework of the 
$su(1,1)$ algebra. 
First, we identify $\ket{m_1}$ with the state (\ref{3-1}):
\begin{equation}\label{4-25}
\ket{m_1}=({\hat b}_1^*)^{2(\sigma_1-\sigma_0)}
({\hat b}^*)^{2\sigma_0}\ket{0}
=\dket{\tau}\otimes \kket{t} \ , 
\end{equation}
\vspace{-0.8cm}
\begin{subequations}\label{4-26}
\begin{eqnarray}
& &\dket{\tau}=({\hat b}_1^*)^{2\tau-3}\ket{0} \ , \qquad
\kket{t}=({\hat b}^*)^{2t-1}\ket{0} \ , 
\label{4-26a}\\
& &2\sigma_0=2t-1 \ , \qquad 
2(\sigma_1-\sigma_0)=2\tau-3 \ . 
\label{4-26b}
\end{eqnarray}
\end{subequations}
A straightforward calculation gives us 
\begin{eqnarray}\label{4-27}
({\hat S}^4)^{2\rho}\ket{m_1}
&=&({\hat S}^4)^{2\rho}\dket{\tau}\otimes \kket{t} \nonumber\\
&=&\frac{(2t-1)!}{(2(t-\rho)-1)!}\cdot 
\frac{(2\tau-1)!}{(2(\tau-\rho)-1)!}\cdot
\frac{(2\tau-3)!}{(2(\tau-\rho)-3)!}\nonumber\\
& &\times \left({\hat O}_+(t-\rho,\tau-\rho)\right)^{(t+\tau)-
((t-\rho)+(\tau-\rho))}\dket{\tau-\rho}\otimes \kket{t-\rho} \ .
\end{eqnarray}
Here, ${\hat O}_+(t-\rho,\tau-\rho)$ is defined in the relation 
(\ref{3-22}) with $t\rightarrow t-\rho$ and $\tau\rightarrow \tau-\rho$. 
The expression (\ref{4-27}) is quite natural. 
We have 
\begin{eqnarray}\label{4-28}
& &{\hat T}_-({\hat S}^4)^{2\rho}\ket{m_1}
=({\hat S}^4)^{2\rho}{\hat T}_-\ket{m_1}=0 \ , 
\nonumber\\
& &{\hat T}_0({\hat S}^4)^{2\rho}\ket{m_1}
=({\hat S}^4)^{2\rho}{\hat T}_0\ket{m_1}
=(t+\tau)({\hat S}^4)^{2\rho}\ket{m_1} \ . 
\end{eqnarray}
The relation (\ref{4-28}) shows that $({\hat S}^4)^{2\rho}\ket{m_1}$ 
is the minimum weight state for $\{{\hat T}_\mu\}$ specified by 
$T=t+\tau$. 
Further, we have 
\begin{eqnarray}\label{4-29}
& &{\hat t}_-\kket{t-\rho}=0 \ , \qquad 
{\hat t}_0\kket{t-\rho}=(t-\rho)\kket{t-\rho}\ , \qquad 
2(t-\rho)-1 \geq 0 \ , \nonumber\\
& &{\hat \tau}_-\dket{\tau-\rho}=0 \ , \qquad 
{\hat \tau}_0\dket{\tau-\rho}=(\tau-\rho)\dket{\tau-\rho}\ , \qquad 
2(\tau-\rho)-3 \geq 0 \ . 
\end{eqnarray}
The relation (\ref{4-29}) shows that $\dket{\tau-\rho}\otimes \kket{t-\rho}$ 
is the minimum weight state for $\{{\hat \tau}_\mu\}$ and 
$\{{\hat t}_\mu\}$ specified by $(\tau-\rho)$ and $(t-\rho)$, 
respectively. 
The operator ${\hat O}_+(t-\rho,\tau-\rho)$ plays the same role 
as the one in the relation (\ref{3-22}). 
This being why the form (\ref{4-27}) is quite reasonable.

The state (\ref{4-27}) can be expressed in the form 
\begin{eqnarray}\label{4-30}
({\hat S}^4)^{2\rho}\ket{m_1}&=&
\sum_{r=0}^{2\rho}(-)^{2\rho-r}D_{t\tau\rho}(r) \nonumber\\
& &\qquad\quad 
\times ({\hat b}^*)^{2t-1-r}({\hat a}^*)^{2\rho-r}({\hat \tau}_+)^r
({\hat b}_1^*)^{2\tau-3-2\rho}\ket{0} \ ,
\end{eqnarray}
where $D_{t\tau\rho}(r)$ denotes an appropriate expansion coefficient, 
its explicit expression being omitted. 
The operator $({\hat S}^3)^{2\lambda}$ is expressed as 
\begin{eqnarray}\label{4-31}
({\hat S}^3)^{2\lambda}&=&
({\hat a}_3^*{\hat b}-{\hat a}^*{\hat b}_3)^{2\lambda} \nonumber\\
&=&\sum_{\kappa=0}^{2\lambda}(-)^{\kappa} 
\left(\begin{array}{@{\,}c@{\,}}
2\lambda \\
\kappa
\end{array}\right)
({\hat a}_3^*)^{2\lambda-\kappa}({\hat a}^*)^{\kappa}
({\hat b})^{2\lambda-\kappa}({\hat b}_3)^{\kappa} \ .
\end{eqnarray}
Then, we find 
\begin{eqnarray}\label{4-32}
& &({\hat S}^3)^{2\lambda}({\hat S}^4)^{2\rho}\ket{m_1} \nonumber\\
&=&\biggl[ \sum_{\kappa=0}^{2\lambda}\sum_{r=0}^{2\rho}(-)^{\rho-(r-\kappa)}
D_{t\tau\rho}(r)
\left(\begin{array}{@{\,}c@{\,}}
2t-1-r \\
2\lambda-\kappa
\end{array}\right)
\left(\begin{array}{@{\,}c@{\,}}
r \\
\kappa
\end{array}\right)\nonumber\\
& &\ \ \times ({\hat b}^*)^{2t-1-2\lambda-(r-\kappa)}
({\hat a}^*)^{2\rho-(r-\kappa)}
({\hat \tau}_+)^{r-\kappa}\biggl]
({\hat a}_3^*)^{2\lambda}({\hat b}_1^*)^{2\tau-3-2\rho}\ket{0} \ .
\end{eqnarray}
Since $r-\kappa\geq 0$, the state (\ref{4-12}) vanishes if $2t-1-2\lambda <0$ 
and also if $2\tau-3-2\rho <0$. 
Therefore, combining with the inequality (\ref{4-29}), 
the state (\ref{4-32}) is defined for 
\begin{subequations}\label{4-33}
\begin{eqnarray}
& &\hbox{\rm if}\ \ 2t-1 \geq 2\tau-3 \ , \qquad 2\lambda \leq 2t-1\ , \qquad 
2\rho\leq 2\tau-3 \ , 
\label{4-33a}\\
& &\hbox{\rm if}\ \ 2t-1 \leq 2\tau-3 \ , \qquad 2\lambda \leq 2t-1\ , \qquad 
2\rho\leq 2t-1 \ . 
\label{4-33b}
\end{eqnarray}
\end{subequations}
The above is the condition governing $\lambda$ and $\rho$ and, 
in the next subsection, this condition will be used.

Next, we discuss the point (2). 
In the case $\sigma_1=\sigma_0$, we have 
\begin{equation}\label{4-34}
{\hat S}^4\ket{m_0}=0 \ , 
\end{equation}
so that the orthogonal set constructed on the minimum weight state 
$\ket{m_0}$ is only valid if $\rho=0$. 
In order to obtain further conditions, we change the quantum numbers as 
follows: 
\begin{equation}\label{4-35}
\lambda=\nu_1+\nu_2+\nu_3\ , \qquad 
\lambda_0=\nu_2-\nu_3+l\ , \qquad 
\mu=\nu_1 \ , \qquad 
\mu_0=-l \ . 
\end{equation}
Then, the angular momentum coupling rule leads us to 
\begin{equation}\label{4-36}
\nu=\nu_2+\nu_3 \ , \  \nu_2+\nu_3+1\ , \cdots 
,\ 2\nu_1+\nu_2+\nu_3\ , \qquad
\nu_0=\nu_2-\nu_3 \ . 
\end{equation}
In the new notations, the state (\ref{4-23}) with $T_0=T$ is 
replaced by 
\begin{eqnarray}\label{4-37}
\ket{\nu_1\nu_2\nu_3\nu\rho;\sigma_0\sigma_1}
&=&\sum_l \langle \nu_1+\nu_2+\nu_3, \nu_2-\nu_3+l, \nu_1,-l
\ket{\nu,\nu_2-\nu_3}\nonumber\\
& &\times (-)^{\nu_1+\mu}
\sqrt{\frac{(2\nu_1)!(2\nu_1+\nu_2+\nu_3)!}{(\nu_1-l)!(\nu_1+l)!
(\nu_1+2\nu_2+l)!(\nu_1+2\nu_2-l)!}} \nonumber\\
& &\times ({\hat S}_1^3)^{\nu_1-l}({\hat S}_1^2)^{\nu_1+l}
({\hat S}^2)^{\nu_1+2\nu_2+l}({\hat S}^3)^{\nu_1+2\nu_2-l}
({\hat S}^4)^{2\rho}\ket{\sigma_0,\sigma_1} \ , \qquad
\end{eqnarray}
where $\ket{\sigma_0,\sigma_1}=\ket{m_1}$. 
For the discussion of the case $\sigma_1=\sigma_0$, the 
following relation is useful: 
\begin{eqnarray}\label{4-38}
({\hat S}_1^i)^r({\hat S}^i)^n
&=&\frac{n!}{(n-r)!}({\hat S}^1)^r({\hat S}^i)^{n-r}\nonumber\\
& &+\sum_{p=1}^{r}\frac{r!}{p!(r-p)!}\cdot\frac{n!}{(n-r+p)!}({\hat S}^1)^{r-p}
({\hat S}^i)^{n-r+p}({\hat S}_1^i)^p \ . \nonumber\\
& &\qquad\qquad\qquad\qquad\qquad\qquad\qquad\qquad\qquad
(i=2,3)
\end{eqnarray}
Since ${\hat S}_1^i\ket{m_0}=0$ for $i=2$,3, we have 
\begin{equation}\label{4-39}
({\hat S}_1^i)^r({\hat S}^i)^n\ket{m_0}=
\frac{n!}{(n-r)!}({\hat S}^1)^r({\hat S}^i)^{n-r}\ket{m_0} \ .
\end{equation}
Thus, after straightforward calculation, we have 
\begin{eqnarray}\label{4-40}
& &\ket{\nu_1\nu_2\nu_3r\rho;\sigma_0\sigma_1}|_{\rho=0,\sigma_1=\sigma_0}
\nonumber\\
&=&\delta_{\nu_1,\nu_2+\nu_3}
\frac{\sqrt{(2\nu_1)!(2\nu_1+\nu_2+\nu_3)!}}{(2\nu_2)!(2\nu_3)!}\cdot
\sqrt{\frac{2(\nu_1+\nu_2+\nu_3)+1}{2(\nu_2+\nu_3)+1}} \nonumber\\
& &\times ({\hat S}^1)^{2\nu_1}({\hat S}^2)^{2\nu_2}({\hat S}^3)^{2\nu_3}
\ket{m_0} \ .
\end{eqnarray}
The state (\ref{4-40}) describes the symmetric representation specified by 
$\nu_1$, $\nu_2$, $\nu_3$ and $\sigma_0$.

\subsection{Energy eigenvalue}

We are now able to determine the energy eigenvalue of the Hamiltonian 
${\hat H}_m={\hat H}+\chi{\hat {\mib Q}}^2$. 
First, we note that ${\hat H}$ can be expressed in the form 
\begin{equation}\label{4-41}
{\hat H}=-\frac{1}{2}({\hat {\mib P}}^2-{\hat {\mib Q}}^2-{\hat \Sigma})\ .
\end{equation}
Here, ${\hat {\mib P}}^2$ and ${\hat {\mib Q}}^2$ denote the Casimir 
operators of the $su(4)$ and the $su(3)$ algebras shown in the relations 
(\ref{2-3}) and (\ref{2-7}), respectively. 
The operator ${\hat \Sigma}$ is defined as 
\begin{equation}\label{4-42}
{\hat \Sigma}=\frac{1}{12}({\hat S}_1^1+{\hat S}_2^2+{\hat S}_3^3)^2
-({\hat S}_1^1+{\hat S}_2^2+{\hat S}_3^3) \ .
\end{equation}
Therefore, the eigenstate of ${\hat H}$ is given by the state (\ref{4-11}). 
The eigenvalue of ${\hat {\mib P}}^2$ is given in the relation (\ref{2-25}). 
Noting the relations (\ref{4-13}) and (\ref{4-14}), the eigenvalues of 
${\hat {\mib Q}}^2$ and ${\hat \Sigma}$ are expressed as 
\begin{subequations}\label{4-43}
\begin{eqnarray}
& &{\hat {\mib Q}}^2\ket{\lambda\rho\sigma_0\sigma_1}
=F_{\sigma_1\sigma_0\rho\lambda}^{(p)}\ket{\lambda\rho\sigma_0\sigma_1} \ , 
\label{4-43a}\\
& &F_{\sigma_1\sigma_0\rho\lambda}^{(p)}=2\lambda(\lambda+1)
+\frac{2}{3}\Bigl(2(\sigma_1-\sigma_0)-2\rho+\lambda\Bigl)\Bigl(
2(\sigma_1-\sigma_0)-2\rho+\lambda+3\Bigl)\ , \  \qquad
\label{4-43b}
\end{eqnarray}
\end{subequations}
\vspace{-0.8cm}
\begin{subequations}\label{4-44}
\begin{eqnarray}
& &{\hat {\Sigma}}\ket{\lambda\rho\sigma_0\sigma_1}
=\Sigma_{\sigma_1\sigma_0\rho\lambda}\ket{\lambda\rho\sigma_0\sigma_1} \ , 
\label{4-44a}\\
& &\Sigma_{\sigma_1\sigma_0\rho\lambda}=
\frac{1}{3}(\sigma_1+2\sigma_0-4\rho-4\lambda)(
\sigma_1+2\sigma_0-4\rho-4\lambda+6)\ .
\label{4-44b}
\end{eqnarray}
\end{subequations}
With the use of the form (\ref{4-41}), the eigenvalue of ${\hat H}$ is 
given as 
\begin{equation}\label{4-45}
E_{\sigma_1\sigma_0\rho\lambda}=-\Bigl(2\lambda(2\sigma_0+1
-2\rho-2\lambda)+2\rho(2\sigma_1+3-2\rho)\Bigl)
\end{equation}
Therefore, the eigenvalue of ${\hat H}_m$ is obtained in the form 
\begin{equation}\label{4-46}
E_{\sigma_1\sigma_0\rho\lambda}^{(m)}=E_{\sigma_1\sigma_0\rho\lambda}+
\chi F_{\sigma_1\sigma_0\rho\lambda}^{(p)} \ . 
\end{equation}

In the same idea as that in \S 3, we rewrite 
$E_{\sigma_1\sigma_0\rho\lambda}$ and 
$F_{\sigma_1\sigma_0\rho\lambda}^{(p)}$ in terms of the fermion number $N$. 
The relations (\ref{2-53}) and (\ref{2-55}) can be rewritten as 
\begin{eqnarray}
& &{\hat \Omega}=n_0-2+{\hat T}_0 \ , 
\label{4-47}\\
& &{\hat N}=3n_0-6+3{\hat T}_0+\frac{1}{2}({\hat S}_1^1+{\hat S}_2^2
+{\hat S}_3^3) \ . 
\label{4-48}
\end{eqnarray}
The relation (\ref{4-47}) was already used in the relations 
(\ref{3-32}) and (\ref{3-35}). 
Operation of ${\hat \Omega}$ 
and ${\hat N}$ on $\ket{\lambda\rho\sigma_0\sigma_1}$ leads us to 
\begin{eqnarray}
& &\Omega=n_0-2+T=n_0+\sigma_1 \ , 
\label{4-49}\\
& &N=3n_0+4(\lambda+\rho)+2\sigma_1-2\sigma_0=2n_0+n_1+4(\lambda+\rho) \ .
\label{4-50}
\end{eqnarray}
Substituting the forms (\ref{4-49}) and (\ref{4-50}) to the relations 
(\ref{4-45}) and (\ref{4-43b}), we can express the energy eigenvalue 
in terms of $N$:
\begin{eqnarray}
E_{Nn_0n_1\rho}&=&
-\left(\frac{1}{2}N-\frac{1}{2}(2n_0+n_1)-2\rho\right)
\left(2\Omega+1-\frac{1}{2}n_1-\frac{1}{2}N\right) \nonumber\\
& &-2\rho(2\Omega+3-2n_0-2\rho) \ , 
\label{4-51}\\
F_{Nn_0n_1\rho}^{(p)}&=&
G_{Nn_0n_1}+2E_{Nn_0n_1\rho} \ , \nonumber\\
G_{Nn_0n_1}&=&2(\Omega-n_0)(\Omega-n_0+3)+(\Omega-n_1)^2
-\frac{1}{3}(3\Omega-N)(3\Omega-N+6) \ . \nonumber\\
& &
\label{4-52}
\end{eqnarray}
The case ($n_1=n_0,\ \rho=0$) gives the expression for the symmetric 
representation. 
From the relation (\ref{4-50}), it follows that $N-(2n_0+n_1)$ must 
be a positive even integer, the reason being that $4(\lambda+\rho)$ 
is positive even integer. 
This means that the change in the fermion number relatively to that 
of the minimum weight state is restricted to even number, i.e., 
it is of the pairing-type.

In \S 3, we knew that there exists a range where $N$ can change. 
In the present case, we will discuss this problem. 
The relation (\ref{2-20}) gives us 
\begin{equation}\label{4-53}
n_0 \leq n_1 \leq 2\Omega-n_0 \ .
\end{equation}
From the condition $2\Omega-n_0 \geq n_0$, we have 
\begin{equation}\label{4-54}
n_0 \leq \Omega \ .
\end{equation}
In the present case, the relation (\ref{4-26b}) gives us 
$2t-1=2\Omega-(n_0+n_1)$ and 
$2\tau-3=n_1-n_0$. 
The condition $2t-1 \geq 2\tau-3$ in the relation (\ref{4-33a}) is 
equivalent to 
\begin{equation}\label{4-55}
n_1 \leq \Omega \ .
\end{equation}
In this case, $2\lambda \leq 2\Omega-(n_0+n_1)$ and 
$2\rho \leq n_1-n_0$. 
Then, we have 
\begin{equation}\label{4-56}
4(\lambda+\rho) \leq 4(\Omega-n_0) \ .
\end{equation}
From the expression (\ref{4-50}) for $N$, we obtain 
\begin{equation}\label{4-57}
2n_0+n_1 \leq N \leq 4\Omega-2n_0+n_1 \ .
\end{equation}
The condition $2t-1 \leq 2\tau-3$ is equivalent to 
\begin{equation}\label{4-58}
\Omega \leq n_1 \ , \qquad \hbox{\rm i.e.,}\qquad 
\Omega \leq n_1 \leq 2\Omega-n_0 \ .
\end{equation}
In this case, the relation (\ref{4-33b}) gives us 
$2\lambda \leq 2\Omega-(n_0+n_1)$, $2\rho \leq 2\Omega-
(n_0+n_1)$, which leads to 
\begin{equation}\label{4-59}
4(\lambda +\rho) \leq 8\Omega-4(n_0+n_1) \ .
\end{equation}
The expression (\ref{4-50}) gives us 
\begin{equation}\label{4-60}
2n_0+n_1 \leq N \leq 8\Omega-2n_0-3n_1 \ .
\end{equation}

Next, we investigate the range where $\rho$ can change for a given $N$. 
In the case where $n_0\leq \Omega$ and $n_1 \leq \Omega$, 
in addition to $2\rho \leq n_1-n_0$, we have 
$2\lambda \leq 2\Omega-(n_0+n_1)$, which gives 
\begin{eqnarray}\label{4-61}
& &0\leq 2\lambda =\frac{1}{2}(N-(2n_0+n_1))-2\rho \leq 
2\Omega-(n_0+n_1) \ ,\nonumber\\
& &\quad \hbox{\rm i.e.,}\qquad 
\frac{1}{2}(N-4\Omega+n_1) \leq 2\rho \leq \frac{1}{2}\left(
N-(2n_0+n_1)\right)\ .
\end{eqnarray}
Then, by comparing $(n_1-n_0)$ with 
$(N-(2n_0+n_1))/2$, we have 
\begin{subequations}\label{4-62}
\begin{eqnarray}
& &\frac{1}{2}(N-4\Omega+n_1)\leq 2\rho \leq n_1-n_0 \ , \qquad
(3n_1 \leq N)
\label{4-62a}\\
& &\frac{1}{2}(N-4\Omega+n_1) \leq 2\rho \leq \frac{1}{2}
\left(N-(2n_0+n_1)\right) \ . \qquad (3n_1 \geq N)
\label{4-62b}
\end{eqnarray}
\end{subequations}
Further, we notice the sign of 
$(N-4\Omega+n_1)/2$. 
Then, 
from the condition (\ref{4-62}), we obtain 
\begin{subequations}\label{4-63}
\begin{eqnarray}
& &\hbox{\rm if}\ \ 2n_0+n_1 \leq N \leq 3n_1 \ , \quad 
0\leq 2\rho \leq \frac{1}{2}\left(N-(2n_0+n_1)\right) \ , 
\label{4-63a}\\
& &\hbox{\rm if}\ \ 3n_1 \leq N \leq 4\Omega-n_1 \ , \quad 
0\leq 2\rho \leq n_1-n_0 \ , 
\label{4-63b}\\
& &\hbox{\rm if}\ \ 4\Omega-n_1 \leq N \leq 4\Omega-2n_0+n_1  \ , \quad 
\frac{1}{2}(N-4\Omega+n_1) \leq 2\rho \leq n_1-n_0 \ . \nonumber\\
& &
\label{4-63c}
\end{eqnarray}
\end{subequations}
Of course, we can prove $4\Omega-n_1\geq 3n_1 \geq 
2n_0+n_1$ and $4\Omega-2n_0+n_1 
\geq 4\Omega-n_1$. 
The case where $n_0 \leq \Omega$ and $\Omega \leq n_1 \leq 2\Omega-n_0$ 
is also treated in the same idea as that in the previous case. 
Instead of the relation $2\rho\leq n_1-n_0$, we use 
$2\rho \leq 2\Omega-(n_0+n_1)$ 
and compare $(2\Omega-(n_0+n_1))$ with $(N-4\Omega+n_1)/2$. 
The result is as follows: 
\begin{subequations}\label{4-64}
\begin{eqnarray}
& &\hbox{\rm if}\ \ 2n_0+n_1\! \leq N \!\leq 4\Omega-n_1 \ , \quad 
0\leq 2\rho\! \leq\! \frac{1}{2}\left(N-(2n_0+n_1)\right) \ , 
\label{4-64a}\\
& &\hbox{\rm if}\ \ 4\Omega-n_1\! \leq N\! \leq 8\Omega-2n_0-3n_1  \ , \quad 
\frac{1}{2}(N-4\Omega+n_1)\! \leq\! 2\rho \leq 
\frac{1}{2}\left(N-(2n_0+n_1)\right) \ . \nonumber\\
& &
\label{4-64b}
\end{eqnarray}
\end{subequations}
Of course, we have $4\Omega-n_1 \geq 2n_0+n_1$ and 
$8\Omega-2n_0-3n_1 \geq 4\Omega-n_1$.

Finally, we should contact with the treatment from the side of $N=6\Omega$. 
This case can be treated by replacing $N$ with $(6\Omega-N)$. 
The detail will be given in \S 5.

\section{Discussion}

\subsection{The triplet formation}

The most important and interesting result of the Bonn model may be to 
describe the triplet formation quite naturally in spite of two-body 
interaction. 
In this subsection, first, we will analyze our result for the case of 
triplet formation, which is obtained in the framework of the original 
Bonn model Hamiltonian (\ref{2-8}) ($\chi=0)$.

The energy eigenvalue (\ref{3-55a}), $E_{Nn_0\tau\lambda}$, and 
the associated form ${\cal E}_{6\Omega-Nn_0\tau\lambda}$ 
($=E_{6\Omega-Nn_0\tau\lambda}+2(3\Omega-N)$) may be written as 
\begin{subequations}\label{5-1}
\begin{eqnarray}
& &E_{Nn_0\tau\lambda}=E_0(N,n_0)+{\cal E}(N,\tau,\lambda) \ , 
\label{5-1a}\\
& &{\cal E}_{6\Omega-Nn_0\tau\lambda}
=E_0(N,n_0)+{\cal E}(6\Omega-N,\tau,\lambda) \ , 
\label{5-1b}
\end{eqnarray}
\end{subequations}
where 
\begin{eqnarray}
E_0(N,n_0)&=&
-\frac{1}{3}N\left(2\Omega+3-\frac{1}{3}N\right) +n_0(2\Omega+3-n_0) \ , 
\label{5-2}\\
{\cal E}(N,\tau,\lambda)&=&
\begin{cases}
-\frac{4}{9}\left(\tau-\frac{3}{2}-2\lambda\right)(N-N_c) \ , & 
\left(2\lambda \neq \tau-\frac{3}{2}\right)\\
\left(\tau-\frac{3}{2}\right)\left(\tau+\frac{1}{2}\right) \ , & 
\left(2\lambda=\tau-\frac{3}{2}\right)
\end{cases}
\label{5-3}\\
N_c&=&
3\Omega-\frac{5}{4}\left(\tau-\frac{3}{2}-2\lambda\right)+
\frac{9}{4}\cdot
\frac{\left(\tau-\frac{3}{2}\right)\left(\tau+\frac{1}{2}\right)}
{\tau-\frac{3}{2}-2\lambda} \ . 
\label{5-4}
\end{eqnarray}
We observe that $E_0(N,n_0)$ is of the form of the expression 
given by Petry et al.\cite{a} 
In this sense, our results contain the cases which 
Petry et al. did not discuss. 
Moreover, in the case $0\leq 2\lambda < \tau-3/2$, for 
$N<3\Omega$, $N=3\Omega$ and $N>3\Omega$, we have, 
respectively, $E_{Nn_0\tau\lambda}>{\cal E}_{6\Omega-Nn_0\tau\lambda}$, 
$E_{Nn_0\tau\lambda}={\cal E}_{6\Omega-Nn_0\tau\lambda}$ and 
$E_{Nn_0\tau\lambda} <{\cal E}_{6\Omega-Nn_0\tau\lambda}$. 
In the case $\tau-3/2 < 2\lambda \leq 2(\tau-3/2)$, 
vice versa. 
In the case $2\lambda=\tau-3/2$, 
$E_{Nn_0\tau\lambda}={\cal E}_{6\Omega-Nn_0\tau\lambda}$. 
Of course, $N$ in these cases should obey the condition (\ref{3-59}). 
For the case $\tau=3/2$, only $\lambda=0$ is permitted and 
${\cal E}(N,\tau,\lambda)$ vanishes, that is, 
$E_{Nn_0\tau\lambda}>{\cal E}_{6\Omega-Nn_0\tau\lambda}=E_0(N,n)$. 
Therefore, the case $\tau=3/2$ is in a special position, and later, its 
physical meaning will be discussed. 
From the above consideration, it is interesting to investigate 
the deviation from $E_0(N,n_0)$ for $\tau>3/2$.

We are led to study a set of the inequalities 
\b\label{5-5}
{\cal E}(N,\tau,\lambda) <0 \ , \qquad {\cal E}(6\Omega-N,\tau,\lambda)<0 \ ,
\end{equation}
which imply $E_{Nn_0\tau\lambda}<E_0(N,n_0)$ and 
${\cal E}_{6\Omega-Nn_0\tau\lambda} <E_0(N,n_0)$, 
respectively. 
We discuss the case ${\cal E}(N,\tau,\lambda)<0$ in rather detail. 
The relation (\ref{5-3}) shows that in the case $2\lambda=\tau-3/2$, 
${\cal E}(N,\tau,\lambda)>0$ and in the case 
$2\lambda \neq \tau-3/2$, we have the inequality 
\b\label{5-6}
\left(\tau-\frac{3}{2}-2\lambda\right)\left(N-N_c\right) > 0 \ .
\end{equation}
Therefore, we have 
\beq
& &N_c < N \quad \hbox{\rm for} \quad 0 \leq 2\lambda <\tau-\frac{3}{2} \ , 
\label{5-7}\\
& &N_c > N \quad \hbox{\rm for} \quad \tau-\frac{3}{2} < 2\lambda \leq 
2\left(\tau-\frac{3}{2}\right) \ . 
\label{5-8}
\eeq
For the cases (\ref{5-7}) and (\ref{5-8}), $N$ should satisfy, respectively, 
\beq
& &N_c < N_{\rm max} \ , \qquad
N_{\rm max}=6\Omega-3n_0-4\left(\tau-\frac{3}{2}\right)+2\lambda \ , 
\label{5-9}\\
& &N_c > N_{\rm min} \ , \qquad
N_{\rm min}=3n_0+2\left(\tau-\frac{3}{2}\right)+2\lambda \ . 
\label{5-10}
\eeq
The conditions (\ref{5-7}) and (\ref{5-8}) should be compatible with the 
relations (\ref{5-9}) and (\ref{5-10}), respectively. 
We search condition which satisfies the above-mentioned compatibility. 
After a rather lengthy consideration, we have the following: 
The inequality ${\cal E}(N,\tau,\lambda)<0$ realizes in the set 
$(N,\tau,\lambda)$ satisfying 
\beq
& &N_c < N \leq N_{\rm max} \ , 
\label{5-11}\\
& &0 < \tau-\frac{3}{2} < \frac{3}{5}\left(\Omega -n_0 -\frac{3}{2}
\right) \ , 
\label{5-12}\\
& &0 \leq 2\lambda < 2\lambda_c \ , 
\label{5-13}
\eeq
where $N_c$ and $N_{\rm max}$ are given in the relations (\ref{5-4}) and 
(\ref{5-9}), respectively, and $2\lambda_c$ is defined as 
\beq\label{5-14}
2\lambda_c&=&
\left(\tau-\frac{3}{2}\right)(1-\delta_c) \ , \nonumber\\
\delta_c&=&
\frac{3}{2}\cdot \frac{\tau+\frac{1}{2}}{(\Omega-n_0)-(\tau-\frac{3}{2})
+\sqrt{((\Omega-n_0)-(\tau-\frac{3}{2}))^2+\frac{1}{4}(\tau-\frac{3}{2})
(\tau+\frac{1}{2})}} \ .\qquad
\eeq
The case ${\cal E}(6\Omega-N,\tau,\lambda)<0$ is treated by replacing $N$ 
in ${\cal E}(N,\tau,\lambda)$ with $(6\Omega-N)$. 
Then, only the relation (\ref{5-11}) changes to 
\b\label{5-15}
6\Omega-N_{\rm max} \leq N <6\Omega-N_c \ , 
\end{equation}
where 
\beq\label{5-16}
& &6\Omega-N_{\rm max}=3n_0+4\left(\tau-\frac{3}{2}\right)-2\lambda \ , 
\nonumber\\
& &6\Omega-N_c = 3\Omega+\frac{5}{4}\left(\tau-\frac{3}{2}-2\lambda\right)
-\frac{9}{4}\cdot\frac{\left(\tau-\frac{3}{2}\right)\left(\tau+\frac{1}{2}
\right)}{\tau-\frac{3}{2}-2\lambda} \ .
\eeq
The above is the condition which leads us to the inequality (\ref{5-5}).

Now, we will investigate the case $\tau=3/2$ in more detail and 
through this discussion, the physical meaning of this case will be clarified. 
The relations (\ref{3-16}) and (\ref{3-17}) give us 
\b\label{5-17}
\lambda=\mu=\nu=\nu_0=0 \quad \hbox{\rm for}\quad \tau=\frac{3}{2} \ .
\end{equation}
The states with these quantum numbers can be expressed as 
\beq
& &\dkket{tT;b}=\sum_{t_0+\tau_0=T}\!\!\!\!\!{}'\ \ 
C_{tt_0\frac{3}{2}\tau_0}(T)
({\hat \tau}_+)^{\tau_0-\frac{3}{2}}\ket{0}
\otimes ({\hat t}_+)^{t_0-t}({\hat b}^*)^{2t-1}\ket{0} \ , 
\label{5-18}\\
& &\dkket{tT;a}=\sum_{t_0+\tau_0=T}\!\!\!\!\!{}'\ \ 
C_{tt_0\frac{3}{2}\tau_0}(T)
({\hat \tau}_+)^{\tau_0-\frac{3}{2}}\ket{0}
\otimes ({\hat t}_+)^{t_0-t}({\hat a}^*)^{2t-1}\ket{0} \ , 
\label{5-19}
\eeq
The state (\ref{5-18}) is of the abbreviated form of the state (\ref{3-23}) 
and the state (\ref{5-19}) is obtained by replacing ${\hat b}^*$ with 
${\hat a}^*$ in the state (\ref{5-18}). 
The energy eigenvalue is $E_0(N,n_0)$ (cf. Eq.(\ref{5-2})). 
Of course, $N$ is a multiplet of 3 and satisfies 
\b\label{5-20-0}
3n_0 \leq N \leq 6\Omega-3n_0 \ .
\end{equation}
The states (\ref{5-18}) and (\ref{5-19}) are constructed from the 
operators ${\hat t}_+$, ${\hat \tau}_-$, ${\hat b}^*$ and ${\hat a}^*$. 
They are color-neutral in the sense that they are invariant under the 
group $SU(3)$, since the operators ${\hat b}^*$, ${\hat a}^*$, 
$\{{\hat t}_\mu\}$ and $\{{\hat \tau}_\mu\}$ commute with the $su(3)$ 
generators (\ref{2-6}). 
The states with $\tau >3/2$ are no longer invariant under the 
group $SU(3)$ and so they are not color-neutral. 
As was already discussed, in the original Bonn model Hamiltonian, 
colored states are lower in energy than the color-neutral ones if 
$N$, $\tau$ and $\lambda$ satisfy the conditions 
(\ref{5-11})$\sim$(\ref{5-13}) and (\ref{5-15}). 
This result is characteristic of the original Bonn model.

However, position of the colored state can be controlled by adopting the 
Hamiltonian modified from that in the original Bonn model, that is, 
${\hat H}_m={\hat H}+\chi{\hat {\mib Q}}^2$ given in the relation 
(\ref{2-10}). 
As was already shown, the eigenstates of ${\hat H}_m$ does not change 
from those of ${\hat H}$. 
The eigenvalue of ${\hat {\mib Q}}^2$ is given in the relation 
(\ref{3-55b}), i.e., $F_{\tau\lambda}^{(t)}$ ($=F^{(t)}(\tau,\lambda)$). 
We can observe the following relation in $F^{(t)}(\tau,\lambda)$: 
\bsub\label{5-20}
\beq
& &F^{(t)}(\tau,\lambda)=0 \ , \qquad {\rm if}\qquad \tau=\frac{3}{2} \quad
(\lambda=0) \ , 
\label{5-20a}\\
& &F^{(t)}(\tau,\lambda)>0 \ , \qquad {\rm if}\qquad \tau>\frac{3}{2} \quad
(0 < 2\lambda \leq 2\tau-3) \ . 
\label{5-20b}
\eeq
\esub
Therefore, the energy eigenvalue of ${\hat H}_m$, $E_{Nn_0\tau\lambda}^{(m)}$, 
is changed from the form (\ref{5-1a}). (Later, we will 
discuss the case (\ref{5-1b})): 
\beq
& &E_{Nn_0\frac{3}{2}0}^{(m)}=E_0(N,n_0) \ , 
\label{5-21}\\
& &E_{Nn_0\tau\lambda}^{(m)}=E_0(N,n_0)+{\cal E}^{(m)}(N,\tau,\lambda) \ , 
\quad \left(\tau>\frac{3}{2}\right)
\label{5-22}\\
& &{\cal E}^{(m)}(N,\tau,\lambda)={\cal E}(N, \tau, \lambda)+
\chi F^{(t)}(\tau,\lambda) \ . 
\label{5-23}
\eeq
The relation (\ref{5-21}) tells us that even if $\chi {\hat {\mib Q}}^2$ 
is switched on, the energy of the color-neutral states does not change 
from $E_0(N,n_0)$ obtained in the original Bonn model Hamiltonian. 
On the other hand, as is shown in the form (\ref{5-22}), 
the energy of the colored state $(\tau>3/2)$ is influenced 
by $\chi{\cal {\mib Q}}^2$. 
Therefore, even if ${\cal E}(N,\tau,\lambda) <0$, by choosing 
$\chi$ appropriately, we can make 
\b\label{5-24}
{\cal E}^{(m)}(N,\tau,\lambda)>0 \ , \qquad 
\hbox{\rm i.e.,}\qquad \chi > 
\frac{-{\cal E}(N,\tau,\lambda)}{F^{(t)}(\tau,\lambda)} \ .
\end{equation}
In the case (\ref{5-1b}), even if ${\cal E}(6\Omega-N,\tau,\lambda)<0$, 
under an appropriate choice of 
$\chi$, ${\cal E}^{(m)}(6\Omega-N,\tau,\lambda)$ 
($={\cal E}(6\Omega-N,\tau,\lambda)+\chi F^{(t)}(\tau,\lambda)$) 
becomes 
\b\label{5-25}
{\cal E}^{(m)}(6\Omega-N,\tau,\lambda) > 0 \ , \qquad
\hbox{\rm i.e.,}\qquad 
\chi > \frac{-{\cal E}(6\Omega-N,\tau,\lambda)}{F^{(t)}(\tau,\lambda)} \ .
\end{equation}
Of course, in the cases ${\cal E}(N,\tau,\lambda)>0$ and 
${\cal E}(6\Omega-N,\tau,\lambda)>0$, the signs 
of the inequalities (\ref{5-24}) and (\ref{5-25}) are inverted. 
The above is a new feature which cannot be observed in the original Bonn 
model.

\subsection{The pairing correlation}

Before investigating the effect of $\chi{\hat {\mib Q}}^2$ in the present 
case, we discuss the case $\chi=0$. 
In this case, the most interesting discussion may be related to the 
comparison of the symmetric and the partial symmetric representation 
with each other. 
The former representation leads to $n_1=n_0$ and $\rho=0$. 
The energy 
eigenvalue (\ref{4-51}) reduces to 
\b\label{5-28}
E^{(s)}(N,n_0)=-\left(\frac{1}{2}N-\frac{3}{2}n_0\right)
\left(2\Omega +1-\frac{1}{2}n_0-\frac{1}{2}N\right) \ .
\end{equation}
Further, the relation (\ref{4-57}) gives us 
\b\label{5-29}
3n_0 \leq N \leq 4\Omega-n_0 \ .
\end{equation}
By replacing $N$ in the relations (\ref{5-28}) and (\ref{5-29}) with 
$(6\Omega-N)$ and by adding $2(3\Omega-N)$, we have 
\beq\label{5-30}
& &{\cal E}^{(s)}(6\Omega-N,n_0)=-\left(\frac{1}{2}N-\Omega-\frac{1}{2}n_0
\right)
\left(3\Omega +3-\frac{3}{2}n_0-\frac{1}{2}N\right) \ ,
\nonumber\\
& &2\Omega+n_0 \leq N \leq 6\Omega-3n_0 \ .
\eeq
The relation (\ref{4-54}) gives us 
\b\label{5-31}
0 \leq n_0 \leq \Omega \ . 
\end{equation}

Next, we show the results for the case $n_1 > n_0$, in which new notation 
$\delta$ is used: 
\b\label{5-32}
\delta=n_1-n_0 \ .
\end{equation}
Then, the relation (\ref{4-51}) is rewritten as 
\beq\label{5-33}
E^{(p)}(N,n_0,\delta,\rho)&=&
E^{(s)}(N,n_0)+\Delta E(N,n_0,\delta,\rho) \ , \nonumber\\
\Delta E(N,n_0,\delta,\rho)&=&
\delta\cdot \left(\Omega+\frac{1}{2}-n_0\right) -\frac{1}{4}\delta^2 
\nonumber\\
& &-2\rho\left(\frac{1}{2}N-\frac{3}{2}n_0+2+\frac{\delta}{2}\right)
+(2\rho)^2 \ .
\eeq
The quantities $n_0$, $\delta$, $N$ and $2\rho$ take their values in the 
following ranges: 
\begin{subequations}\label{5-34}
\beq
&(1)&\ \ 0\leq n_0 < \Omega\ , \quad 0<\delta \leq \Omega-n_0 \ , \nonumber\\
& &\ {\rm (i)}\ \ 3n_0+\delta \leq N \leq 3n_0+3\delta \ , \quad 
0\leq 2\rho \leq \frac{1}{2}(N-3n_0-\delta) \ , 
\label{5-34a}\\
& &\ {\rm (ii)}\ \ 3n_0+3\delta \leq N \leq 4\Omega-n_0-\delta \ , \quad 
0\leq 2\rho \leq \delta \ , 
\label{5-34b}\\
& &\ {\rm (iii)}\ 4\Omega-n_0-\delta \leq N \leq 4\Omega -n_0+\delta \ , 
\quad 
\frac{1}{2}(N-4\Omega +n_0+\delta) \leq 2\rho \leq \delta\ , \quad\quad\quad 
\label{5-34c}
\end{eqnarray}
\end{subequations}
\vspace{-0.8cm}
\begin{subequations}\label{5-35}
\begin{eqnarray}
&(2)&\ \ 0\leq n_0 < \Omega\ , \quad \Omega-n_0 \leq 
\delta \leq 2\Omega-2n_0 \ , \qquad\qquad\qquad\qquad\qquad\qquad\qquad\qquad\ 
\nonumber\\
& &\ {\rm (iv)}\ \ 3n_0+\delta \leq N \leq 4\Omega-n_0-\delta \ , \quad 
0\leq 2\rho \leq \frac{1}{2}(N-3n_0-\delta) \ , 
\label{5-35a}\\
& &\ {\rm (v)}\ \ 4\Omega-n_0-\delta \leq N \leq 8\Omega-5n_0-3\delta \ , 
\nonumber\\
& &\qquad\quad 
\frac{1}{2}\left(N-4\Omega+n_0+\delta\right) 
\leq 2\rho \leq \frac{1}{2}(N-3n_0-\delta) \ . 
\label{5-35b}
\eeq
\end{subequations}
The relations (\ref{5-34}) and (\ref{5-35}) come from the conditions 
(\ref{4-54}), (\ref{4-55}) and (\ref{4-63}) and the conditions 
(\ref{4-54}), (\ref{4-58}) and (\ref{4-64}), respectively. 
By replacing $N$ with $(6\Omega-N)$, we obtain the expressions 
calculated from the side $N=6\Omega$: 
\beq\label{5-36}
{\cal E}^{(p)}(6\Omega-N,n_0,\delta,\rho)&=&
{\cal E}^{(s)}(6\Omega-N,n_0)+\Delta E(6\Omega-N,n_0,\delta,\rho) \ , 
\nonumber\\
\Delta E(6\Omega-N,n_0,\delta,\rho)&=&
\delta\cdot \left(\Omega+\frac{1}{2}-n_0\right) -\frac{1}{4}\delta^2 
\nonumber\\
& &-2\rho\left(3\Omega-\frac{1}{2}N-\frac{3}{2}n_0+2+\frac{\delta}{2}\right)
+(2\rho)^2 \ . 
\eeq
\vspace{-0.8cm}
\begin{subequations}\label{5-37}
\beq
&(1)'&\ \ 0\leq n_0 < \Omega\ , \quad 0<\delta \leq \Omega-n_0 \ , \nonumber\\
& &\ {\rm (i)'}\ \ 6\Omega-3n_0-3\delta \leq N 
\leq 6\Omega-3n_0-\delta \ , \quad 
0\leq 2\rho \leq \frac{1}{2}(6\Omega-N-3n_0-\delta)\ , \nonumber\\
& &\label{5-37a}\\
& &\ {\rm (ii)'}\ 2\Omega+n_0+\delta \leq N \leq 6\Omega-3n_0-\delta \ , 
\quad 
0 \leq 2\rho \leq \delta\ , \nonumber\\
& &\label{5-37b}\\
& &\ {\rm (iii)'}\ \ 2\Omega+n_0-\delta \leq N 
\leq 2\Omega+n_0+\delta \ , \quad 
\frac{1}{2}(2\Omega-N+n_0+\delta)\leq 2\rho \leq \delta \ , \nonumber\\
& &\label{5-37c}
\end{eqnarray}
\end{subequations}
\vspace{-0.8cm}
\begin{subequations}\label{5-38}
\begin{eqnarray}
&(2)'&\ \ 0\leq n_0 < \Omega\ , \quad \Omega-n_0 \leq 
\delta \leq 2\Omega-2n_0 \ , \ \nonumber\\
& &\ {\rm (iv)'}\ \ 2\Omega+n_0+\delta \leq N \leq 6\Omega-3n_0-\delta \ , 
\quad 
0 \leq 2\rho \leq \frac{1}{2}(6\Omega-N-3n_0-\delta)\ , \nonumber\\
& &\label{5-38-a}\\
& &\ {\rm (v)'}\ \ -2\Omega+5n_0+3\delta \leq N 
\leq 2\Omega+n_0+\delta \ , \nonumber\\
& &\qquad\qquad
\frac{1}{2}(2\Omega-N+n_0+\delta) \leq 2\rho \leq 
\frac{1}{2}(6\Omega-N-3n_0-\delta) 
\ . 
\label{5-38b}
\eeq
\end{subequations}
%
\begin{figure}[b]
\begin{center}
\includegraphics[height=8cm]{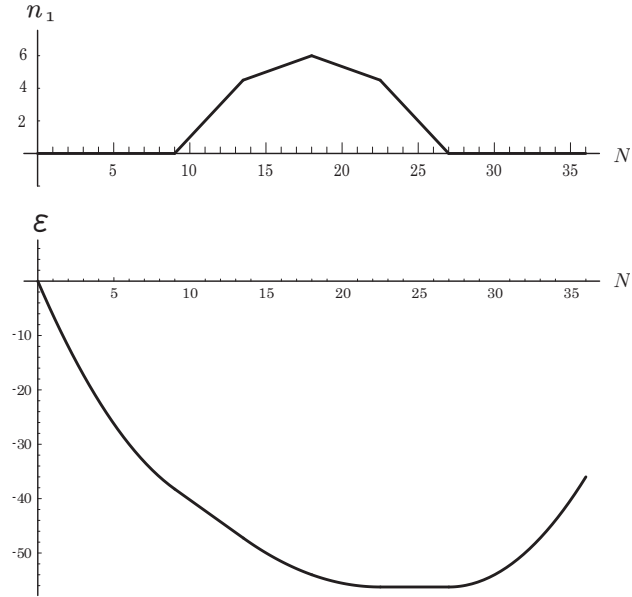}
\caption{The total energy (lower panel) and $n_1$ (upper panel) 
are shown in a function of $N$. The parameters are taken as 
$\Omega=6$ and $\chi=0$. 
}
\label{fig:5-1}
\end{center}
\end{figure}

We can choose the minimum energy state. 
In the case $\chi=0$, 
the total energy and the quantity $n_1$ in the 
energy minimum states 
are depicted as functions of particle 
number $N$ in the lower and upper panels, respectively, in Fig.\ref{fig:5-1}, 
in which we fix $n_0=0$. 
The parameter $\Omega$ is taken as 6 and in this numerical calculation, 
we treat $N$ as a continuous variable. 
Of course, $N$ is meaningful when $N$ is integer. 
In this case, it should be noted that $n_1$ changes continuously, and then, 
the behavior of energy as a function of $N$ corresponds to 
the change of $n_1$.

\begin{figure}[t]
\begin{center}
\includegraphics[height=8cm]{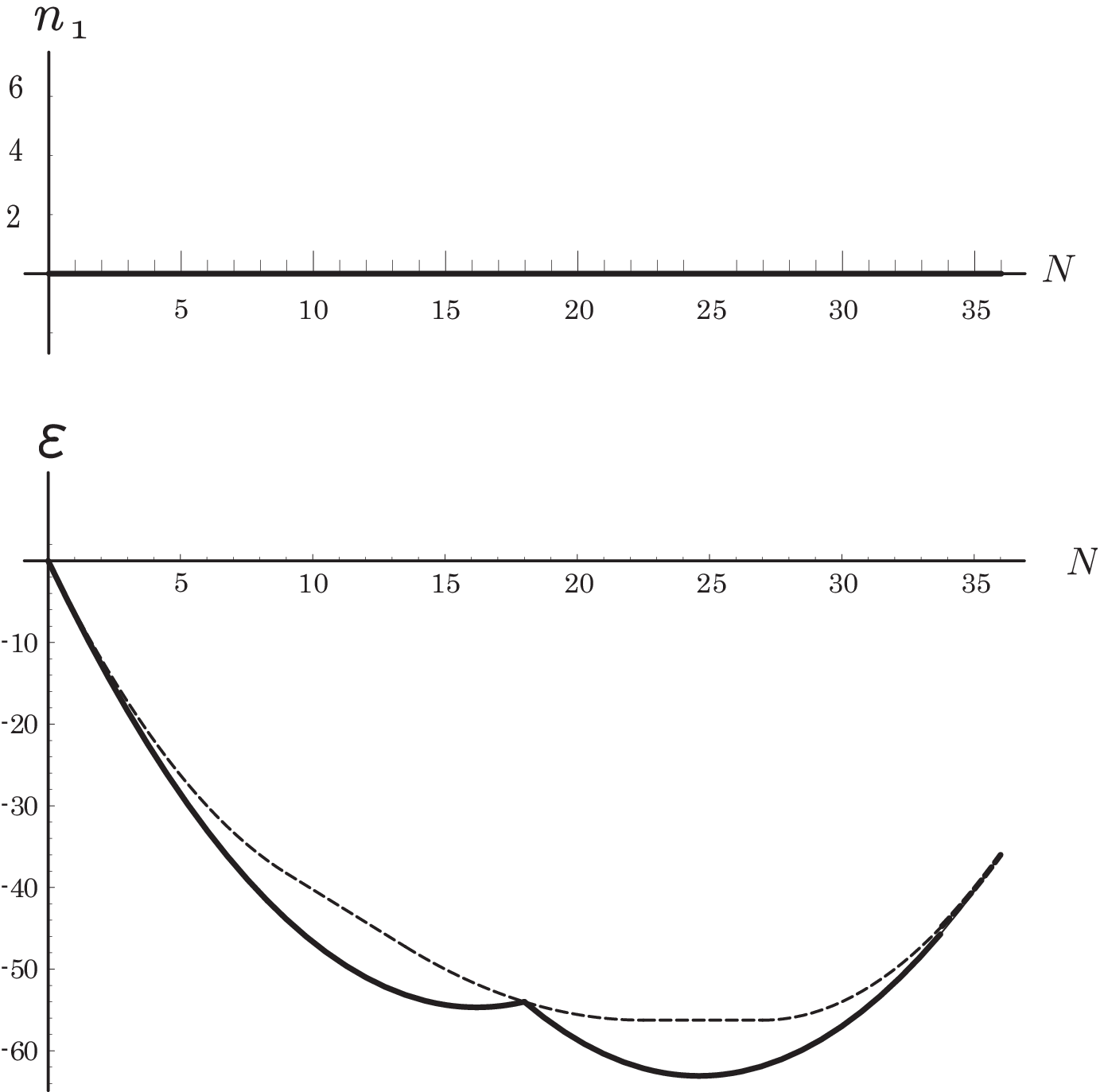}
\caption{The total energy (lower panel) and $n_1$ (upper panel) 
are shown in a function of $N$. The parameters are taken as 
$\Omega=6$ and $\chi=-1/4$. In the lower panel, the dashed curve represents 
the energy in the case $\chi=0$.}
\label{fig:5-2}
\end{center}
\end{figure}

\begin{figure}[t]
\begin{center}
\includegraphics[height=8cm]{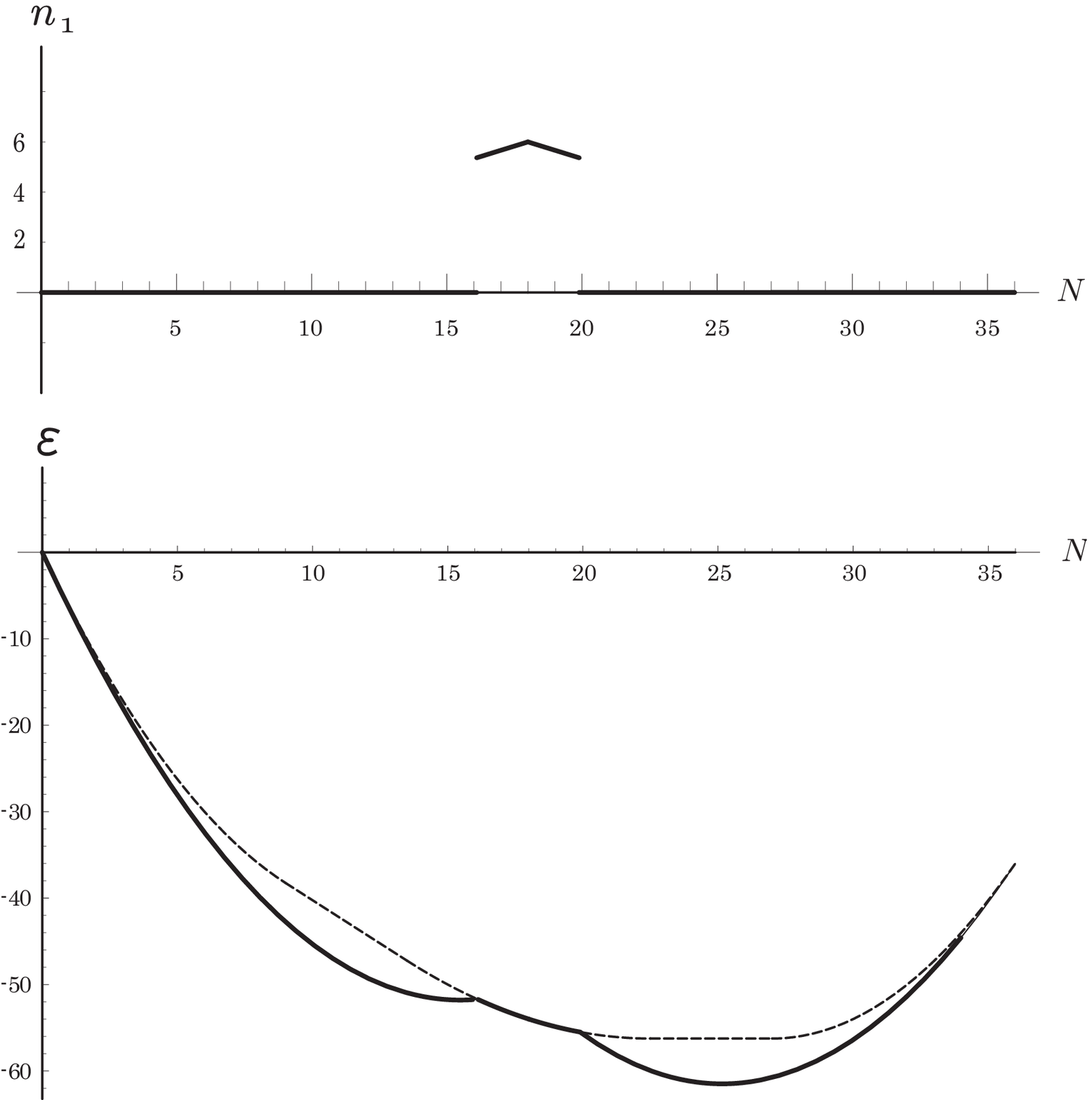}
\caption{The total energy (lower panel) and $n_1$ (upper panel) 
are shown in a function of $N$. The parameters are taken as 
$\Omega=6$ and $\chi=-1/5$. In the lower panel, the dashed curve represents 
the energy in the case $\chi=0$.
}
\label{fig:5-3}
\end{center}
\end{figure}
\begin{figure}[t]
\begin{center}
\includegraphics[height=8cm]{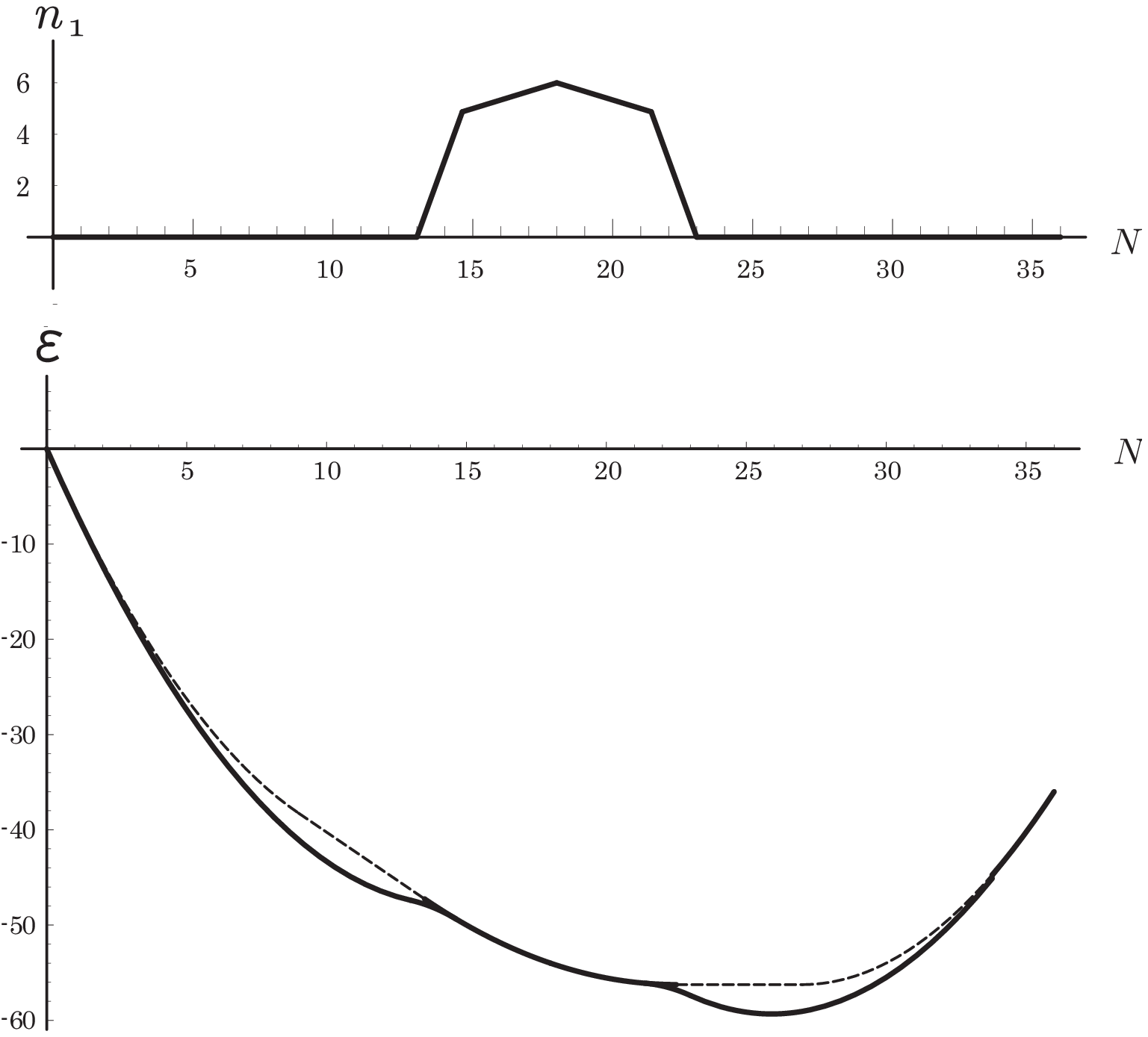}
\caption{The total energy (lower panel) and $n_1$ (upper panel) 
are shown in a function of $N$. The parameters are taken as 
$\Omega=6$ and $\chi=-1/8$. In the lower panel, the dashed curve represents 
the energy in the case $\chi=0$.
}
\label{fig:5-4}
\end{center}
\end{figure}

Our next task is to investigate the effect of $\chi{\hat {\mib Q}}^2$. 
In \S 5.1, we learned that in the case of the triplet formation there exist 
the color-neutral states, which are not influenced by this effect. 
In the present case, we can show also that there exist the states which 
are not influenced by $\chi{\hat {\mib Q}}^2$ and they are identical to the 
states (\ref{5-18}) and (\ref{5-19}). 
First, we set up the condition that 
${\hat Q}_+(\lambda\mu\nu\nu_0)$ defined in the relation (\ref{a8}) 
should be a unit operator. 
This can be realized in the case 
\b\label{5-39}
\lambda=\mu=\nu=\nu_0=0 \ .
\end{equation}
This condition is analogous to the condition (\ref{5-17}). 
Under this condition, the state (\ref{4-24}) is reduced to 
$\ket{0\rho\sigma_1\sigma_0}=({\hat S}^4)^{2\rho}\ket{\sigma_0,\sigma_1}$. 
From the set $\{\ket{0\rho\sigma_1\sigma_0}\}$, we select the 
states in which the energy eigenvalues are not influenced by 
$\chi{\hat {\mib Q}}^2$. 
If we notice $F_{\sigma_1\sigma_0\rho\lambda}^{(p)}$ given in the 
relation (\ref{4-43b}), the condition for selecting the 
states with our aim is given as 
\begin{equation}\label{5-39add}
F_{\sigma_1\sigma_0\rho 0}^{(p)}=\frac{2}{3}
\left(2(\sigma_1-\sigma_0)-2\rho\right)
\left(2(\sigma_1-\sigma_0)-2\rho+3\right)=0 \ .
\end{equation}
Noting $2(\sigma_1-\sigma_0)=n_1-n_0$, the relation (\ref{5-39add}) gives 
us 
\begin{equation}\label{5-40add}
2\rho=n_1-n_0 \ . 
\end{equation}
Then, the total quark number $N$ shown in the relation (\ref{4-50}) is 
given as 
\begin{equation}\label{5-41add}
N=2n_0+n_1+2(n_1-n_0)=3n_1\ , \quad \hbox{\rm i.e.,}\quad
n_1=\frac{N}{3} \ . 
\end{equation}
The energy eigenvalue (\ref{4-51}) is reduced to 
\begin{equation}\label{5-42add}
E_{Nn_0n_1\rho}=-\left(\frac{N}{3}-n_0\right)
\left(2\Omega+3-\frac{N}{3}-n_0\right) \ .
\end{equation}
This form is identical to the energy eigenvalue for the color-neutral 
triplet state (\ref{5-2}). 
With the use of the expression (\ref{4-30}), the eigenstates are obtained in 
the form 
\begin{eqnarray}
& &\ket{0\rho\sigma_1\sigma_0}
=\left(\sum_{r=0}^{2\rho}(-)^{2\rho-r}D_{t\tau\rho}(r)
({\hat \tau}_+)^{r}({\hat t}_+)^{n_1-n_0-r}\right)
({\hat b}^*)^{2(\Omega-n_1)}\ket{0}\ , \nonumber\\ 
& &
\qquad\qquad\qquad\qquad\qquad\qquad\qquad\qquad
\qquad\qquad\qquad\qquad\qquad
(n_1 \leq \Omega) 
\label{5-43add}\\
& &\ket{0\rho\sigma_1\sigma_0}
=\left(\sum_{r=0}^{2\rho}(-)^{2\rho-r}D_{t\tau\rho}(r)
({\hat \tau}_+)^{r}({\hat t}_+)^{2\Omega-n_0-n_1-r}\right)
({\hat a}^*)^{2(n_1-\Omega)}\ket{0}\ .  \nonumber\\
& &
\qquad\qquad\qquad\qquad\qquad\qquad\qquad\qquad
\qquad\qquad\qquad\qquad\qquad
(n_1 \geq \Omega) 
\label{5-44add}
\end{eqnarray}
We can show that the states (\ref{5-43add}) and (\ref{5-44add}) are reduced 
to the states (\ref{5-18}) and (\ref{5-19}), respectively, through 
the relations 
$\tau_0=3/2+r$, $t_0=\Omega+1/2-n_0-r$, $t=\Omega-n_1+1/2$ ($n_1\leq \Omega$) 
and 
$t=n_1-\Omega+1/2$ $(n_1\geq \Omega)$. 
Thus, in the framework of the pair correlation, we could derive 
the color-neutral triplet states.

The state $\ket{0\rho\sigma_1\sigma_0}$ shown in the forms 
(\ref{5-43add}) and (\ref{5-44add}) can be rewritten as 
\begin{subequations}\label{5-45add}
\begin{eqnarray}
& &\ket{0\rho\sigma_1\sigma_0}=({\hat S}^4)^{n_1-n_0}
({\hat b}_1^*{\hat b})^{n_1-n_0}\ket{\Omega(n_0)} \ , 
\label{5-45aadd}\\
& &\ket{\Omega(n_0)}=({\hat b}^*)^{2(\Omega-n_0)}\ket{0} \  .
\label{5-45addb}
\end{eqnarray}
\end{subequations}
The state $\ket{\Omega(n_0)}$ is the eigenstate of ${\hat \Omega}$ and 
${\hat N}_i$ defined in the relations (\ref{2-53}) and (\ref{2-54}) 
with the eigenvalues $\Omega$ and $n_0$, respectively. 
Further, we have ${\hat b}^*{\hat b}_1\ket{\Omega(n_0)}
={\hat S}_4\ket{\Omega(n_0)}=0$. 
Therefore, the color-neutral triplet state $\ket{0\rho\sigma_1\sigma_0}$ 
is generated by operating $({\hat S}^4)^{n_1-n_0}
({\hat b}_1^*{\hat b})^{n_1-n_0}$ on $\ket{\Omega(n_0)}$, 
which is also color-neutral. 
The above argument teaches us that the operator 
$({\hat S}^4)^{n_1-n_0}({\hat b}_1^*{\hat b})^{n_1-n_0}$ plays a role of 
generating $(n_1-n_0)$ color-neutral triplets on $\ket{\Omega(n_0)}$, 
i.e., $(n_1-n_0)$ nucleons. 
Then, $n_0$ nucleons may be in the $\Delta$-excitation. 
The above is nothing but the picture presented in the 
original Bonn model.\cite{a} 
The above argument can be generalized to the case $\lambda \neq 0$. 
The state $\ket{\lambda\rho\sigma_1\sigma_0}$ is rewritten as 
\begin{equation}\label{5-46-2}
\ket{\lambda\rho\sigma_1\sigma_0}=
({\hat S}^3)^{2\lambda}\cdot ({\hat S}^4)^{2\rho}
({\hat b}_1^*{\hat b})^{2\rho}\cdot 
({\hat b}_1^*)^{2(\sigma_1-\sigma_0-\rho)}({\hat b}^*)^{2(\sigma_0-\rho)}
\ket{0} \ . 
\end{equation}
By operating $({\hat S}^4)^{2\rho}({\hat b}_1^*{\hat b})^{2\rho}$ on the state 
$({\hat b}_1^*)^{2(\sigma_1-\sigma_0-\rho)}({\hat b}^*)^{2(\sigma_0-\rho}
\ket{0}$, the triplet state is formed, and further, by operating 
$({\hat S}^3)^{2\lambda}$ on this state, we obtain 
$\ket{\lambda\rho\sigma_1\sigma_0}$. 
The above is very similar to the case of the $su(2)$-pairing model.

%
%

Under the above circumstance, we analyze the effect of 
$\chi{\hat {\mib Q}}^2$. 
The eigenvalue of ${\hat {\mib Q}}^2$ is given in the relation 
(\ref{4-52}) and it is positive-definite. 
Therefore, if $\chi>0$ or $\chi <0$, the energy eigenvalue becomes larger or 
smaller than the original value $E_{Nn_0n_1\rho}$, respectively, and 
its value $E_{Nn_0n_1\rho}^{(m)}$ can be expressed as 
\beq\label{5-40}
E_{Nn_0n_1\rho}^{(m)}&=&
E_{Nn_0n_1\rho}+\chi F_{Nn_0n_1\rho}^{(p)} \nonumber\\
&=&(1+2\chi)E_{Nn_0n_1\rho}+\chi G_{Nn_0n_1} \ . 
\eeq
Of course, we have also 
\b\label{5-41}
{\cal E}_{6\Omega-Nn_0n_1\rho}^{(m)}
=(1+2\chi){\cal E}_{6\Omega-Nn_0n_1\rho}+\chi G_{6\Omega-Nn_0n_1} \ . 
\end{equation}
In the relations (\ref{5-40}) and (\ref{5-41}), we can see interesting 
features. 
The effects of $\chi{\hat {\mib Q}}^2$ are divided to two types. 
First is related to $\chi G_{Nn_0n_1}$. 
In this case, $F_{Nn_0n_1\rho}^{(p)}>0$ and $E_{Nn_0n_1\rho}<0$, 
and then, $G_{Nn_0n_1}>0$. 
Therefore, depending the sign of $\chi$, all the energy eigenvalues 
specified by ($N,n_0,n_1$) are raised or lowered. 
Second is related to $(1+2\chi)E_{Nn_0n_1\rho}$. 
By the factor $(1+2\chi)$, $E_{Nn_0n_1\rho}$ is scaled up or down. 
In the triplet formation, all the energy eigenvalues specified by 
$(t, \lambda)$ are raised or lowered, but, $E_{Nn_0\tau\lambda}$ 
itself does not change. 

\begin{figure}[t]
\begin{center}
\includegraphics[height=8cm]{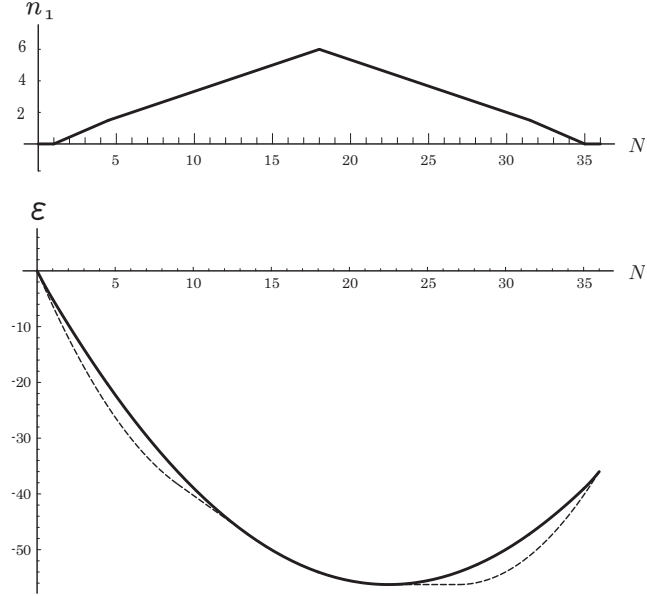}
\caption{The total energy (lower panel) and $n_1$ (upper panel) 
are shown in a function of $N$. The parameters are taken as 
$\Omega=6$ and $\chi=1$. In the lower panel, the dashed curve represents 
the energy in the case $\chi=0$.
}
\label{fig:5-5}
\end{center}
\end{figure}

\begin{figure}[t]
\begin{center}
\includegraphics[height=8cm]{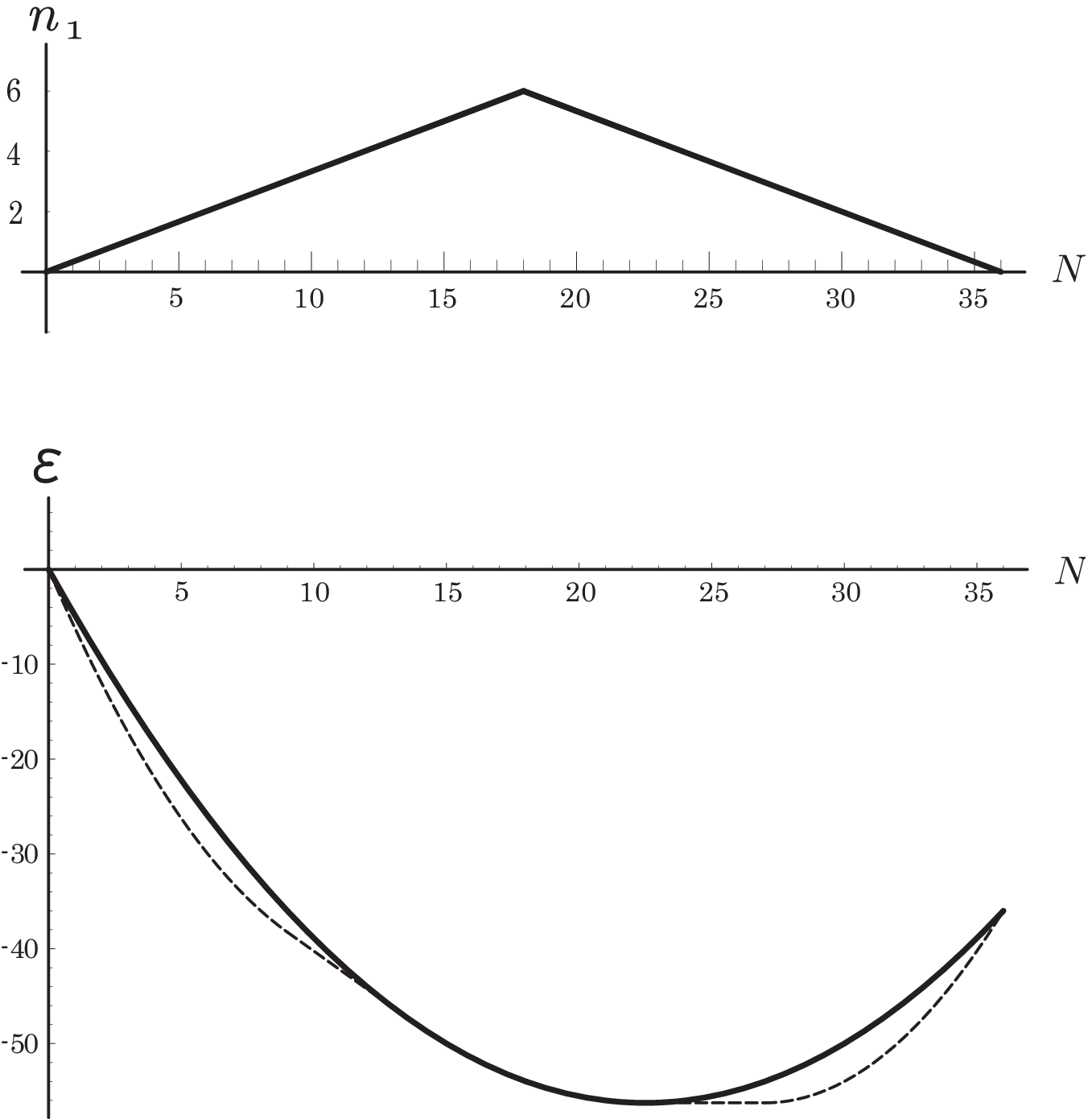}
\caption{The total energy (lower panel) and $n_1$ (upper panel) 
are shown in a function of $N$. The parameters are taken as 
$\Omega=6$ and $\chi=3/2$. In the lower panel, the dashed curve represents 
the energy in the case $\chi=0$.
}
\label{fig:5-6}
\end{center}
\end{figure}

The effects of $\chi{\hat {\mib Q}}^2$ are investigated numerically. 
As is mentioned in \S 5.1., there is the situation in which 
the energy of colored state is lower than that of the color neutral one 
in the original Bonn model with $\chi=0$. 
However, if the $\chi{\hat {\mib Q}}^2$ term, which also retains the 
color $su(3)$ symmetry, is switched on, then the energy of colored state 
is influenced while the energy of color neutral one has no effect. 
If $\chi$ is positive, the energy of colored state raises in comparison 
with that 
of color neutral one. 
Thus, the color neutrality of the physical state is not broken. 
The same situation occurs because the color neutral 
state has no effect for the $\chi{\hat {\mib Q}}^2$ term due to 
the honor of Eq.(\ref{5-40add}) in the pairing correlation. 
Thus, we should take $\chi$ being positive in order to retain the 
color neutrality of the physical state. 
However, for the sake of the instruction, first, we take $\chi$ as negative 
values. 

In Fig.\ref{fig:5-2}, the total energy (lower panel) and $n_1$ 
(upper panel) are shown for the 
energy minimum state in the case $\chi=-1/4$ and $\Omega=6$. 
We fix the same value $n_0=0$ as the case $\chi=0$ in Fig.\ref{fig:5-1}.  
The dashed curve represents the minimum 
energy in the case $\chi=0$ for the comparison. 
For all the regions of $N \ (=0 \sim 6\Omega)$, 
the quantity $n_1$ is zero and the 
total energy with $\chi=-1/4$ is lower than that with $\chi=0$. 
Figure \ref{fig:5-3} shows the same quantities (total energy and $n_1$) 
in the case $\chi=-1/5$. 
It should be noted that the structure for $n_1$ appears, 
that is, the quantity $n_1$ shows discontinuity. 
Namely, the $n_1$ has gap. 
In the regions $N\leq 16$ and $N\geq 20$, the quantity 
$n_1$ is zero. 
In these regions, the total energy is lower than that 
in the color neutral case, $\chi=0$. 
However, in the region $16<N<20$, $n_1$ is equal to $N/3$ or $2\Omega-N/3$. 
In this region, the minimum energy is equivalent to that of the color neutral 
case, $\chi=0$. 
Thus, according to $n_1=0$ or $n_1 \neq 0$, the structure of the 
minimum energy states changes from that of the 
colored state to the color neutral one.

Contrarily, in the case $\chi=-1/8$, the quantity $n_1$ has no gap and 
changes continuously as is shown in Fig.\ref{fig:5-4}. 
But, the situation is similar to the case $\chi=-1/5$. 
In the regions $0 \leq N \leq 13$ and $23 \leq N \leq 36$, 
$n_1$ is zero. 
On the other hand, in the regions $14\leq N\leq 18$ and $18 \leq N 
\leq 22$, the quantity $n_1$ is $N/3$ and $2\Omega-N/3$, respectively. 
In the regions $13\leq N \leq 14$ and $22\leq N \leq 23$, $n_1$ has
the value where all regions are connected continuously. 
In correspondence with the change of $n_1$, 
the state with minimum energy also changes. 

The above facts pointed out in Figs.\ref{fig:5-2}$\sim$\ref{fig:5-4}, 
which show the total energy and $n_1$ with negative values for $\chi$, 
indicate that 
the quantity $n_1$ can be regarded as a order parameter of the 
phase transition. 
In the case $\chi=-1/5$, $n_1$ has gap, and then, the transition 
may be regarded as the first order. 
On the contrary, in the case $\chi=-1/8$, $n_1$ has no gap, and then, 
the transition may be regarded as the second order in terms of the 
usual phase transition. We can show that, in these parameterizations 
$\Omega=6$ and $n_0=0$, $\chi=-1/6$ gives a critical point, that is, 
$\chi \leq -1/6$ gives the first order, and $\chi >-1/6$ gives 
the second order.

Let us return to the physical situation, that is, $\chi >0$. 
In Figs.\ref{fig:5-5} and \ref{fig:5-6} with $\Omega=6$ and $n_0=0$, 
the same physical quantities as those previously presented in 
Figs.\ref{fig:5-2}$\sim$\ref{fig:5-4} are shown 
in the case $\chi=1$ and $\chi=3/2$, respectively. 
In the case $\chi >0$, the energy is pushed up in comparison with 
that in the case $\chi=0$. 
This situation is the same as that of the triplet formation. 
In the case $\chi=3/2$, there is no structure as for $n_1$. 
Namely, $n_1$ increases monotonically in the region $0\leq N\leq 18\ (=
3\Omega)$ with $n_1=N/3$ and decreases monotonically in the 
region $18 \leq N\leq 36\ (=6\Omega$) with $n_1=2\Omega-N/3$. 
On the other hand, in the case $\chi=1$, the structure appears as for 
$n_1$. 
If we take $N$ as a continuous variable, for example, 
$n_1$ is equal to 0, $3(N-1)/7$ and $N/3$ 
in the regions 
$0\leq N \leq 1$, $1\leq N\leq 4.5$ and $4.5 \leq N\leq 18$, 
respectively.

\section{Final remark --- Possibility of the asymmetric representation}

Until the present, we have described the Bonn model and its modification 
in the framework of the Schwinger boson representation of the 
$su(4)$ algebra. 
The basic idea is the use of the relation (\ref{2-27}) and its associated 
relation (\ref{2-28}), and we could treat many-quark system with the 
$su(4)$ algebraic structure in the case of the symmetric representation. 
However, one open question still remains to be answered. 
It is the description of the present many-quark system in the framework 
of the asymmetric representation. 
In this section, we will sketch this problem.

As was already shown, the form (\ref{2-27}) which comes from the case 
($M=3, N=1$) in Ref.\citen{c} cannot generate any asymmetric representation. 
Then, as a next step, we consider the case $(M=3, N=2)$: 
\beq\label{6-1}
& &{\hat S}^i=({\hat a}_i^*{\hat b}-{\hat a}^*{\hat b}_i)
-{\hat \alpha}^*{\hat \beta}_i \ , \qquad
{\hat S}_i=({\hat b}^*{\hat a}_i-{\hat b}_i^*{\hat a})
-{\hat \beta}_i^*{\hat \alpha} \ , \nonumber\\
& &{\hat S}_i^j=({\hat a}_i^*{\hat a}_j-{\hat b}_j^*{\hat b}_i
+\delta_{ij}({\hat a}^*{\hat a}-{\hat b}^*{\hat b}))
-{\hat \beta}_j^*{\hat \beta}_i+\delta_{ij}{\hat \alpha}^*{\hat \alpha} \ .
\eeq
Here, ${\hat \alpha}$, ${\hat \alpha}^*$, ${\hat \beta}_i$, ${\hat \beta}_i^*$ 
($i=1,2,3$) denote newly added boson operators. 
The terms in the brackets on the right-hand sides 
are nothing but the terms in the form (\ref{2-27}). 
The expression (\ref{2-27}) is associated with the $su(1,1)$ algebra which 
is shown in the relation (\ref{2-28}). 
The form (\ref{6-1}) is associated with the $su(2,1)$ algebra shown in the 
relation 
\beq\label{6-2}
& &{\hat T}^1={\hat b}^*{\hat a}^*+\sum_i {\hat b}_i^*{\hat a}_i^* \ , 
\qquad
{\hat T}_1={\hat a}{\hat b}+\sum_i {\hat a}_i{\hat b}_i \ , \nonumber\\
& &{\hat T}^2={\hat b}^*{\hat \alpha}^*+\sum_i {\hat \beta}_i^*{\hat a}_i^* 
\ , 
\qquad
{\hat T}_2={\hat \alpha}{\hat b}+\sum_i {\hat a}_i{\hat \beta}_i \ , \nonumber\\& &{\hat T}_2^1={\hat a}^*{\hat \alpha}+\sum_i {\hat b}_i^*{\hat \beta}_i 
\ , 
\qquad
{\hat T}_1^2={\hat \alpha}^*{\hat a}+\sum_i {\hat \beta}_i^*{\hat b}_i \ , 
\nonumber\\
& &{\hat T}_1^1={\hat b}^*{\hat b}+{\hat a}^*{\hat a}
+\sum_i ({\hat b}_i^*{\hat b}_i +{\hat a}_i^*{\hat a}_i) +4 \ , 
\nonumber\\
& &{\hat T}_2^2={\hat b}^*{\hat b}+{\hat \alpha}^*{\hat \alpha}
+\sum_i ({\hat \beta}_i^*{\hat \beta}_i +{\hat a}_i^*{\hat a}_i) +4 \ . 
\eeq
In the same sense as the relation (\ref{2-30}), we have 
\begin{equation}\label{6-3}
[\ \hbox{\rm any\ of\ the\ generators\ (\ref{6-2})\ , \ 
any\ of\ the\ generators\ (\ref{6-1})}\ ]=0 \ .
\end{equation}

The minimum weight state $\ket{m}$ which satisfies the conditions 
corresponding to the relations (\ref{2-12}) and (\ref{2-14}) is 
given in the form 
\begin{equation}\label{6-4}
\ket{m}=({\hat b}_1^*)^{2(\sigma_1-\sigma_2)}
({\hat b}_1^*{\hat \beta}_2^*-{\hat b}_2^*{\hat \beta}_1^*)^{2(
\sigma_2-\sigma_3)}({\hat b}^*)^{2\sigma_3}\ket{0} \ .
\end{equation}
This form was already discussed in Ref.\citen{6} for the case of 
the Lipkin model. 
Certainly, $\ket{m}$ is in the asymmetric representation and if $\sigma_2
=\sigma_3$ ($=\sigma_0$), $\ket{m}$ reduces to $\ket{m_1}$ or $\ket{m_0}$. 
The state $\ket{m}$ satisfies 
\beq\label{6-5}
& &{\hat T}_1\ket{m}={\hat T}_2\ket{m}={\hat T}_2^1\ket{m}=0 \ , \nonumber\\
& &{\hat T}_1^1\ket{m}=(2\sigma_1+4)\ket{m} \ , \qquad
{\hat T}_2^2\ket{m}=(2\sigma_2+4)\ket{m} \ .
\eeq
The above indicates that $\ket{m}$ is also the minimum weight state of the 
$su(2,1)$ algebra. 
Combining this fact with the relation (\ref{6-3}), we can conclude 
that in the present Schwinger boson representation the orthogonal set 
for the $su(4)\otimes su(2,1)$ algebra is given by appropriate 
operation of six generators 
$({\hat S}^1, {\hat S}^2, {\hat S}^3, {\hat S}_1^2, {\hat S}_1^3, 
{\hat S}_2^3$) and three generators $({\hat T}^1, {\hat T}^2, {\hat T}_1^2)$. 
As a general argument, we recognize that the minimum weight state for the 
$su(4)$ and the $su(2,1)$ algebra is specified by three quantum numbers 
and except them, six and three quantum numbers for the $su(4)$ and 
the $su(2,1)$ algebra are necessary, respectively, to specify the 
orthogonal set for the $su(4)\otimes su(2,1)$ algebra. 
Our present Schwinger boson representation is composed of twelve kinds 
of bosons, and then, the orthogonal set of the present boson space is 
specified by twelve quantum numbers. 
For the above argument, we can conjecture that the form (\ref{6-1}) 
can present any of the orthogonal set for the $su(4)$ algebra. 
This conjecture seems for us to arrive at a conclusion that our 
Schwinger boson representation generates the asymmetric representation for 
the Bonn model.

However, the above argument lacks an important factor of the Bonn model, 
which was taken up in \S 2.3. 
We must investigate this factor. 
In parallel with the relation (\ref{2-12}), we set up the following 
relation under the hole picture: 
\beq
& &{\breve S}^i|\tilde m)=0 \ , \quad 
{\breve S}_2^1|\tilde m)={\breve S}_3^1|\tilde m)={\breve S}_3^2|\tilde m)=0
\ , 
\label{6-6}\\
& &{\breve S}_i^i|\tilde m)=-2\sigma_i|\tilde m) \ . 
\label{6-7}
\eeq
Here, $|\tilde m)$ denotes the minimum weight state in the hole picture 
and ${\breve S}_i$ etc. are defined in the relation (\ref{2-56}). 
With the use of this relation, Eqs.(\ref{6-6}) and (\ref{6-7}) 
can be rewritten as 
\beq
& &{\wtilde S}^i|\tilde m)=0 \ , \quad 
{\wtilde S}_1^2|\tilde m)={\wtilde S}_1^3|\tilde m)={\wtilde S}_2^3|\tilde m)=0
\ , 
\label{6-8}\\
& &{\wtilde S}_i^i|\tilde m)=2\sigma_i|\tilde m) \ . 
\label{6-9}
\eeq
We investigate the asymmetric representation related to the counterparts of 
the relations (\ref{6-8}) and (\ref{6-9}) in the frame of the 
Schwinger boson representation (\ref{6-1}). 
In parallel with $\ket{m}$ shown in the relation (\ref{6-4}), we obtain 
the following form: 
\beq
& &\ket{\tilde m}=({\hat a}_1^*)^{2(\sigma_1-\sigma_0)}
({\hat \alpha}^*)^{2\sigma_0'}({\hat a}^*)^{2(\sigma_0-\sigma_0')}\ket{0} \ , 
\label{6-10}\\
& &\qquad
\sigma_2=\sigma_3=\sigma_0 \ , \qquad \sigma_0'\ : \ 
\hbox{\rm arbitrary\ in\ the\ range}\ 
0\leq \sigma_0'\leq \sigma_0 \ . 
\label{6-11}
\eeq
The state $\ket{\tilde m}$ belongs to the symmetric representation. 
If $\sigma_0'=0$, $\ket{\tilde m}$ reduces to the form replaced 
${\hat b}_1^*$ and ${\hat b}^*$ in $\ket{m_1}$ or $\ket{m_0}$ by 
${\hat a}_1^*$ and ${\hat a}^*$. 
The above teaches us that the Schwinger boson representation cannot 
generate the asymmetric representation. 
Thus, we have two conclusions which are contradictory to each other. 
The reason why such contradiction occurs is simple: 
The expression (\ref{6-1}) does not adapt to the property such as shown in the 
relation (\ref{2-68}). 
If our argument is based on the conjecture mentioned in the last paragraph, 
we have to conclude that for the Bonn model the treatment by the asymmetric 
representation may be meaningless.

If it is permitted to introduce further new boson operators 
$({\hat \alpha}_i, {\hat \alpha}_i^*, {\hat \beta}, {\hat \beta}^*)$, we 
can give the property such as shown in the relation (\ref{2-68}) 
to the present system. 
It is performed through the process that the terms 
${\hat \alpha}_i^*{\hat \beta}$, ${\hat \beta}^*{\hat \alpha}_i$ and 
$({\hat \alpha}_i^*{\hat \alpha}_j-\delta_{ij}{\hat \beta}^*{\hat \beta})$ 
are added to ${\hat S}^i$, ${\hat S}_i$ and ${\hat S}_i^j$ shown in the 
relation (\ref{6-1}). 
The result is as follows: 
\beq\label{6-12}
{\hat S}^i&=&({\hat a}_i^*{\hat b}-{\hat a}^*{\hat b}_i)
+({\hat \alpha}_i^*{\hat \beta}-{\hat \alpha}^*{\hat \beta}_i) \ , \nonumber\\
{\hat S}_i&=&({\hat b}^*{\hat a}_i-{\hat b}_i^*{\hat a})
+({\hat \beta}^*{\hat \alpha}_i-{\hat \beta}_i^*{\hat \alpha}) \ , \nonumber\\
{\hat S}_i^j&=&({\hat a}_i^*{\hat a}_j-{\hat b}_j^*{\hat b}_i
+\delta_{ij}({\hat a}^*{\hat a}-{\hat b}^*{\hat b})) \nonumber\\
& &+({\hat \alpha}_i^*{\hat \alpha}_j-{\hat \beta}_j^*{\hat \beta}_i
+\delta_{ij}({\hat \alpha}^*{\hat \alpha}-{\hat \beta}^*{\hat \beta})) 
\ . 
\eeq
The above is merely a simple sum of two independent bags and each bag can be 
treated by the method presented in this paper. 
This consideration also supports that in the Bonn model obeying the $su(4)$ 
algebra, the asymmetric representation may be meaningless.

\section{Concluding remarks}

In this paper, we described the Bonn model and its modification for many-quark system 
in the framework of the Schwinger boson representation of the $su(4)$ algebra 
in the full and the partial symmetric representation. 
The asymmetric representation may be meaningless. 
All the results are given analytically in the exact form. 
Further, we could show up various features hidden in this model. 
However, we must point out that there is an unsolved problem. 
In this paper, we discussed the Bonn model by separating into two 
cases under the name of the triplet formation and the pairing 
correlation. 
Both descriptions seem to be apparently very different from each other. 
Are both essentially equivalent to each other or not ? 
This is our open problem and it is our future problem.

As the concluding remarks, we will mention another future problem. 
For the present development, we found it most convenient to 
consider the Schwinger boson representation of the $su(4)$ algebra. 
It is well known that the dynamics of many-fermion system 
may be described in terms of bosons. 
As an example in old time, in the collective model of Bohr and 
Mottelson, bosons were introduced through the quantization of the 
oscillations of a liquid drop to describe excitations of nuclei. 
In the theory of plasma oscillation, excited states of the 
electron gas are described by the so-called random phase 
approximation which is nothing else but the mapping of the 
particle-hole pairs onto bosons. 
In either cases, the physical meaning of the boson operators is quite clear. 
However, the bosons used in the present investigation seem to give us 
an impression to be no more than the tools for describing 
many-fermion system in spite of the success. 
This statement tells us that the physical interpretation of the bosons 
in the Schwinger representation remains a challenging question. 
In relation to the above-mentioned question, inevitably, 
we must investigate the present form in the original fermion space. 
This is also our future problem.

\section*{Acknowledgements} 
One of the authors (Y.T.) would like to express his sincere thanks to 
Professor J. da Provid\^encia, one of co-authors of this paper, 
for his warm hospitality during his visit to Coimbra in spring of 2008. 
The author (M.Y.) would like to express sincere thanks to Professor 
T. Kunihiro for suggesting him the importance of the $so(5)$- and 
the $su(4)$-algebra in the description of many-fermion dynamics, 
which was presented in Refs.\citen{b} and \citen{13}. 
Further, he would like to acknowledge to the members of 
the Department of Pure and Applied Physics of Kansai University 
opened in 2007, especially, to Professors N. Ohigashi, M. Sugihara-Seki and 
T. Wada for their cordial encouragements. 
He is also indebted to Dr. T. Urade of Osaka Prefectural University for 
helping him to exchange the research information between J. da P. and 
M. Y. 
One of the authors (Y.T.) 
is partially supported by the Grants-in-Aid of the Scientific Research 
No.18540278 from the Ministry of Education, Culture, Sports, Science and 
Technology in Japan.

\appendix
\section{The orthogonal set for the $su(3)$ algebra}

In this Appendix, we will summarize the orthogonal set for the $su(3)$ algebra 
in the form suitable in the treatment of \S\S 3 and 4. 
The present $su(3)$ generators and its Casimir operator are shown in 
the relations (\ref{2-6}) and (\ref{2-7}), respectively. 
In order to clarify the tensor properties of these generators, 
we introduce new notations: 
\begin{eqnarray}\label{a1}
& &{\hat Q}_{-\frac{1}{2}}^*=-{\hat S}_1^2 \ , \qquad
{\hat Q}_{\frac{1}{2}}^*={\hat S}_1^3 \ , \qquad
{\hat Q}_{-\frac{1}{2}}=-{\hat S}_2^1 \ , \qquad
{\hat Q}_{\frac{1}{2}}={\hat S}_3^1 \ , 
\nonumber\\
& &{\hat R}_+={\hat S}_2^3 \ , \qquad
{\hat R}_-={\hat S}_3^2 \ , \qquad
{\hat R}_0=\frac{1}{2}({\hat S}_2^2-{\hat S}_3^3) \ , 
\nonumber\\
& &{\hat Q}_0={\hat S}_1^1-\frac{1}{2}({\hat S}_2^2+{\hat S}_3^3) \ . 
\end{eqnarray}
It should be noted that $({\hat R}_+,\ {\hat R}_-,\ {\hat R}_0)$ 
forms the $su(2)$ algebra and 
$({\hat Q}_{-\frac{1}{2}}^*,\ {\hat Q}_{\frac{1}{2}}^*)$ indicates the tensor 
operator with rank being 1/2 and $z$-component $(-1/2, \ 1/2)$.

The minimum weight state $\ket{\lambda,\kappa,\alpha}$ obeys the 
conditions 
\begin{eqnarray}
& &{\hat Q}_{-\frac{1}{2}}\ket{\lambda,\kappa,\alpha}
={\hat Q}_{\frac{1}{2}}\ket{\lambda,\kappa,\alpha}
={\hat R}_-\ket{\lambda,\kappa,\alpha}=0 \ , 
\label{a2}\\
& &{\hat R}_0\ket{\lambda,\kappa,\alpha}
=-\lambda\ket{\lambda,\kappa,\alpha} \ , \qquad
{\hat Q}_0\ket{\lambda,\kappa,\alpha}
=-\kappa\ket{\lambda,\kappa,\alpha} \ , 
\nonumber\\
& &\qquad\qquad
\lambda=0,\ \frac{1}{2},\ 1, \cdots ,\qquad
\alpha: \hbox{\rm a\ set\ of\ extra\ quantum\ numbers}. 
\label{a3}
\end{eqnarray}
The eigenvalue $-\kappa$ is given in individual case.

The state $\ket{\lambda,\kappa,\alpha}$ gives us 
\begin{eqnarray}\label{a4}
& &\ket{\lambda\lambda_0,\kappa,\alpha}=
\sqrt{\frac{(\lambda-\lambda_0)!}{(2\lambda)!(\lambda+\lambda_0)!}}
({\hat R}_+)^{\lambda+\lambda_0}\ket{\lambda,\kappa,\alpha} \ , 
\nonumber\\
& &{\hat R}_0\ket{\lambda\lambda_0,\kappa,\alpha}
=\lambda_0\ket{\lambda\lambda_0,\kappa,\alpha} \ , \qquad
\lambda_0=-\lambda,\ -\lambda+1,\cdots ,\lambda-1,\ \lambda\ .
\end{eqnarray}
With the use of the relations 
$[{\hat Q}_{-\frac{1}{2}}\ ,\ {\hat R}_+]=0$ and 
$[{\hat Q}_{\frac{1}{2}}\ ,\ {\hat R}_+]={\hat Q}_{-\frac{1}{2}}$ 
and the condition (\ref{a2}), we have 
\begin{equation}\label{a5}
{\hat Q}_{-\frac{1}{2}}\ket{\lambda\lambda_0,\kappa,\alpha}
={\hat Q}_{\frac{1}{2}}\ket{\lambda\lambda_0,\kappa,\alpha}=0 \ .
\end{equation}
The relation (\ref{a5}) teaches us that the state 
$\ket{\lambda\lambda_0,\kappa,\alpha}$ plays a role of the vacuum 
for $({\hat Q}_{-\frac{1}{2}}^* , {\hat Q}_{\frac{1}{2}}^*)$. 
We can construct the tensor operator with rank being $\mu$ in the form 
\begin{equation}\label{a6}
{\hat Q}_{\mu\mu_0}^*=
\sqrt{\frac{(2\mu)!}{(\mu+\mu_0)!(\mu-\mu_0)!}}({\hat Q}_{\frac{1}{2}}^*)^{
\mu+\mu_0}({\hat Q}_{-\frac{1}{2}}^*)^{\mu-\mu_0} \ .
\end{equation}
Of course, ${\hat Q}_{\mu\mu_0}\ket{\lambda\lambda_0,\kappa,\alpha}=0$. 
With the use of the operator (\ref{a6}), we define the state 
\begin{eqnarray}
& &\ket{\lambda\mu\nu\nu_0,\kappa,\alpha}
={\hat Q}_+(\lambda\mu\nu\nu_0)\ket{\lambda,\kappa,\alpha} \ , 
\label{a7}\\
& &{\hat Q}_+(\lambda\mu\nu\nu_0)
=\sum_{\lambda_0\mu_0}
\langle \lambda\lambda_0\mu\mu_0\ket{\nu\nu_0}
\sqrt{\frac{(\lambda-\lambda_0)!}{(2\lambda)!(\lambda+\lambda_0)!}}
{\hat Q}_{\mu\mu_0}^*({\hat R}_+)^{\lambda+\lambda_0} \ .
\label{a8}
\end{eqnarray}
The state (\ref{a7}) satisfies 
\begin{eqnarray}
& &{\hat {\mib R}}^2\ket{\lambda\mu\nu\nu_0,\kappa,\alpha}
=\nu(\nu+1)\ket{\lambda\mu\nu\nu_0,\kappa,\alpha} \ , 
\label{a9}\\
& &{\hat {R}}_0\ket{\lambda\mu\nu\nu_0,\kappa,\alpha}
=\nu_0\ket{\lambda\mu\nu\nu_0,\kappa,\alpha} \ , 
\label{a10}\\
& &{\hat {\mib Q}}^2\ket{\lambda\mu\nu\nu_0,\kappa,\alpha}
=\left(2\lambda(\lambda+1)+\frac{2}{3}\kappa(\kappa+3)\right)
\ket{\lambda\mu\nu\nu_0,\kappa,\alpha} \ , 
\label{a11}\\
& &{\hat {Q}}_0\ket{\lambda\mu\nu\nu_0,\kappa,\alpha}
=(3\mu-\kappa)\ket{\lambda\mu\nu\nu_0,\kappa,\alpha} \ . 
\label{a12}
\end{eqnarray}
Here, ${\hat {\mib R}}^2$ and ${\hat {\mib Q}}^2$ denote the Casimir 
operators of the $su(2)$ and the $su(3)$ algebras. 
The relations (\ref{a9})$\sim$(\ref{a11}) may be trivial and for obtaining 
the relation (\ref{a12}), the following formula is useful: 
\begin{equation}\label{a13}
[\ {\hat Q}_0\ , \ {\hat Q}_\mu^*\ ]
=\frac{3}{2}{\hat Q}_\mu^* \ , \quad
(\mu=\pm \frac{1}{2}) \ , \qquad
[\ {\hat Q}_0\ , \ {\hat R}_+\ ]=0 \ . 
\end{equation}
The quantities $\kappa$ and $\alpha$ depend on the individual case. 
If the Hamiltonian under consideration commutes with ${\hat S}_j^i$, 
the energy eigenvalue problem may be enough to consider in the frame of 
the state $\ket{\lambda,\kappa,\alpha}$.

\section{The proof of the relation (\ref{3-17})}

The state $\dket{\lambda\mu\nu\nu_0,\tau\tau_0}$ can be rewritten in 
the form 
\begin{eqnarray}
& &\dket{\lambda\mu\nu\nu_0,\tau\tau_0}
=({\hat \tau}_+)^{\tau_0-\tau}{\hat Q}_+(\lambda\mu\nu\nu_0)\dket{\lambda,\tau}
\ , 
\label{b1}\\
& &{\hat Q}_+(\lambda\mu\nu\nu_0)\dket{\lambda,\tau} \nonumber\\
&=&
\sum_{\lambda_0\mu_0}(-)^{\mu+\mu_0}
\langle \lambda\lambda_0\mu\mu_0 \ket{\nu\nu_0}
\sqrt{\frac{(2\lambda)!}{(\lambda+\lambda_0)!(\lambda-\lambda_0)!}}
\sqrt{\frac{(2\mu)!}{(\mu+\mu_0)!(\mu-\mu_0)!}}
\nonumber\\
& &\times \sum_{kl}(-)^{k+l}k!l! 
\left(\begin{array}{@{\,}c@{\,}}
\lambda+\lambda_0 \\
l
\end{array}\right)
\left(\begin{array}{@{\,}c@{\,}}
\lambda-\lambda_0 \\
k
\end{array}\right)
\left(\begin{array}{@{\,}c@{\,}}
\mu+\mu_0 \\
k
\end{array}\right)
\left(\begin{array}{@{\,}c@{\,}}
\mu-\mu_0 \\
l
\end{array}\right)
\nonumber\\
& &\times 
({\hat a}_1^*)^{k+l}({\hat a}_2^*)^{\lambda+\lambda_0-l}
({\hat a}_3^*)^{\lambda-\lambda_0-k}({\hat b}_3^*)^{\mu+\mu_0-k}
({\hat b}_2^*)^{\mu-\mu_0-l}({\hat b}_1^*)^{(2\tau-3)-2(\lambda+\mu)+(k+l)}
\ket{0} \ . \nonumber\\
& &
\label{b2}
\end{eqnarray}
In the above expression, the exponents of ${\hat a}_2^*$, etc. should be 
positive. From this condition, we have the following inequalities: 
\begin{eqnarray}
& &k \leq \lambda-\lambda_0 \ , \qquad l\leq \mu-\mu_0 \ , 
\qquad 
k\leq \mu+\mu_0 \ , \qquad
l \leq \lambda+\lambda_0 \ , 
\label{b3}\\
& & 2(\lambda +\mu)-(2\tau -3)\leq k+l \ . 
\label{b4}
\end{eqnarray}
The relation (\ref{b3}) leads us to 
\b\label{b5}
k+l \leq (\lambda+\mu)-|\nu_0| \ . \qquad
(\nu_0=\lambda_0+\mu_0)
\end{equation}
Therefore, the term $(2(\lambda+\mu)-(2\tau-3))$ in the relation (\ref{b4}) 
is smaller than $((\lambda+\mu)-|\nu_0|)$: 
\b\label{b6}
2(\lambda+\mu)-(2\tau-3) \leq 
(\lambda+\mu)-|\nu_0| \ , \quad
{\rm i.e.,} \quad
\lambda+\mu+|\nu_0| \leq 2\tau-3 \ . 
\end{equation}
The maximum value of $|\nu_0|$ is equal to $(\lambda+\mu)$, and then, 
by replacing $|\nu_0|$ in the relation (\ref{b6}) with $(\lambda+\mu)$, 
we obtain 
\b\label{b7}
2(\lambda+\mu) \leq 2\tau-3 \ .
\end{equation}

\end{document}